\documentclass{jfm}
\usepackage{rotating,graphicx}
\usepackage{graphicx}
\usepackage{epstopdf, epsfig}

\usepackage{amsmath}
\usepackage{extpfeil}
\usepackage{amssymb}
\usepackage{lscape}
\usepackage{graphicx}

\usepackage[utf8]{inputenc}
\usepackage[T1]{fontenc}
\usepackage{mathptmx}
\usepackage{soul}
\usepackage{mathtools} 

\usepackage{microtype}

\usepackage{booktabs} % To thicken table lines

\usepackage{graphicx}% Include figure files
\usepackage{caption}
\usepackage{hyperref}
\usepackage{subcaption}
\usepackage{bm} % in preamble
\usepackage{algpseudocode}
\usepackage{algorithm}
\usepackage{amsmath}
\usepackage{bm} % for bold math symbols
\let\Psi\varPsi
\usepackage{placeins}
\usepackage{listings}
\usepackage[dvipsnames]{xcolor}
\usepackage{soul}
\usepackage{placeins} % To sue floatBarrier

\usepackage[toc,page]{appendix}

\usepackage{cleveref}
\newcommand\myshade{85}
\colorlet{mylinkcolor}{violet}
\colorlet{mycitecolor}{Aquamarine}
\colorlet{myurlcolor}{Aquamarine}

\hypersetup{
  linkcolor  = mylinkcolor!\myshade!black,
  citecolor  = mycitecolor!\myshade!black,
  urlcolor   = myurlcolor!\myshade!black,
  colorlinks = true,
}

\definecolor{codegreen}{rgb}{0,0.6,0}
\definecolor{codegray}{rgb}{0.5,0.5,0.5}
\definecolor{codepurple}{rgb}{0.58,0,0.82}
\definecolor{backcolour}{rgb}{0.95,0.95,0.92}

\lstdefinestyle{mystyle}{
    backgroundcolor=\color{backcolour},   
    commentstyle=\color{codegreen},
    keywordstyle=\color{magenta},
    numberstyle=\tiny\color{codegray},
    stringstyle=\color{codepurple},
    basicstyle=\ttfamily\footnotesize,
    breakatwhitespace=false,         
    breaklines=true,                 
    captionpos=b,                    
    keepspaces=true,                 
    numbers=left,                    
    numbersep=5pt,                  
    showspaces=false,                
    showstringspaces=false,
    showtabs=false,                  
    tabsize=2
}

\newcommand\R{\mbox{\textit{Re}}}

\newcommand\Ca{\mbox{\textit{Ca}}}
\newcommand\We{\mbox{\textit{$Ca^{-1}$}}}
\newcommand\Ka{\mbox{\textit{Ka}}}

\usepackage{amsmath}
\usepackage{comment}

\usepackage[left]{lineno}
\shorttitle{Linear feedback control of liquid film on a moving substrate via free-surface stresses}
\shortauthor{F. Pino, B. Scheid, M.A. Mendez and D.T.
Papageorgiou}

\title{Linear feedback control of liquid film on moving substrate via free-surface stresses}

\author{Fabio Pino\aff{1,2,5}
  \corresp{Currently at Department of Applied Mathematics and Theoretical Physics (DAMTP), University of Cambridge, Cambridge, United Kingdom, \email{fp448@cam.ac.uk}},
  Benoit Scheid \aff{2},
 Miguel A. Mendez\aff{1,3,4} \\\and 
Demetrios T.
Papageorgiou\aff{5}}
 
\affiliation{\aff{1}von Karman Institute for Fluid Dynamics, EA Department, Sint Genesius Rode, Belgium
\aff{2} Transfers, Interfaces and Processes (TIPs), Université libre de Bruxelles, Brussels, Belgium
\aff{3} Aero-Thermo-Mechanics Laboratory, Université Libre de Bruxelles, Elsene, Brussels, Belgium
\aff{4} Aerospace Engineering Research Group, Universidad Carlos III de Madrid, Leganés, Spain
\aff{5} Department of Mathematics, Imperial College London, London, United Kingdom
}

\begin{document}

\maketitle

\begin{abstract} % 250 words max
Liquid films on moving substrates are used in dip-coating processes to form uniform protective layers. Controlling free-surface waves is essential due to the film's inherent linear instability. Therefore, we develop a linear feedback controller to regulate the film toward a desired flat state by modulating the free-surface shear and pressure, with feedback gains derived analytically from linearised equations. Control performance is assessed for finite-amplitude waves using a Weighted Integral Boundary-Layer (WIBL) model at reduced Reynolds number $\delta = 8$. We identify parameter regimes in which pressure feedback is linearly destabilising while shear is stabilising, and vice versa, with the control mechanisms determined by the balance between the kinematic and dynamic wave velocities. Both stabilising and destabilising combinations of feedback coefficients can drive finite-amplitude waves toward the flat state $\bar{h}=1.1$ in finite time. In pressure-unstable regimes, the control induces a limit-cycle behaviour, in which long waves decay slowly due to the interplay between thickness and slope terms. The travelling-wave solution, although it decays slowly, moves against gravity, whereas other combinations reduce the wave amplitude in the direction of uncontrolled propagation. These results provide a foundation for higher-Reynolds-number studies and the design of industrially feasible actuator layouts.
\end{abstract}

% TODO Before submission, we will have to remove these
\begin{keywords}
liquid film, linear stability analysis, feedback control, dip-coating
\end{keywords}

\section{Introduction}
Controlling the free surface of a liquid film toward a prescribed flat state is a central challenge in dip-coating processes, where achieving a uniform coating thickness is a primary quality requirement. In a typical dip-coating configuration, a solid substrate is immersed in a liquid bath and subsequently withdrawn vertically at a constant speed \citep{weinstein2004coating,jose2020advances}. As the substrate moves upward, it entrains a thin layer of liquid that later solidifies to form a protective or functional coating \citep{scriven1988physics,landau1988dragging}.

The formed liquid films admit different nondimensional flat-film solutions $\bar{h}$, which comprise thin films ($\bar{h}\leq 1$) and thick films ($1<\bar{h}<\sqrt{3}$) \citep{wilson1982drag,snoeijer2008thick}. Linear stability analyses have demonstrated that both branches are unstable for all Reynolds numbers \citep{pino2024linear,tu1986stability,Gosset2007-ew}, with two-dimensional streamwise perturbations typically dominating over three-dimensional modes \citep{BarreiroVillaverde2023a}. The growth of 2D perturbations degrades the surface smoothness and ultimately compromises coating quality. This motivates the development of feedback control strategies that regulate the free surface to a desired flat state.

For falling-film flows, a well-established approach to feedback control design is based on pole-placement techniques \citep[Chapter~5]{franklin2002feedback}. This method involves designing the feedback coefficient to ensure that the eigenvalues of any flat-film solutions are stable. In this context, \citet{armaou2000feedback} developed a stability-based feedback control law for the Kuramoto--Sivashinsky (KS) equation and demonstrated stabilisation of the flat-film solution against perturbations of arbitrary wavelength. Building on the KS framework, \citet{gomes2017stabilizing} extended this approach to the control of non-trivial film states, including travelling waves, using a finite number of point actuators. Moving to Benney and weighted integral boundary-layer (WIBL) models, \citet{thompson2016stabilising} investigated mass-conserving blowing and suction actuation at the solid substrate. Their linear analysis showed that even simple proportional feedback can stabilise the flat state against finite-amplitude disturbances. More recently, \citet{holroyd2023linear} further generalised these ideas by solving a linear-quadratic regulator (LQR) problem with continuous actuation, enabling the stabilisation of both flat-film and prescribed travelling-wave solutions.

Despite these results, actuation strategies based on mass injection at the solid boundary are not well-suited to industrial applications. A less intrusive yet largely unexplored approach is to modulate the shear and pressure distributions at the free surface by imposing a gas flow. \citet{samanta2014shear} showed that a uniform interfacial shear stress may either stabilise or destabilise a liquid film, depending on its orientation relative to the base flow. In confined laminar settings, \citet{lavalle2019suppression} demonstrated that interfacial shear couples directly to film displacements, whereas in turbulent gas flows, \citet{tseluiko2011nonlinear} argued that gas-induced shear and pressure perturbations depend linearly on small-amplitude interfacial deformations. Taken together, these studies highlight the potential of a closed-loop control strategy in which free-surface stresses are coupled to local variations in film thickness relative to a prescribed base state.

In this work, we investigate the design of stability-based linear feedback control laws for a two-dimensional liquid film on a vertically moving substrate by modulating the shear and pressure distributions at the free surface. The film dynamics are modelled using a newly derived Benney (BE) equation and the Weighted Integral Boundary Layer (WIBL) model of \citep{mendez2021dynamics}, which accounts for imposed free-surface stresses (see §\ref{sec:problem_description}). Shear and pressure are assumed to be independent, and each is modelled as a linear function of the film-thickness deviation from the flat-film solution. The feedback regulator is designed based on analytical stability characteristics of the governing equations.

The proposed feedback laws are tested to suppress growing finite-amplitude perturbations and to drive the film toward a prescribed flat state with thickness $\bar{h}=1.1$. Attention is restricted to the thick-film regime, which is known to exhibit stronger linear instabilities and therefore presents a more demanding control problem. Feedback gains are selected to stabilise all eigenvalues of the linearised system. Control design is carried out for both reduced-order models and the full Navier–Stokes equations, the latter requiring the solution of the generalised Orr–Sommerfeld eigenvalue problem using a Chebyshev–Tau spectral method \citep[Chapter~VII]{johnson1996chebyshev,lanczos1988applied}. The regions of linear stability and instability of the feedback coefficients are identified. The stabilisation mechanisms are interpreted using wave-hierarchy arguments \citep[Chapter~10]{whitham}. Finally, the effectiveness of the proposed control strategies is assessed through nonlinear simulations of finite-amplitude wave suppression at reduced Reynolds $\delta=8$ and $\bar{h}=1.1$ in a periodic domain, using the WIBL model with a Fourier pseudo-spectral implementation \citep{Fornberg_1996} (§\ref{sec:numerical_sims}). Results are presented in §\ref{sec:results}, followed by conclusions and perspectives in §\ref{sec:conclusion}.

\section{Problem Description}
\label{sec:problem_description}
\begin{figure}
    \centering
    \includegraphics[height=6.5cm]{Fig1.pdf}
    \caption{Scheme of a wavy liquid film with thickness $h(x,t)$, developing over a substrate moving against gravity $g$ at constant speed $U_p$ with imposed shear $\tau_g$ and pressure $p_g$ distribution at the free surface.}
    \label{fig:scheme_white_box_control}
\end{figure}
Figure~\ref{fig:scheme_white_box_control} shows the 2D liquid film with thickness $h(x,t)$ flowing over a flat solid substrate, which moves against gravity $g$ at constant speed $U_p$. The liquid film has a density $\rho$, dynamic viscosity $\mu$, kinematic viscosity $\nu$, and surface tension $\sigma$. The liquid film is in contact with air, which is considered to have a density $\rho_g$ and dynamic viscosity $\mu_g$. 

We consider a Cartesian reference system $\mathcal{O}(x, y, z)$ centred on the substrate, with the $x$-axis aligned with gravity $g$ oriented downwards and the $y$-axis normal to the substrate and oriented toward the liquid film’s free surface. At the free surface ($y=h$), we define a local orthogonal reference system $\mathcal{O}(\bm{n}, \bm{t})$ composed of a normal vector $\bm{n}$ and a tangential vector $\bm{t}$, given by: 
\begin{equation} \label{eq:local_ref_frame} 
\bm{n} = \frac{(-\partial_x h, 1)^T}{\sqrt{(\partial_x h)^2  + 1}}, 
\qquad\qquad\qquad\qquad \bm{t} = \frac{(1, \partial_x h)^T}{\sqrt{(\partial_x h)^2+1 }}.
\end{equation} 
with $\partial_xh = \partial h/\partial x$.

The liquid film is characterized by a velocity field $\bm{u} = (u, v)^T$, a pressure field $p(x, y, t)$ with the flow rate $q(x,t)$ and a in-depth averaged kinetic energy $e(x,t)$ defined as:
\begin{equation}
\label{eq:def_flow_rate_kin_en}
    q(x,t) = \int_{0}^{h(x,t)}\, u(x,y,t)\, dy, \qquad\qquad e(x,t)=\int_{0}^{h(x,t)}\, \frac{1}{2}\rho u(x,y,t)^2\,dy.
\end{equation}

At the free surface, the liquid film is subjected to externally imposed pressure $p_g(x,t)$ and shear stress $\tau_g(x,t)$ distributions. We assume the film dynamics are one-way coupled to the air, meaning that the imposed stresses are independent of the film evolution unless explicitly specified by a feedback control law.

% ------------- Mathematical Formulation -------------  
\subsection{Scaling quantities and nondimensional groups}
\label{sec:scaling}
Following the scaling introduced by \citet{mendez2021dynamics} for coating flows, the reference velocity $u_{ref}$ and film thickness $h_{ref}$ are defined as:
\begin{equation}
\label{eq:ref_quantity_0}
    u_{ref} = U_p, \qquad\qquad\qquad\qquad h_{ref} = \sqrt{\frac{\nu U_p}{g}}\,,
\end{equation}
where the subscript $ref$ denotes reference quantities.

Based on \eqref{eq:ref_quantity_0}, the remaining dependent and independent variables are scaled accordingly:
\begin{subequations}
\label{eq:ref_quantities}
\begin{equation}
    (u,v)=u_{ref}\,(\hat{u},\hat{v}), \quad\quad (x,y)= h_{ref}\,(\hat{x},\hat{y}), \quad\quad h=h_{ref}\,\hat{h},
\end{equation}
\begin{equation}
    t=(h_{ref}/u_{ref})\,\hat{t},\quad\quad p= p_{\infty} + (\rho g h_{ref})\,\hat{p}, \quad\quad q=(u_{\rm ref}h_{\rm ref})\,\hat{q}\quad\quad e = (\rho u_{ref}^2)\,\hat{e},
\end{equation}
\end{subequations}
where the hat $\hat{\bullet}$ denotes the non-dimensional quantities and $p_{\infty}$ the atmospheric pressure. To avoid cumbersome notation, we do not use $\hat{\cdot}$ in the subscript of the partial derivative for the nondimensional quantities.

To derive the reduced order models in Subsection \ref{subsec:reduced_order_models}, we introduce an additional slow time and space scales $\hat{X}$ and $\hat{T}$, given by:
\begin{equation}
\label{eq:slow_scales}
\hat{X}=\varepsilon\,\hat{x},\qquad\qquad\qquad\qquad\hat{T}=\varepsilon\,\hat{t},
\end{equation}
where $\varepsilon\ll 1$. To be consistent   with the continuity equations at $O(\varepsilon)$, the wall normal velocity component $\hat{v}$ is expressed as:
\begin{equation}
\label{eq:v_wall_normal_slow}
    \hat{v}=\varepsilon\,\hat{V}.
\end{equation}

Based on the reference quantities \eqref{eq:ref_quantity_0}, \eqref{eq:ref_quantities} and $\varepsilon$, the Reynolds number $\R$ and the reduced Reynolds number $\delta$ are defined as:
\begin{equation}
    \R = \frac{u_{ref}h_{ref}}{\nu} = \sqrt{\frac{U_p^3}{g\nu}}= \sqrt{(\Ka\,\Ca)^3}\,,\qquad\qquad\qquad\qquad \delta=\varepsilon\R,
\end{equation}
where $\Ca$ is the Capillary number and $\Ka$ is the Kapitza number, expressed as:
\begin{equation}
    \Ca = \frac{U_p \mu}{\sigma},\,\qquad\qquad\qquad\qquad \Ka = \frac{\sigma}{\rho g^{1/3} \nu^{4/3}}.
\end{equation}

In the following, we are going to refer to the scaling presented in this section as the \textit{Nusselt-like scaling}, while the scaling involving the slowly varying variables \eqref{eq:slow_scales} with the wall-normal velocity \eqref{eq:v_wall_normal_slow} as \textit{slow scaling}.

% Governing Equations
\subsection{Governing Equations}
\label{sec_apx:governing_eq}
The liquid film is governed by the 2D continuity and Navier-Stokes equations, which in nondimensional form using the scaling quantities \eqref{eq:ref_quantity_0} and \eqref{eq:ref_quantities}, read:
\begin{subequations}
\label{eq:nondim_gov_eq}
\begin{equation}
\label{eq:cont_adim}
    \nabla\cdot\hat{\bm{u}} = 0
\end{equation}
\begin{equation}
\label{eq:Navier-Stokes} 
\R(\partial_t\hat{\bm{u}} + \hat{\bm{u}}\cdot\nabla\hat{\bm{u}}) = -\nabla\hat{p} + \nabla^2\hat{\bm{u}} + \hat{\bm{g}}  
\end{equation}
\end{subequations}
where the nondimensional gravitational acceleration vector $\hat{\bm{g}}$ is given by:
\begin{equation}
    \hat{\bm{g}} = (1,0) ^T.
\end{equation}

At the substrate ($\hat{y}=0$), the no-slip boundary condition reads:
\begin{equation}
\label{eq:bc_strip}
\hat{\bm{u}}|_0 = (-1,0)^T.
\end{equation}

At the free surface ($\hat{y}=\hat{h}$), the kinematic boundary condition reads:
\begin{equation}
\label{eq:bc-free_sruface_0}
	\hat{v} = \partial_t\hat{h} + \hat{u}\partial_x\hat{h},
\end{equation}

and the normal and tangential stress balance conditions read:
\begin{equation}
\label{eq:bc-free_sruface_stress_orojected}
    \hat{p}_g-\hat{p} + (2\,\mathsfbi{\hat{E}}\cdot\bm{\hat{n}})\cdot\bm{\hat{n}} =  \frac{\We\partial_{xx}\hat{h}}{(1 + (\partial_x\hat{h}))^{3/2}},\qquad\qquad
    (2\,\mathsfbi{\hat{E}}\cdot\bm{\hat{n}})\cdot\bm{\hat{t}} = \hat{\tau}_{g}, 
\end{equation}
where $\mathsfbi{\hat{E}}=(1/2)(\nabla\bm{\hat{u}}+(\nabla\bm{\hat{u}})^T)$ is the nondimensional rate-of-strain tensor and $\hat{p}_g$ and $\hat{\tau}_{g}$ are the controlled free-surface stresses.

% Steady-state solution
\subsection{Steady-state solution}
\label{subsec:steady_state}
The governing equations in Subsection \ref{sec_apx:governing_eq} admit a steady-state solution given by a flat interface $\overline{h}$ with velocity $(\bar{u},\bar{v})^T$ and pressure $\bar{p}$ fields reading:
\begin{equation}
\label{eq:base_state}
\bar{u}(\hat{y}) = -\frac{1}{2}\hat{y}^2 + (\hat{\tau}_g+\Bar{h})\hat{y} - 1,\qquad\qquad
\bar{v}(\hat{y}) = 0,\qquad\qquad
\bar{p}(\hat{y}) = 0,
\end{equation}
where $\Bar{\cdot}$ denotes the base state quantities. 

Integrating the streamwise velocity profile $\bar{u}$ over the film thickness and using the flow rate definition \eqref{eq:def_flow_rate_kin_en} gives a non-monotonic relation between $\bar{q}$ and $\bar{h}$, which reads:
\begin{equation}
\label{eq:steady_state_sols}
    \bar{q} = \frac{\bar{h}^3}{3} + \hat{\tau}_g\frac{\bar{h}^2}{2} - \bar{h}.
\end{equation}

Considering the case without imposed shear-stress $\hat{\tau}_g=0$, the relation \eqref{eq:steady_state_sols} entails thin film solutions for $\bar{h} \leq 1$ and thick film solutions $\bar{h} > 1$. The condition $\bar{h}=1$ corresponds to the maximum thickness attainable with simply extracting the substrate from the bath and is called Derjaguin's flat film solution \citep{derjaguin1993thickness}. The solution for $\bar{h}=\sqrt{3}$ defines the limit above which the flow rate becomes positive, and the liquid film enters the falling film regime. 

In our investigation, we are interested in controlling the wavy liquid film towards a thick flat-film solution $\hat{h}=1.1$ with $\hat{\tau}_g=0$ as it will be presented in subsection \ref{subsec:test_case_des}.

% Derivation of the reduced order models
\section{Description Reduced order models}
\label{subsec:reduced_order_models}

In the derivation of the Benney and WIBL reduced order models, we make the following assumptions:
\begin{equation}
\label{eq:ass_ROM}
    \hat{p}_g = \varepsilon^{-1}\hat{P}_g,\qquad\qquad \We = O(\varepsilon^3),\qquad\qquad \delta=O(1),\qquad\qquad \partial_T\hat{\tau}_g = O(\varepsilon^2).
\end{equation}

The derivation of the reduced order models rests on the first order boundary layer equations obtained using the \textit{slow scaling} \eqref{eq:slow_scales} with \eqref{eq:v_wall_normal_slow} and retaining terms up to $O(\varepsilon)$.

The evolution of the film thickness $\hat{h}$ is given by the integral continuity equation, obtained by integrating the continuity equation over the film thickness and using the kinematic boundary condition at the free surface. The equation reads:
\begin{equation}
\label{eq:integral_continuity}
    \partial_{T}\hat{h} + \partial_{X}\hat{q} = 0.
\end{equation}

The difference between the Benney and WIBL models resides in how they represent the flow rate $\hat{q}$. In the Benney model, $\hat{q}$ is enslaved to $\hat{h}$ via an analytical expression, whereas in the WIBL formulation, $\hat{q}$ is treated as an independent variable determined by solving its evolution equation.
% -----------------
%       Benney equation
% -----------------
\subsection{Benney type Equation}
\label{subsubsection:Benney_equation}
The slaving relation between $\hat{q}$ and $\hat{h}$ is obtained by solving the boundary layer equations with a gradient expansion in $\varepsilon$ and then integrating the velocity profile to obtain the flow rate at various levels of approximation. 

The velocity components $\hat{u}$ and $\hat{v}$ are approximated in terms of $\varepsilon$ as:
\begin{equation}
    \label{eq:gradient_exp}
    \hat{u} \approx \hat{u}^{(0)} + \varepsilon\hat{u}^{(1)}, \qquad\qquad\qquad \hat{V} \approx \hat{V}^{(0)} + \varepsilon\hat{V}^{(1)}.
\end{equation}

Inserting \eqref{eq:gradient_exp} in the boundary layer equations, solving at various orders and integrating the streamwise velocity gives the leading order $\hat{q}^{0}$ and the first order $\hat{q}^{1}$ flow rate solutions, which read:
\begin{subequations}
\label{eq:q_grad_expansion_Benney}
\begin{align}
    \hat{q}^{(0)} \coloneqq& \int_{0}^{\hat{h}}\,\hat{u}^{(0)}\,d\hat{y} 
    =\frac{1}{3}\hat{h}^3 + \hat{\tau}_g\frac{1}{2}\hat{h}^2 - \hat{h}  + \frac{\hat{h}^3\partial_{XXX}\hat{h}}{3}
    - \frac{1}{3}\hat{h}^3\partial_X\hat{P}_g, \label{eq:q0_benney}\\
    \hat{q}^{(1)} \coloneqq& \int_{0}^{\hat{h}}\,\hat{u}^{(1)}\,d\hat{y} = \R\,\hat{h}^4\left[\frac{2}{15}\hat{h}\hat{\tau}_g\partial_X\hat{h}
    + \frac{2}{15}\hat{h}^2\partial_X\hat{h}
     + \partial_X\hat{\tau}_g\left(\frac{7\hat{h}^2}{240}+
    - \frac{3}{40}\hat{h} \hat{\tau}_g + \frac{5}{24}\right)\right].\label{eq:q1_benney}
\end{align}
\end{subequations}

Inserting the flow rates relations \eqref{eq:q_grad_expansion_Benney} in the integral continuity equation \eqref{eq:integral_continuity} yields the Benney-type equation for the evolution of the film thickness $\hat{h}$.
% -----------------
%       WIBL model
% -----------------
\subsection{Weighted Integral Boundary Layer (WIBL) model}
\label{subsubsection:WIBL}
The first-order WIBL equations, derived by \citet{mendez2021dynamics}, consist of the integral continuity equation \eqref{eq:integral_continuity} and an evolution equation for the flow rate $\hat{q}$. The latter is obtained by combining a leading-order gradient expansion of the velocity field in the small parameter $\varepsilon$ with a Galerkin projection.

The resulting evolution equation for the flow rate $\hat{q}$ reads:
\begin{equation}
\label{eq:WIBL_qt}
    \partial_T\hat{q} = \mathcal{H}_{\text{WIBL}}(\hat{h},\hat{q},\delta,\hat{\tau}_g,\hat{P}_g),
\end{equation}
where the nonlinear operator $\mathcal{H}_{\text{WIBL}}(\hat{h},\hat{q},\delta,\hat{\tau}_g,\hat{P}_g)$ is given by:
\begin{equation}
\begin{aligned}
\label{eq:WIBL_operator_momentum}
    \mathcal{H}_{\text{WIBL}}(\hat{h},\hat{q},\delta,\hat{\tau}_g,\hat{p}_g) = F_{ine} + F_{grv} + F_{pgr} + F_{shr} + F_{dif},
\end{aligned}
\end{equation}
with the different terms reading:
\begin{align}
\label{eq:terms_WIBL}
    F_{ine}\,\coloneq&\,\hat{\tau}_g\partial_X\hat{h}\Big(-\frac{5\hat{q}}{112} -\frac{19\hat{\tau}_g\partial_X\hat{h}}{672} -\frac{17}{168}\Big) 
    -\frac{19\hat{h}\hat{\tau}_g\partial_X\hat{q}}{336} -\frac{15\hat{h}\hat{q} \partial_X\hat{\tau}_g}{224} -\frac{25\hat{h}^3\hat{\tau}_g \partial_X\hat{\tau}_g}{1344}+\\& -\frac{31\hat{h}^2\partial_X\hat{\tau}_g}{672} + \frac{9\hat{q}^2\partial_X\hat{h} + \hat{h}\hat{q}\partial_X\hat{h}}{7\hat{h}^2} -\frac{\partial_X\hat{h}}{7} -\frac{17\hat{q}\partial_X\hat{q}}{7\hat{h}}-\frac{3\partial_X\hat{q}}{7},\label{eq:ine_WIBL} \\
    F_{grv}\,\coloneq&\,\frac{5\hat{h}}{6 \delta },\quad F_{pgr}\,\coloneq\, \frac{5 \hat{h}(\partial_{XXX}\hat{h}-\partial_X\hat{P}_g)}{6\delta}, \quad F_{shr}\,\coloneq\,\frac{5(\hat{h}\hat{\tau}_g-2)}{4\delta\hat{h}}, \quad F_{dif}\,\coloneq\,-\frac{5\hat{q}}{2\delta\hat{h}^2},
\end{align}
% term neglected in Fine -\frac{\hat{h}^2\partial_T\hat{\tau}_g}{48}

where $F_{\text{ine}}$ accounts for inertial effects, $F_{\text{grv}}$ for gravitational effects, and $F_{\text{pgr}}$ for the pressure gradient, which includes both the gradient of the jet pressure distribution $\partial_X \hat{P}_g$ and the surface tension contribution $\partial_{XXX} \hat{h}$. The term $F_{\text{shr}}$ represents the difference between the shear stress at the free surface and that at the substrate, while $F_{\text{dif}}$ corresponds to the diffusive effects in the wall-normal direction.

For later convenience in analysing instability mechanisms, we introduce the nondimensional, in-depth-averaged kinetic energy density, $\hat{e}(\hat{x}, \hat{t})$, in the streamwise direction, which reads:
\begin{equation}
\label{eq:kin_en_def}
    \hat{e}(\hat{x},\hat{t})= \frac{3}{5}\frac{\hat{q}^2}{\hat{h}}+\frac{1}{10}\hat{h}+\frac{1}{5}\hat{q}+\Big(\frac{1}{240} \hat{h}^3\hat{\tau}_g+\frac{1}{40}\hat{h}^2+\frac{1}{40} \hat{h}\hat{q}\Big)\hat{\tau}_g
\end{equation}
The imposed shear stress at the free surface, $\hat{\tau}_g$, has a significant impact on the average kinetic energy of the liquid film. This influence increases with the film thickness and can become the dominant contribution in thick films ($1 < \hat{h} < \sqrt{3}$).

\section{Methodology}
\label{sec:control_design}
This section presents the linearised reduced-order models (ROMs) and the Navier-Stokes equations in Subsection~\ref{subsec:design_feedback_laws_linearised}. The feedback coefficients for the free-surface stresses are designed based on the analytical solution of the linearised equations and are presented in Subsection~\ref{subsec:design_feedback_laws}. Eventually, Subsection~\ref{sec:numerical_sims} describes the numerical setup for the nonlinear control test case.

% ----- Linearised governing equations -----
\subsection{Linearised governing equations}
\label{subsec:design_feedback_laws_linearised}
The dependent variables ($\hat{u},\hat{v},\hat{p},\hat{h}$) for the Navier-Stokes equations and ($\hat{h},\hat{q}$) for the simplified models are decomposed into a steady state and small perturbation components, reading:
\begin{subequations}
\label{eq:expansion_linear_stab_1}    
\begin{equation}
    \hat{u} = \bar{u} + \zeta\tilde{u},\qquad\qquad \hat{v}= \bar{v}  + \zeta\tilde{v},\qquad\qquad \hat{p} = \bar{p}  + \zeta\tilde{p},
\end{equation}
\begin{equation}
 \hat{h} = \bar{h}  + \zeta\tilde{h}, \qquad\qquad \hat{q} = \bar{q} + \zeta\tilde{q},
\end{equation}
\end{subequations}
where $\zeta \ll 1$ and $\tilde{\bullet}$ denotes the perturbations quantities. 

The shear stress $\hat{\tau}$ and pressure $\hat{p}$ distributions at the free surface are treated as independent control inputs, each coupled to the perturbed film thickness $\tilde{h}$ via the feedback coefficients $\alpha\in\mathbb{R}$ and $\beta\in\mathbb{R}$, reading:
\begin{equation}
\label{eq:stress_order_ass}
\hat{\tau}=-\zeta\alpha\tilde{h},\qquad\qquad\qquad\qquad \hat{p}_g=-\zeta\beta\tilde{h}.
\end{equation}

The linearised equations are derived by injecting \eqref{eq:expansion_linear_stab_1} with \eqref{eq:stress_order_ass} into the governing equations and retaining terms up to $O(\zeta)$.

In the case of the Benney equation, the resulting linearised equations read:
\begin{equation}
\label{eq:lin_Benney}
\begin{gathered}
240\,\partial_T\tilde{h} + \bar{h}^2(\bar{h}\,\varepsilon\,(\bar{h}\R\,(32\,\bar{h}^2\partial_{XX}\tilde{h}+(7\,\bar{h}^2+50)\partial_{XX}\tilde{\tau}_g-50\,\partial_{XT}\tilde{\tau}_g)-80\,\partial_{XX}\tilde{p}_g)+\\+80\,\bar{h} \partial_{XXXX}\tilde{h}+120\,\partial_X\tilde{\tau}_g)+240\,(\bar{h}^2-1)\partial_X\tilde{h} = 0.
\end{gathered}
\end{equation}

The linearised WIBL equations read:
\begin{equation}
\label{eq:linear_opeartor_WIBL}
    \partial_{T}\begin{pmatrix}
           \tilde{h} \\
           \tilde{q}
         \end{pmatrix} = \mathcal{L}_{D} \begin{pmatrix}
           \tilde{h} \\
           \tilde{q} \end{pmatrix}  + \mathcal{L}_{C}  \begin{pmatrix}
          \tilde{p}_g \\
           \tilde{\tau}_g \end{pmatrix},
\end{equation}
where $\mathcal{L}_D$ is a linear operator acting component-wise on the state variables $(\tilde{h}, \tilde{q})$, and $\mathcal{L}_C$ is a linear operator acting component-wise on the control variables $(\tilde{p}_g, \tilde{\tau}_g)$, reading:
\begin{equation}
\label{eq:lin_opt_def_WIBL}
    \mathcal{L}_{D} \coloneqq \begin{pmatrix} 
           0 & -\partial_X \\
           \mathcal{L}_{D,\tilde{h}} & \mathcal{L}_{D,\tilde{q}}
            \end{pmatrix}, \qquad\qquad\qquad\qquad
    \mathcal{L}_{C} \coloneqq \begin{pmatrix} 
           0 & 0 \\
           \mathcal{L}_{C,\tilde{p}_g} & \mathcal{L}_{C,\tilde{\tau}_g}
            \end{pmatrix},
\end{equation}
with the functions $\mathcal{L}_{D,\tilde{h}},\mathcal{L}_{D,\tilde{q}},\mathcal{L}_{C,\tilde{p}_g}$ and $\mathcal{L}_{C,\tilde{\tau}_g}$ defined as:
\begin{equation}
\begin{aligned}
    \mathcal{L}_{D,\tilde{h}}(\tilde{h}) \coloneqq&\ \left(\frac{1}{7} \bar{h}^6 -\frac{17}{21} \bar{h}^4+\bar{h}^2\right) \partial_X\tilde{h}+\frac{5 \bar{h}^3\partial_{XXX}\tilde{h}}{6 \delta }+( \bar{h}^2 -1)\frac{5\tilde{h}}{2\delta}, \\
    \mathcal{L}_{D,\tilde{q}}(\tilde{q})  \coloneqq&\ -\frac{17}{21}\bar{h}^4\partial_T\tilde{q}+2 \bar{h}^2\partial_X\tilde{q}-\frac{5 \tilde{q}}{2 \delta }, \\ 
    \mathcal{L}_{C,\tilde{p}_g}(\tilde{p}_g) \coloneqq&\ -\frac{5 \bar{h}^3 \partial_X\tilde{p}_g}{6 \delta},\qquad\qquad  \mathcal{L}_{C,\tilde{\tau}_g}(\tilde{\tau}_g)  \coloneqq\ \Big(-\frac{5\bar{h}^6}{224}+\frac{\bar{h}^4}{48} \Big)\partial_X\tilde{\tau}_g+\frac{5 \bar{h}^2 \tilde{\tau}_g}{4\delta}.
\end{aligned}
\end{equation}

Stable conditions arise from the balance between the speeds of the kinematic and dynamic modes, and can be interpreted using wave-hierarchy arguments \citep[Chapter 10]{whitham}. Starting from the linearised system \eqref{eq:linear_opeartor_WIBL}, a wave equation for the small-amplitude free-surface displacement $\tilde{h}$ is obtained by differentiating the linearised momentum equation with respect to $X$. The resulting expression is then simplified by substituting spatial derivatives of the perturbation flow rate $\tilde{q}$ using the linearised continuity equation $\partial_T\tilde{h} = -\partial_X\tilde{q}$. Finally, factoring the expression leads to the following linear operator form:
\begin{equation}
\label{eq:wave_eq_WIBL_linear}
    \mathcal{E}(\tilde{h}) + \mathcal{F}(\tilde{h}) = 0, 
\end{equation}
where:
\begin{subequations}
    \begin{equation}
        \mathcal{E}(\tilde{h}) \coloneqq \Big[\partial_T + \Big(\bar{h}^2-1-\frac{\alpha\bar{h}^2}{2}\Big)\partial_X\Big]\tilde{h}
    \end{equation}
    \begin{equation}
    \begin{aligned}
        \mathcal{F}(\tilde{h}) \coloneqq \frac{2}{5}\,\delta \,\bar{h}^2\Big[\partial_{TT}+\Big(\frac{7\alpha\bar{h}^2+272 \bar{h}^2-672}{336} \Big)\partial_{XT}+\\+
\Big(\frac{3\bar{h}^4-17\bar{h}^2+21}{21}
+\frac{\bar{h}^2\alpha(15\bar{h}^2-14)}{672} + \frac{5\bar{h}(k^2+\beta)}{6\delta}\Big)\partial_{XX}\Big]\tilde{h}
    \end{aligned}
    \end{equation}
\end{subequations}

The operator $\mathcal{E}(\tilde{h})$ represents the evolution of kinematic waves, and $\mathcal{F}(\tilde{h})$ represents the evolution of dynamic waves associated with the transport of kinetic and potential energy. 
The phase speed for the kinematic $c_k$ and dynamic $c_{d\pm}$ waves are defined as:
\begin{subequations}
\label{eq:kin_dyn_waves}
\begin{equation}
\label{eq:kin_waves}
    c_k = \left(\bar{h}^2-1-\frac{\alpha\bar{h}^2}{2}\right),
\end{equation}
\begin{equation}
\label{eq:dyn_waves}
    c_{d\pm} = \frac{\alpha\bar{h}^2}{96}+\frac{17\bar{h}^2}{42}-1\pm \Big(\bar{h}^2\sqrt{\frac{\alpha^2}{9216}-\frac{\alpha}{72}+\frac{37}{1764}}+\sqrt{\frac{5\bar{h}(k^2-\beta)}{6\delta}}\Big) 
\end{equation}
\end{subequations}

Small perturbations decay over time if the speed of the kinematic wave $c_k$ lies within the range defined by the phase speeds of dynamic modes $c_{k\pm}$, as discussed in \citet[Subsection 7.1.3]{kalliadasis2011falling}. This stability condition reads:
\begin{equation}
\label{eq:stab_cond_waves}
    c_{d-}<c_k<c_{d+}.
\end{equation}

For brevity, the linearised Navier-Stokes equations are not reported; following the approach of \cite{pino2024absolute}, these equations are reformulated to remove the streamwise velocity component $\tilde{u}$ and instead expressed in terms of the streamfunction $\Psi$.

The solutions of the linearised reduced order model and Navier-Stokes equations are sought in the form of normal modes, which read:
\begin{subequations}
\label{eq:normal_modes_ROMs}    
\begin{equation}
    \Psi = \,\varphi(\hat{y}) \exp(i(k\hat{x} - \omega\hat{t})) + \text{c.c.},
\end{equation}
\begin{equation}
    \tilde{h} = \eta\exp(i(k\hat{x} - \omega\hat{t})) + \text{c.c.},\qquad\qquad \tilde{q} = \,\gamma\exp(i(k\hat{x} - \omega\hat{t})) + \text{c.c.}
\end{equation}
\end{subequations}
where $k\in\mathbb{R}$ is the wavenumber, $\omega=\omega_r + i\omega_i$ is the complex frequency with $\omega_i$ the temporal growth rate and $\varphi(\hat{y})$, $\eta$ and $\gamma$ are the complex amplitude. The phase speed is given by $c=\omega_r/k$.

Inserting \eqref{eq:normal_modes_ROMs} in the linearised equations gives a relationship between the wavenumber $k$ and the angular frequency $\omega$ known as the \textit{dispersion relation}.

The dispersion relation for the reduced-order model takes the form of an algebraic expression, which, in the case of the Benney equation, reads:
\begin{equation}
\label{eq:disp_rel_Benney}
\begin{gathered}
    k((7\alpha -32) \bar{h}^6 k\delta +50 \alpha  \bar{h}^4\delta(k+\omega)+\\+80 \bar{h}^3 (k^3-\beta k\varepsilon)-120 i (\alpha -2) \bar{h}^2-240 i)-240 i\omega =0,
\end{gathered}
\end{equation}
which has the solution:
\begin{equation}
\label{eq:sol_disp_Benney}
    \omega = \frac{k \left(\bar{h}^2 \left(\bar{h}k\left(80\beta\varepsilon +(32-7\alpha)\delta\bar{h}^3-50\alpha\delta\bar{h}-80 k^2\right)+120 i (\alpha -2)\right)+240 i\right)}{50 \alpha  \delta  \bar{h}^4k-240 i}
\end{equation}

For the WIBL equations, the dispersion relation reads:
\begin{equation}
\label{eq:disp_rel_WIBL}
\begin{aligned}
    672\delta\bar{h}^2\omega^2+2 \omega(\delta\bar{h}^2k(672-(7 \alpha +272)\bar{h}^2)+840i)+k(\bar{h}^2(840i (\alpha-2)+\\+\delta k (3 (5 \alpha +32) \bar{h}^4-2 (7 \alpha +272) \bar{h}^2+672)+560 \bar{h} k (\beta -k^2))+1680 i)=0,
\end{aligned}
\end{equation}
with the solutions:
\begin{equation}
\label{eq:roots_WIBL_linear}
\begin{aligned}
    \omega_{\pm} =& -k-\frac{5i}{4\delta\bar{h}^2}+\frac{\alpha\bar{h}^2k}{96} +\frac{17\bar{h}^2k}{42}\pm\Big(\frac{\sqrt{(49 (\alpha -128) \alpha +9472) \delta ^2 \bar{h}^8k^2}}{672 \delta  \bar{h}^2}+\\
    &+\frac{\sqrt{-100+\frac{5}{21} \delta  \bar{h}^4k \left(-343 i \alpha +224 \bar{h} \left(k^3-\beta k\right)+400 i\right)}}{8 \delta  \bar{h}^2}\Big).
\end{aligned}
\end{equation}

By setting $\mathrm{Im}\,\omega=0$ in \eqref{eq:sol_disp_Benney} and \eqref{eq:roots_WIBL_linear} and solving for $k$, we obtain the critical wavenumber $k_c$  that marks the threshold between growing and decaying perturbations, which is given by:
\begin{equation}
\label{eq:nuetral_k_WIBL}
    k_c = \sqrt{\frac{\delta\bar{h}^3 (\alpha -1)(25\,\alpha -32)}{80}+ \beta}.
\end{equation}

Turning our attention to the linearised Navier--Stokes equations, the dispersion relation is obtained by solving the \textit{Orr-Sommerfeld} (OS) generalised eigenvalue problem with eigenfunction $\varphi$ and eigenvalue $c$, which reads:
\begin{equation}
    \label{eq:Orr_Sommerfeld}
    (D^2-k^2)^2\varphi(\hat{y}) - i\R k[\Bar{u}(D^2-k^2)+1]\varphi(\hat{y}) = -ci\R k(D^2 -k^2)\varphi(\hat{y}),   
\end{equation}
where $D(\bullet)=\partial_{y}(\bullet)$ is the wall-normal differential operator.

The boundary conditions are given by the no-slip condition on the substrate
\begin{subequations}
\label{eq:bcs}
\begin{equation}
\varphi(0)=D\varphi(0)= 0,\label{eq:bcs0}\\
\end{equation}
the kinematics condition on the free surface
\begin{equation}
\eta = \varphi(\hat{h})/(c-a),\label{kinematic_con_eq}
\end{equation}
and the force balance along the normal and tangential directions at the free surface
\begin{equation}
[(D^2 - 3k^2) + i \,\R \,k (c -a)]D\varphi(\hat{h}) - i\eta k[\We k^2 - \beta ] = 0,\label{eq:normal_stress_cond}
\end{equation}
\begin{equation}
(D^2 + k^2)\varphi(\hat{h}) - (1-\alpha)\eta = 0, \label{eq:bc_OS_4}
\end{equation}
\end{subequations}
where $a = \Bar{u}(\Bar{h}) = (\Bar{h}^2/2-1)$ is the undisturbed velocity at the interface.

In the long-wave limit, an approximate solution to the eigenvalue problem is obtained by expanding $\varphi(\hat{y})$ and $c$ in power series of the wavenumber $k$, retaining terms up to $O(k^3)$, yields:
\begin{subequations}
\label{eq:expansion_OS_mod}
\begin{equation}
    \phi(\hat{y}) \approx \phi_0(\hat{y}) + \phi_1(\hat{y})k + \phi_2(\hat{y})k^2 + \phi_3(\hat{y})k^3, \qquad\qquad
    c\approx c_0 + c_1k + c_2k^2 + c_3k^3.
\end{equation}
\end{subequations}

Substituting \eqref{eq:expansion_OS_mod} into the OS problem and solving at the various orders in $k$ yields:
\begin{equation}
\label{eq:expansion_OS_mod_1}
\begin{aligned}
    c_0 =& -\frac{\alpha\bar{h}^2}{2}+\bar{h}^2 - 1,\qquad\qquad c_1 = \frac{1}{240}i\bar{h}^3\Big(80\beta +(\alpha-1) (25\alpha-32)\bar{h}^3 \R\Big),\\
    c_2 =& \frac{(\alpha -1) (\alpha  (3458 \alpha -8415)+5120) \bar{h}^{10}\R^2}{80640}+\frac{(343 \alpha -400) \beta  \bar{h}^7\R}{2520}+(\alpha -1) \bar{h}^4,\\
    c_3 =& -\frac{75872 i\bar{h}^{14} \R^3}{2027025}-\frac{157}{224} i\bar{h}^8 R-\frac{1}{3} i\bar{h}^3\We.
\end{aligned}
\end{equation}

To solve the Orr--Sommerfeld eigenvalue problem~\eqref{eq:Orr_Sommerfeld} beyond the long-wave approximation, we employ the Chebyshev--Tau spectral method \citep[Section~3.1]{johnson1996chebyshev,canuto2012spectral}. This approach approximates the solution to the continuous problem by expanding the eigenfunction as a finite series of $N$ Chebyshev polynomials. As a result, the continuous eigenvalue problem is transformed into a discrete one, with eigenvectors corresponding to the amplitudes of the Chebyshev polynomial. The resulting discrete eigenvalue problem is solved using Python’s \texttt{numpy.linalg.eig} function. To avoid spurious eigenvalues arising from the numerical discretisation \citep{bourne2003hydrodynamic,dawkins1998origin}, the problem is solved using two different spectral resolutions $N=20$ and $N=80$. Only those modes whose eigenvalues differ by less than $0.1$ in Euclidean norm ($\|\cdot\|$) are retained, following the criteria proposed by \citet{gardner1989modified} and \citet{pino2024linear}.

% -------------- Design Feedback coefficient
\subsection{Design stability-based feedback coefficients}
\label{subsec:design_feedback_laws}
The feedback coefficients $\alpha$ and $\beta$ are designed to ensure $\omega_i<0$ for all values of the wavenumber $k$. For the reduced order models, we require the critical wavenumber $k_c$ \eqref{eq:nuetral_k_WIBL} to be purely imaginary. This condition leads to the following inequality:
\begin{equation}
\label{eq:WIBL_disp_11}
80 \beta + 25\alpha^2 \delta \bar{h}^3 - 57\alpha \delta \bar{h}^3 + 32 \delta \bar{h}^3 < 0 .
\end{equation}
Solving \eqref{eq:WIBL_disp_11} with respect to $\alpha$ and $\beta$ yields the stability region in the space of control parameters. For $\alpha=0$, the corresponding stability condition on $\beta$ reduces to
\begin{equation}
\label{eq:stab_cond_beta_WIBL}
    \beta < -\frac{2}{5}\,\delta \bar{h}^3,
\end{equation}
while, for $\beta=0$, the stability condition becomes
\begin{equation}
\label{eq:stab_cond_alpha_WIBL}
    1 < \alpha < \frac{32}{25} .
\end{equation}
When both $\alpha$ and $\beta$ are non-zero, the solution of \eqref{eq:WIBL_disp_11} defines the following range of stable feedback coefficients:
\begin{equation}
\label{eq:stab_cont_WIBL}
    \beta < \frac{49\,\delta \bar{h}^3}{8000}, \qquad
    \frac{57}{50} - \frac{1}{50}\sqrt{\frac{49\,\delta \bar{h}^3 - 8000\beta}{\delta \bar{h}^3}}
    < \alpha <
    \frac{57}{50} + \frac{1}{50}\sqrt{\frac{49\,\delta \bar{h}^3 - 8000\beta}{\delta \bar{h}^3}} .
\end{equation}

For the OS problem, linear stability is ensured by requiring that the growth rate initially decreases with $k$, i.e. the film is long wave stable, 
\begin{equation}
\label{eq:cond_stab_OS_ind_stresses}
    \left.\frac{d\omega_i}{dk}\right|_{k=0} < 0 .
\end{equation}
Substituting the long-wave expansion \eqref{eq:expansion_OS_mod} together with \eqref{eq:expansion_OS_mod_1} into \eqref{eq:cond_stab_OS_ind_stresses} and solving the resulting inequality yields the following stability conditions:
\begin{equation}
\label{eq:cont_OS_feedback}
    \beta < \frac{49\,\bar{h}^3 \R}{8000}, \qquad
    \frac{57}{50} - \frac{1}{50}\sqrt{\frac{49\,\bar{h}^3 \R - 8000\beta}{\bar{h}^3 \R}}
    < \alpha <
    \frac{57}{50} + \frac{1}{50}\sqrt{\frac{49\,\bar{h}^3 \R - 8000\beta}{\bar{h}^3 \R}} .
\end{equation}
Finally, rescaling $\beta$ from the Nusselt-like to the Shkadov-like scaling (Section~\ref{sec:scaling}) according to $\beta \rightarrow \beta/\varepsilon$, and considering $\alpha$ and $\beta$ separately, the stability conditions in \eqref{eq:cont_OS_feedback} recover those obtained from the Benney and WIBL models \eqref{eq:stab_cond_alpha_WIBL}.

The validity of the assumption in \eqref{eq:cond_stab_OS_ind_stresses} is assessed in the results section by comparing the long-wave solution with the numerical solution of the OS.

% ------ Numerical Implementation
\subsection{Setup Numerical simulations}
\label{subsec:sims_descriptions}

\label{subsec:test_case_des}
\label{sec:numerical_sims}
\begin{figure}
  \begin{subfigure}[b]{0.5\textwidth}
      \centering
    \includegraphics[width=\textwidth]{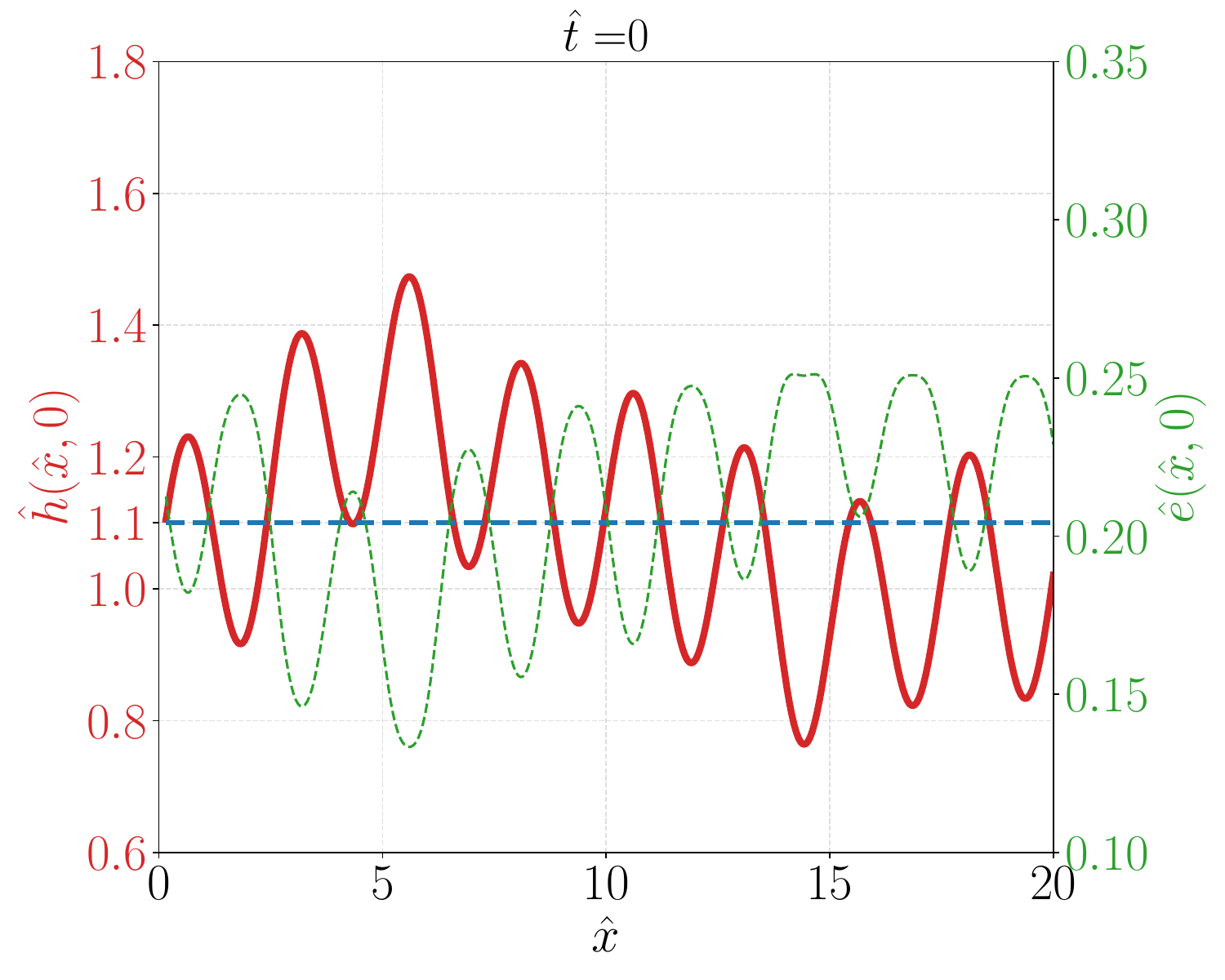}
    \caption{}
  \label{init_cont_WIBL_white}    
    \end{subfigure}
  \hfill
  \begin{subfigure}[b]{0.5\textwidth}
      \centering
    \includegraphics[width=\textwidth]{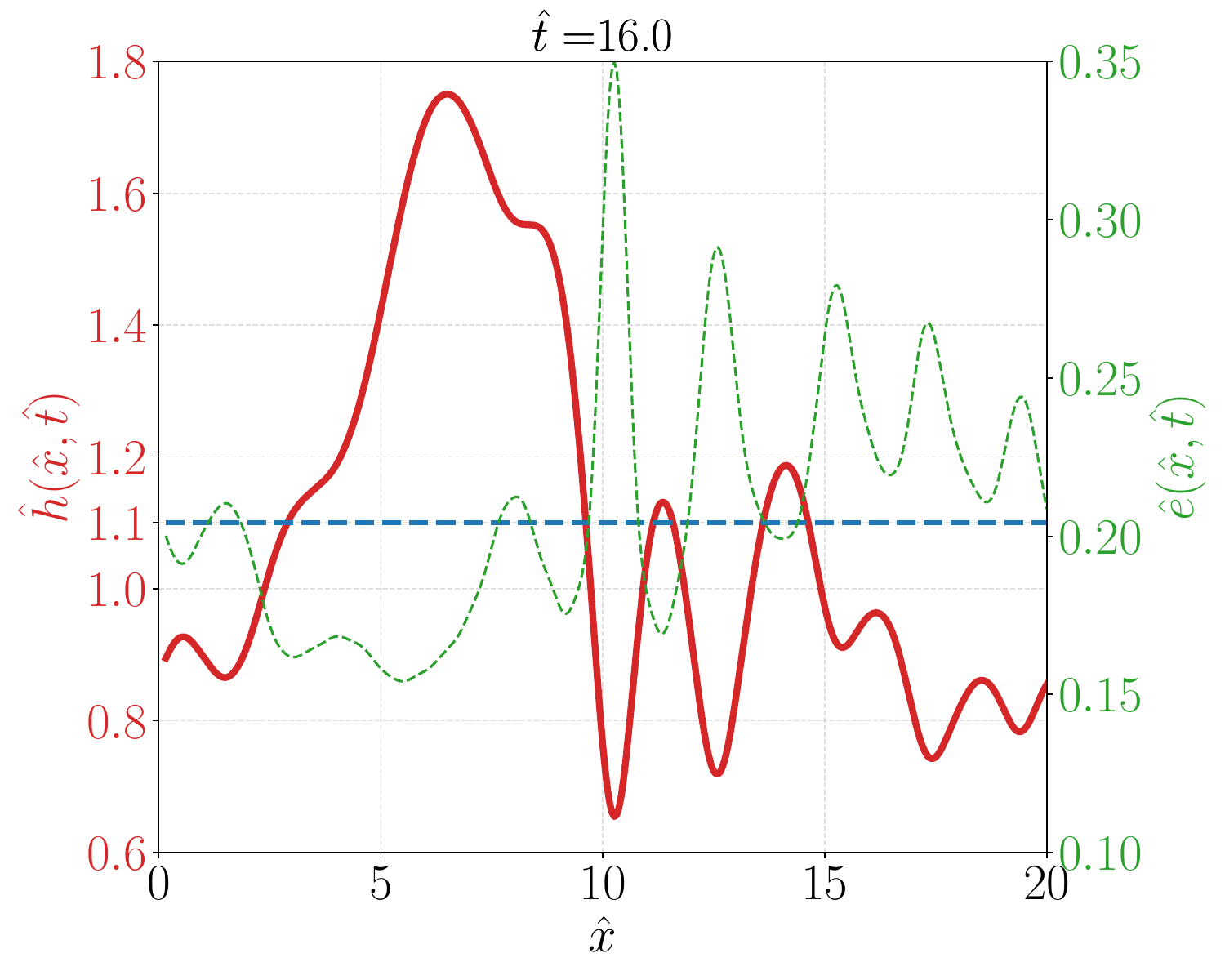}
    \caption{}
  \label{init_cont_WIBL_white_3} 
  \end{subfigure}
    \hfill
  \begin{subfigure}[b]{0.5\textwidth}
      \centering
    \includegraphics[width=\textwidth]{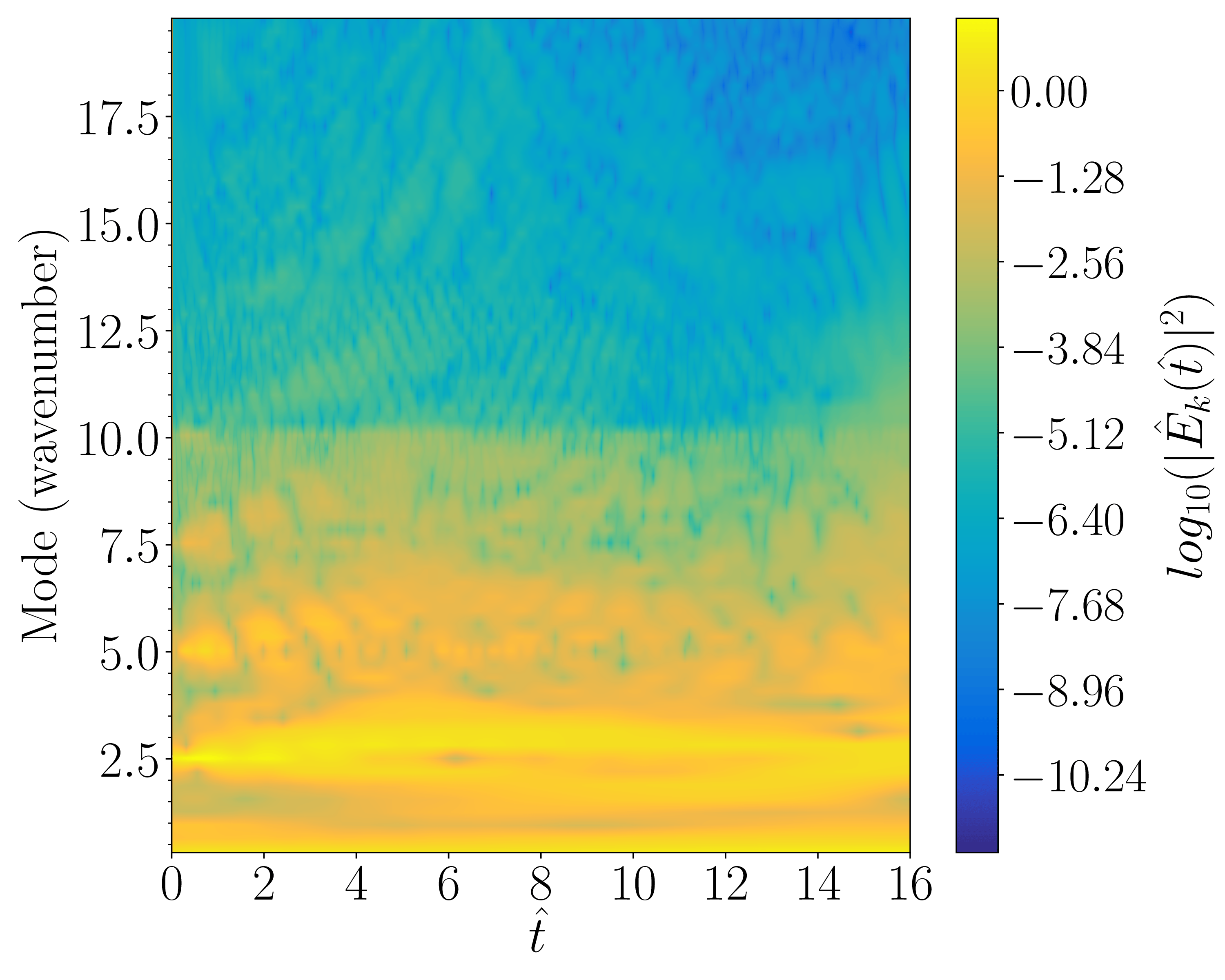}
    \caption{}
    \label{fig_energy_evolution_no_control}
  \end{subfigure}
  \hfill
  \begin{subfigure}[b]{0.5\textwidth}
    \centering
    \includegraphics[width=\textwidth]{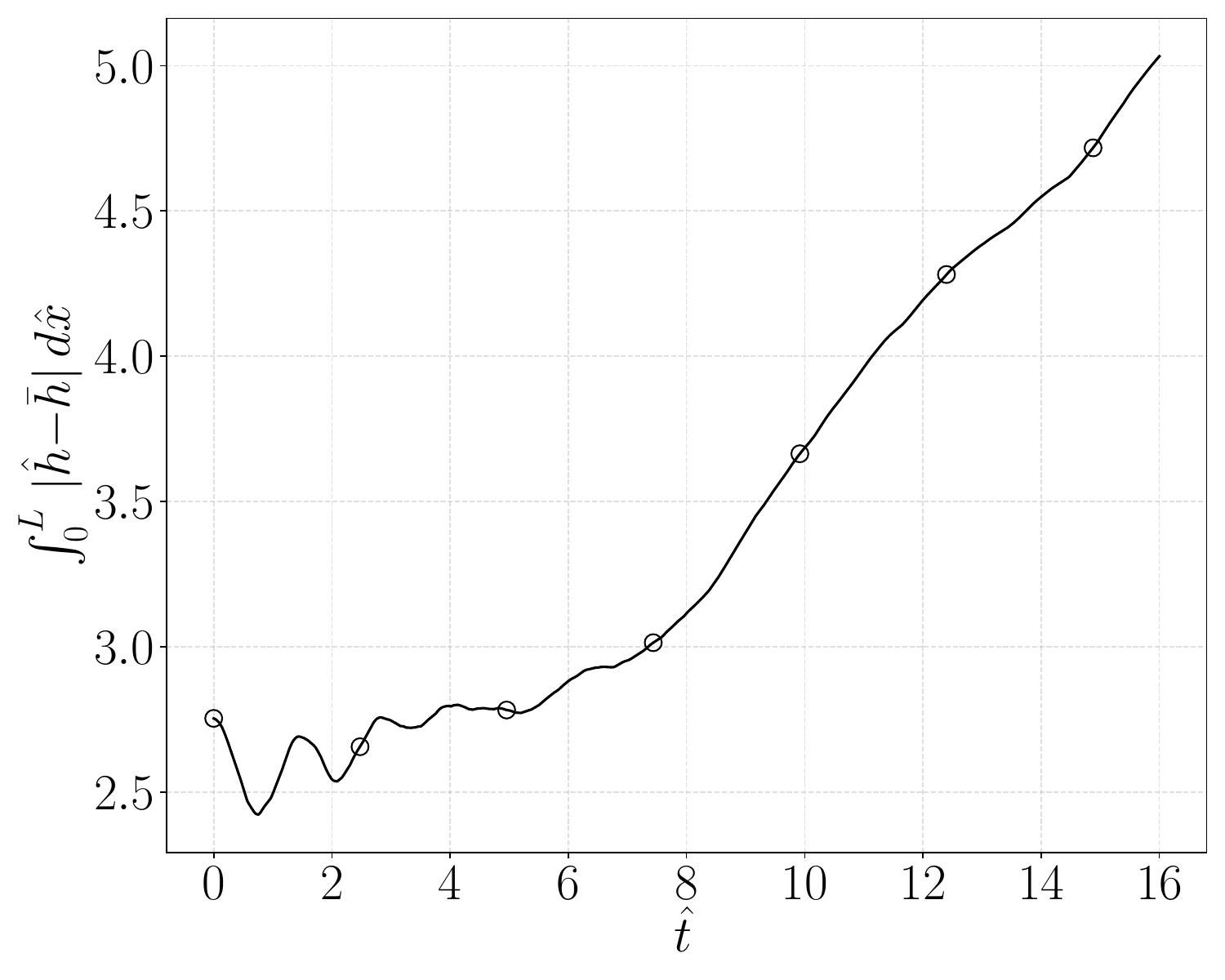}
    \caption{}
    \label{fig_error_evolution_no_control}
  \end{subfigure}
\caption{Evolution of (a,b) the film thickness $\hat{h}$ (solid red line), the corresponding kinetic energy $\hat{e}$ (green dashed line), and the target flat thickness $\bar{h} = 1.1$ (thick blue dashed line) in a periodic domain $\hat{x} \in [0, 20]$ at times (a) $\hat{t} = 0$ and (b) $\hat{t} = 16$, and of (c) the spectral energy and (d) the error.}

Evolution of the unstable initial condition during a simulation with the error with respect to the sought flat state condition. Fourier modes associated to the liquid film thickness (a) their initial conditions amplitudes and (b) their energy evolution throughout the simulation. 
\label{unstable_dev_white_box}
\end{figure}

The feedback control strategies derived using the methods presented in Subsection~\ref{subsec:design_feedback_laws} are tested on the control of growing finite amplitude waves simulated with the WIBL model \eqref{eq:integral_continuity} with \eqref{eq:WIBL_qt} in a periodic domain. The objective of the control strategy is to suppress the growth of surface waves and steer the liquid film toward the flat state $\bar{h} = 1.1$. We measure the control performance throughout the simulation by monitoring the time evolution of the $L^2$ norm of the error between the film thickness $\hat{h}$ and the target flat profile $\bar{h}$.

\begin{table}
\label{tab:scaling_quant}
\small\addtolength{\tabcolsep}{0pt}
\renewcommand{\arraystretch}{1.2}
\begin{tabular}{cc}
% Reference quantities table
\begin{minipage}{0.495\textwidth}
\centering
\begin{tabular}{ccc}
 & Definition & Value\\
$u_{\rm ref}$ & $U_p$ & 0.29 [m/s]    \\
$h_{\rm ref}$ & $\sqrt{(\nu U_p)/g}$ &  172 [$\mu$m]  \\
$p_{\rm ref}$ & $\rho g h_{\rm ref}$ &  10 [Pa] \\
$\tau_{\rm ref}$ & $\mu u_{\rm ref}/ h_{\rm ref}$ & 1.69 [Pa] \\
$q_{\rm ref}$ & $u_{\rm ref}h_{\rm ref}$& 50 [$\mu$kg/m$^3$] \\
$e_{ref}$ & $\rho u_{\rm ref}^2h_{ref}$ & 14.5 [$\mu$J/$m^2$]
\end{tabular}
\end{minipage}
&
% Non-dimensional numbers table
\begin{minipage}{0.495\textwidth}
\centering
\begin{tabular}{ccc}
 & Definition & Value \\
$\R$ & $\sqrt{U_p^3/{(g\nu)}}$ & 53 \\
$\Ca$ & $(U_p\mu)/\sigma$ & 0.0035 \\
$\delta$ & $\R\Ca^{1/3}$ & 8  \\
$\Ka$ & $\sigma/(\rho g^{1/3} \nu^{4/3})$ & 4000 \\
$\varepsilon$ & $\text{Ca}^{-1/3}$ & 0.15
\end{tabular}
\end{minipage}
\end{tabular}
\caption{Definition and value of the reference quantities (left) and the non-dimensional groups (right) used in this analysis, considering water as the working fluid.}
\label{tab:scaling_quant}
\end{table}

Table \ref{tab:scaling_quant} reports the definition and the values of the scaling quantities, the liquid film parameter $\varepsilon$ and the nondimensional groups used in our analysis.

For the numerical solution, we consider a domain with periodic boundary conditions $\Omega = \{ \hat{x} \in \mathbb{R} \mid 0 \leq \hat{x} \leq L \},$ with $L=20$ and a final time $T$ which will be different depending on the used control approach and which will be specified in the results subsections. The governing equations are discretised using a Fourier pseudo-spectral method with 128 discrete modes \citep{dutykh2016brief}. The discretised equations are then integrated in time using the Runge-Kutta fourth-order method with a time step $\Delta\hat{t}=2.6/((Nx/2)^3)=9.9\times 10^{-6}$. To control aliasing errors arising from the truncated Fourier series, we filter out modes with frequencies above half the Nyquist frequency.

The initial liquid film thickness $\hat{h}(\hat{x},0)$, flow rate $\hat{q}(\hat{x},0)$ and kinetic energy $\hat{e}(\hat{x},0)$ read:
\begin{equation}
\label{eq:init_cont_white}
\begin{split}
    \hat{h}(\hat{x},0)\;\coloneq&\;A\Big(\bar{h} + 0.1\exp\Big(-\frac{((2\pi\hat{x}/L)-\pi)^2}{0.8^2}\Big) +  0.3\exp\Big(-\frac{((2\pi\hat{x}/L)-\pi/2)^2}{0.9^2}\Big) +\\&\;- 0.1\exp\Big(-\frac{((2\pi\hat{x}/L)-3\pi/2)^2}{0.4^2}\Big) + 0.2\sin(8(2\pi\hat{x}/L))\Big),\\
    \hat{q}(\hat{x},0)\;\coloneq&\; \frac{\hat{h}(\hat{x},0)^3}{3} - \hat{h}(\hat{x},0), \qquad\qquad
        \hat{e}(\hat{x},0)\;\coloneq\; \frac{3\hat{q}(\hat{x},0)^2}{5\hat{h}(\hat{x},0)}+\frac{1}{10}\hat{h}(\hat{x},0)+\frac{1}{5}\hat{q}(\hat{x},0),
\end{split}  
\end{equation}
where $A$ is the initial amplitude of the film thickness, the flow rate $\hat{q}(\hat{x},0)$ is computed from the steady-state relation \eqref{eq:steady_state_sols} and the kinetic energy from its definition \eqref{eq:kin_en_def} in the absence of the shear stress $\hat{\tau}_g$. To ensure mass conservation with the desired flat film condition, the initial amplitude $A$ is calculated by solving the following equation:
\begin{equation}
   \int_{0}^L\hat{h}(\hat{x},0)\,d\hat{x}=\bar{h}\,L.
\end{equation}

To monitor the evolution of perturbation, the integral of the error between the liquid film thickness and the sought flat state error is defined as: 
\begin{equation}
\label{eq:error_norm}
   \text{error} = \int_{0}^{L}\, |\hat{h}(\hat{X},\hat{T})- \bar{h}_0|\,d\hat{X}
\end{equation}

Figures \ref{init_cont_WIBL_white} and \ref{init_cont_WIBL_white_3} show the film thickness $\hat{h}(\hat{x},0)$ (solid red line), the corresponding kinetic energy $\hat{e}(\hat{x},0)$ (small green dashed line), and the target flat state $\bar{h} = 1.1$ (large blue dashed line) at $\hat{t} = 0$ and 16. The initial film profile is constructed by superimposing Gaussian and harmonic components with varying amplitudes to excite different modes along the domain, promoting nonlinear interactions and energy transfer that drive the system beyond the linear regime. The initial kinetic energy is nearly uniform, ranging from 0.15 to 0.25 with a mean of 0.20, and its maxima and minima align with film thickness troughs and peaks, respectively. As the simulation progresses, the film evolves toward a dynamic state dominated by long-wavelength modes, with a sharp peak forming near $\hat{x} = 7$, while kinetic energy redistributes, peaking just ahead of the film's peak. Mass in the upper part of the peak flows downward under gravity and is partially entrained upward by the substrate motion, steepening the interface and eventually leading to blow-up.

The blow-up of the solution arises from an energy-transfer mechanism involving the dissipation and redistribution of kinetic energy, initially from long-wavelength modes to short-wavelength modes and subsequently back to long wavelengths. Figure \ref{fig_energy_evolution_no_control} shows the evolution of the spectral energy distribution across different wavelengths, and \ref{fig_error_evolution_no_control} shows the evolution of the error. Initially, energy is concentrated in the longest wavelengths, with a dominant peak at mode $k = 9$. As time progresses, energy cascades toward higher wave numbers, reaching approximately $k = 37$, beyond which surface tension effectively dampens wave activity. Energy subsequently transfers back to longer wavelengths around $\hat{t} = 4$, with the final distribution peaking near $k = 2.5$. Energy above $k = 12.5$ becomes negligible, indicating that a resolution of 128 Fourier modes is sufficient to accurately capture the film dynamics.

This energy redistribution process continues until approximately $t = 5$, at which point the energy in mode $k = 8$, which had been gradually declining, drops sharply. This event marks a shift in energy toward smaller temporal and spatial scales, as indicated by rapidly changing patterns at higher wavenumbers. Simultaneously, energy becomes increasingly concentrated in the long-wavelength modes near $k = 8$. A similar phenomenon is observed near $k = 3$, where energy initially grows and then drops abruptly around $t = 6$, followed by redistribution into neighbouring modes. While energy exchanges persist in the intermediate modes, the energy in the lowest mode continues to rise steadily, ultimately triggering the film's rupture at $t = 16$.

This simulation highlights the highly nonlinear nature of film rupture, which is governed by intricate interactions across multiple spatial and temporal scales. The application of feedback control strategies derived from linear theory to such a nonlinear regime provides a stringent test of their robustness and effectiveness in stabilising complex interfacial dynamics.

\section{Results and discussion}
\label{sec:results}
In this subsection, we investigate the effects of stability-based feedback control on the linear (Subsection \ref{subsec:linear_stab_analysis}) and nonlinear (Subsection \ref{subsec:nonlin_control_case}) dynamics of the liquid film. For the linear regime, we assess the impact of feedback on flat-film stability, characterising the response via the growth rate $\omega_i$ and phase speed $c$, and use wave-hierarchy arguments to understand the effects on mass and energy transport. In the nonlinear regime, we evaluate two representative sets of feedback coefficients that stabilise the linearised system across all wavenumbers. Nonlinear simulations are performed using the WIBL model, with actuation initiated at $\hat{T}=10$, after the film evolves freely from the initial condition \eqref{eq:init_cont_white}. Control performance is quantified through the evolution of the film thickness $\hat{h}(\hat{X},\hat{T})$, the logarithmic spectral distribution of kinetic energy $\log_{10}(|\hat{E}_k(\hat{T})|^2)$, and the integral error norm defined in \eqref{eq:error_norm}.

% ---- Linear Stability with feedback control ---- 
\subsection{Linear Stability Analysis}
\label{subsec:linear_stab_analysis}
Selecting values of $\alpha$ and $\beta$ within the stability region \eqref{eq:stab_cont_WIBL} ensures that the growth rates $\omega_i$ are negative for any wavenumber $k$ in the linear regime.
\begin{figure}
  \begin{subfigure}[b]{0.49\textwidth}
    \includegraphics[width=\textwidth]{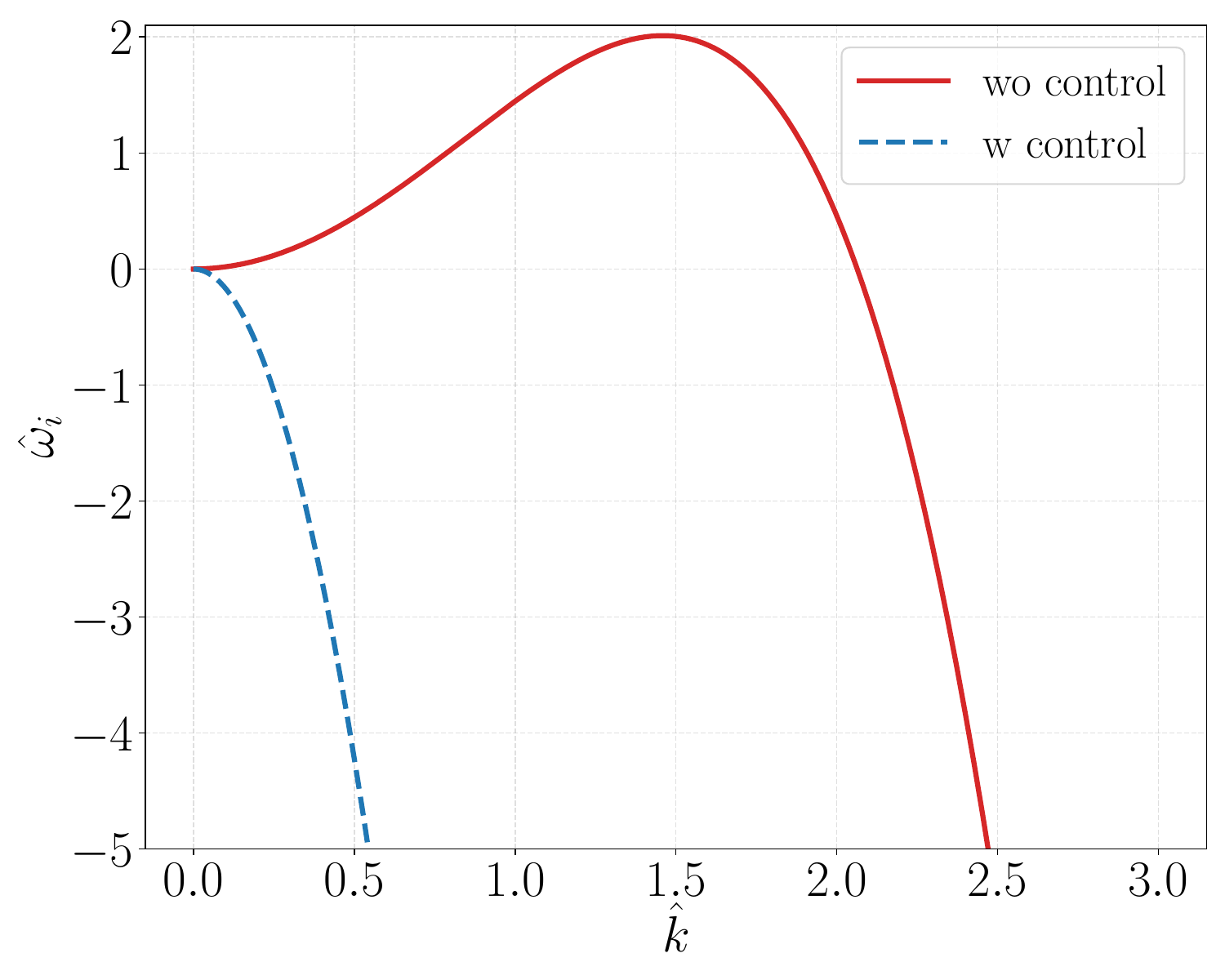}
    \caption{}
  \end{subfigure}
  \hfill
  \begin{subfigure}[b]{0.49\textwidth}
    \includegraphics[width=\textwidth]{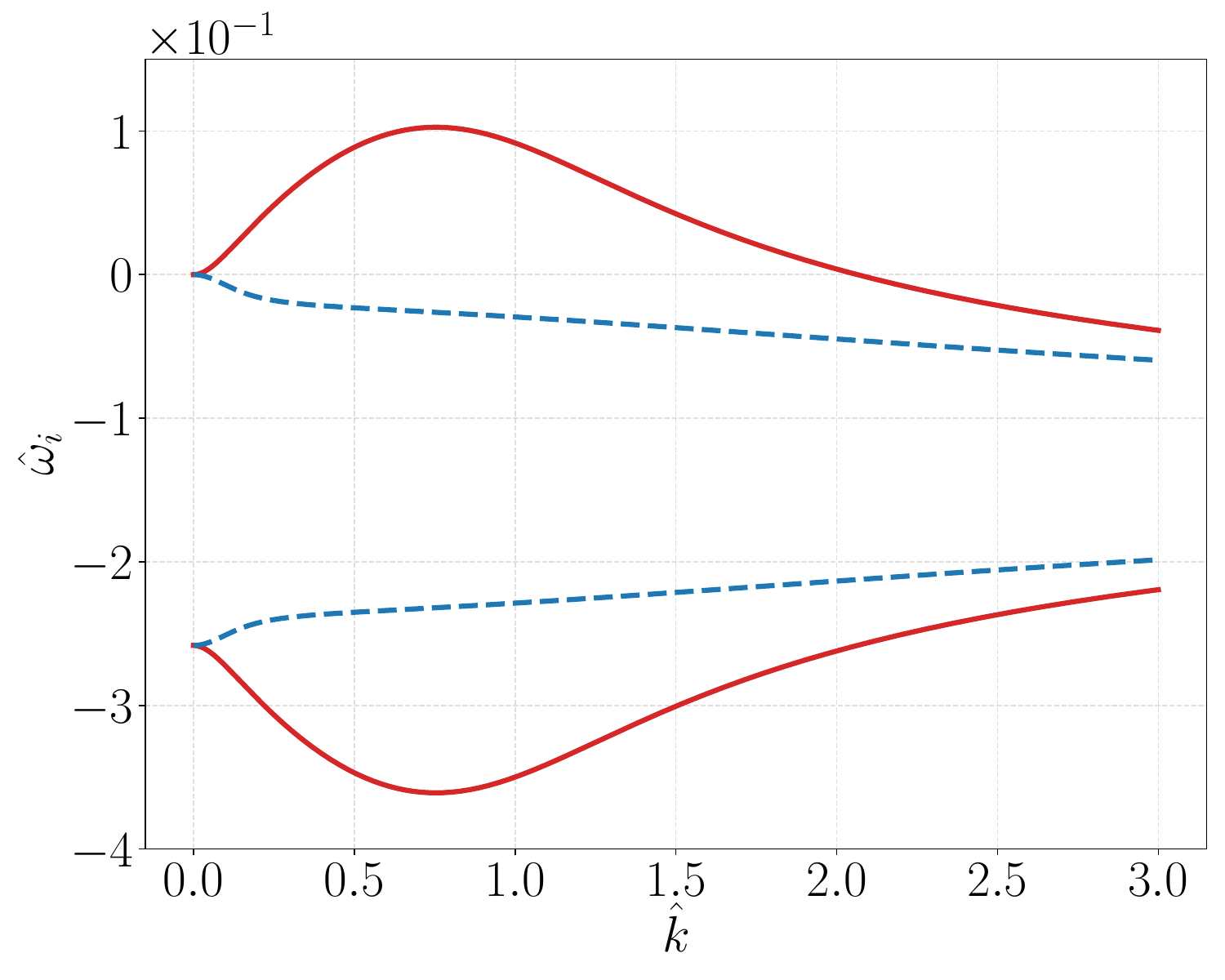}
    \caption{}
  \end{subfigure}
  \caption{Growth rate $\omega_i$ as a function of the wavenumber $k$ for $\bar{h} = 1.1$ and $\delta = 8$ obtained solving the linearised (a) Benney and (b) WIBL equations with $\alpha=\beta=0$ (blue dashed line) and in the case with $\alpha=0$ and $\beta = -6.39$ (blue dashed line).}
  \label{fig:disp_rel_controlled}
\end{figure}
Figure~\ref{fig:disp_rel_controlled} presents $\omega_i$ as a function of $k$ computed solving (a) the Benney model~\eqref{eq:sol_disp_Benney} and (b) the WIBL model~\eqref{eq:roots_WIBL_linear} dispersion relation for a reduced Reynolds number $\delta = \varepsilon \R = 8$ and film thickness $\bar{h} = 1.1$ for the uncontrolled case ($\alpha=\beta=0$) (red solid line) and the controlled case with only feedback pressure ($\alpha=0$ and $\beta=-6.39$) (blue dashed line). In the uncontrolled case, both the WIBL and Benney dispersion relations predict the same cutoff wavenumber but different maximum growth rates. In the controlled case, since $\beta$ lies within the range given in \eqref{eq:stab_cont_WIBL}, $\omega_i$ remains negative for all wavenumbers. The solution to the WIBL dispersion relation has two distinct modes that converge to the same asymptotic behaviour as $k$ increases in both the controlled and uncontrolled cases. Interestingly, the feedback control affects the two modes differently: while the first mode exhibits a reduced growth rate under control, the second mode shows a slight increase. This implies that although the control mechanism stabilises the system overall, it may slow the decay of some modes in the long-time regime.
\begin{figure}
  \begin{subfigure}[b]{0.49\textwidth}
    \includegraphics[width=\textwidth]{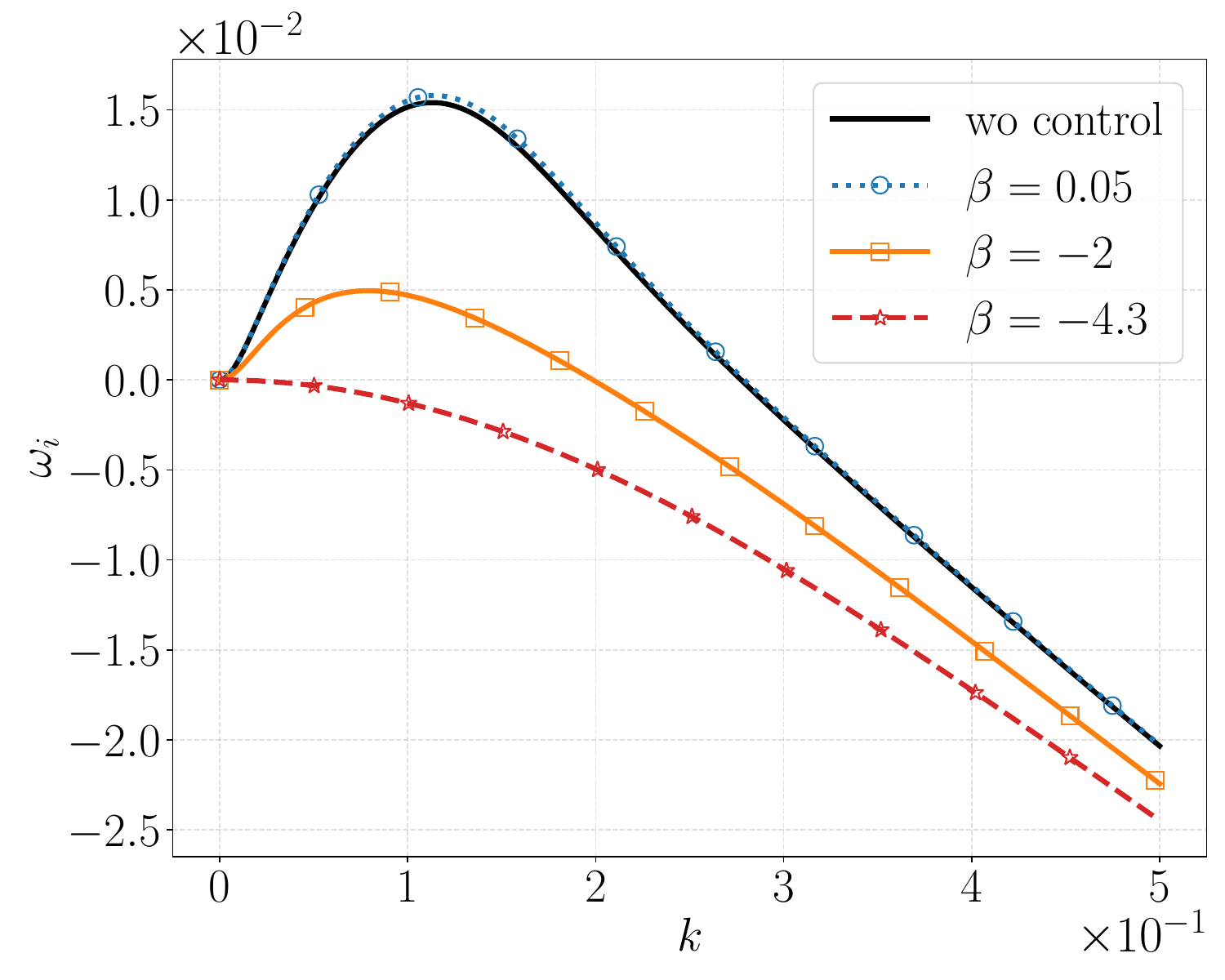}
    \caption{}
  \end{subfigure}
  \hfill
  \begin{subfigure}[b]{0.49\textwidth}
    \includegraphics[width=\textwidth]{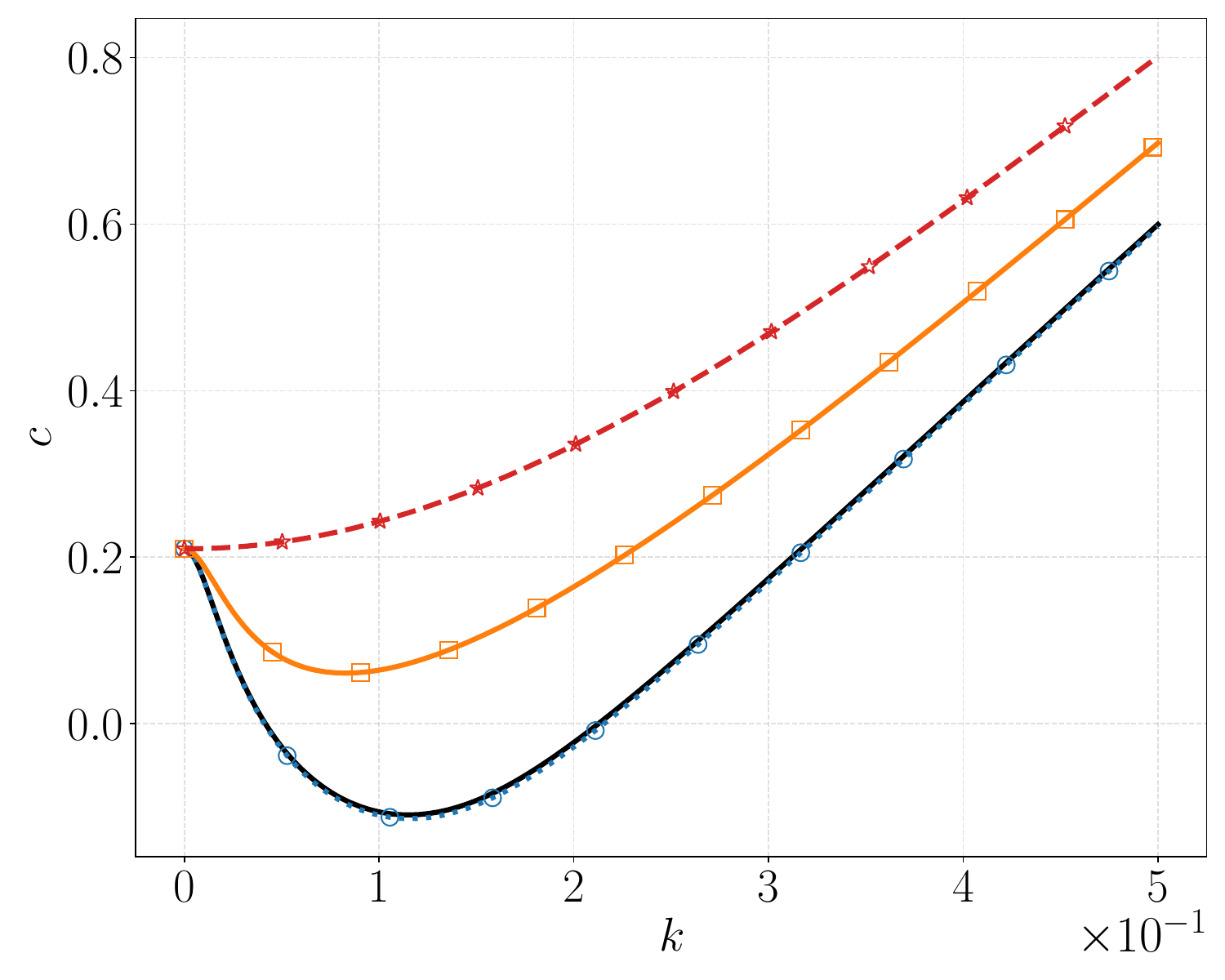}
    \caption{}
  \end{subfigure}
  \hfill
  \begin{subfigure}[b]{0.49\textwidth}
    \includegraphics[width=\textwidth]{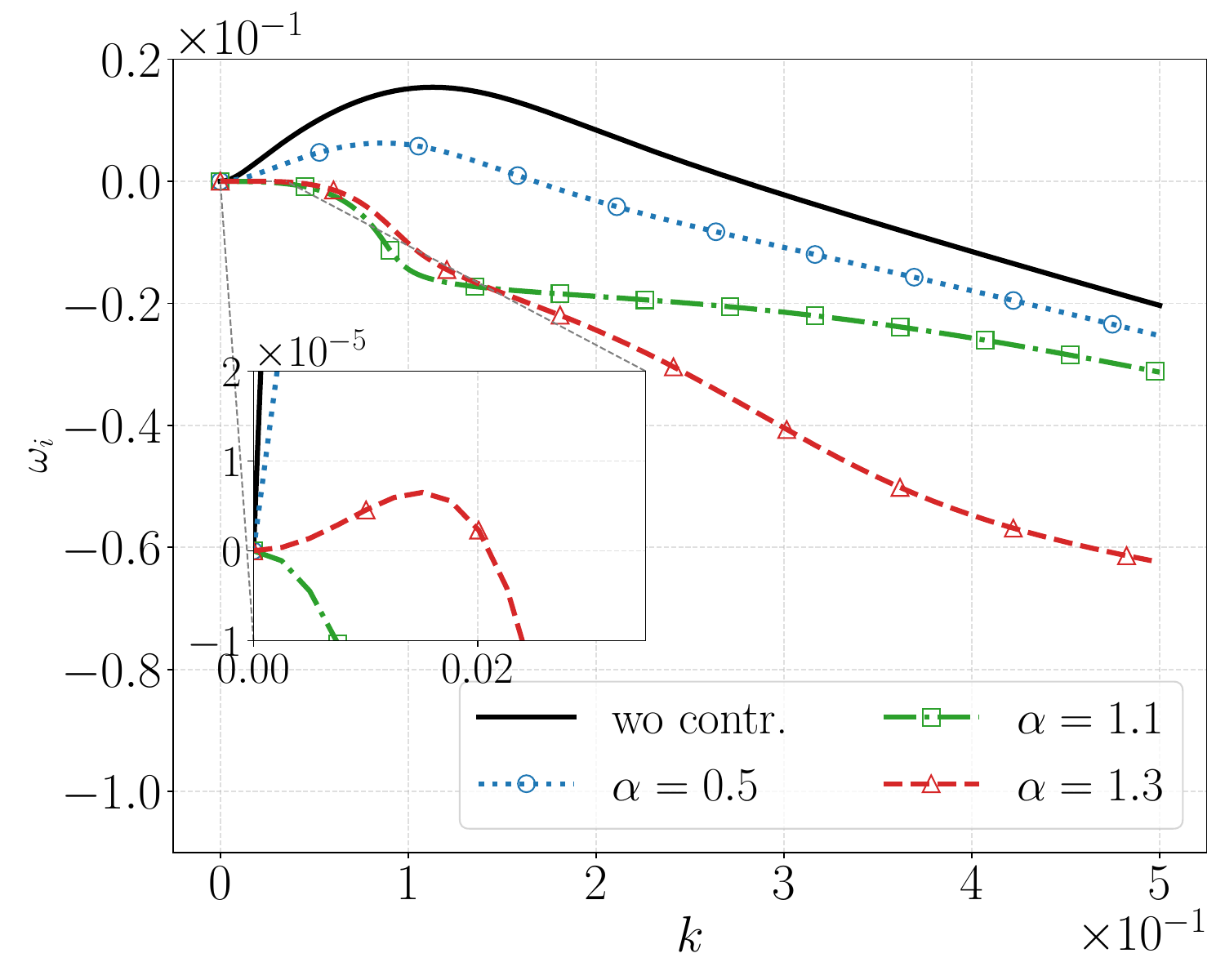}
    \caption{}
  \end{subfigure}
  \hfill
  \begin{subfigure}[b]{0.49\textwidth}
    \includegraphics[width=\textwidth]{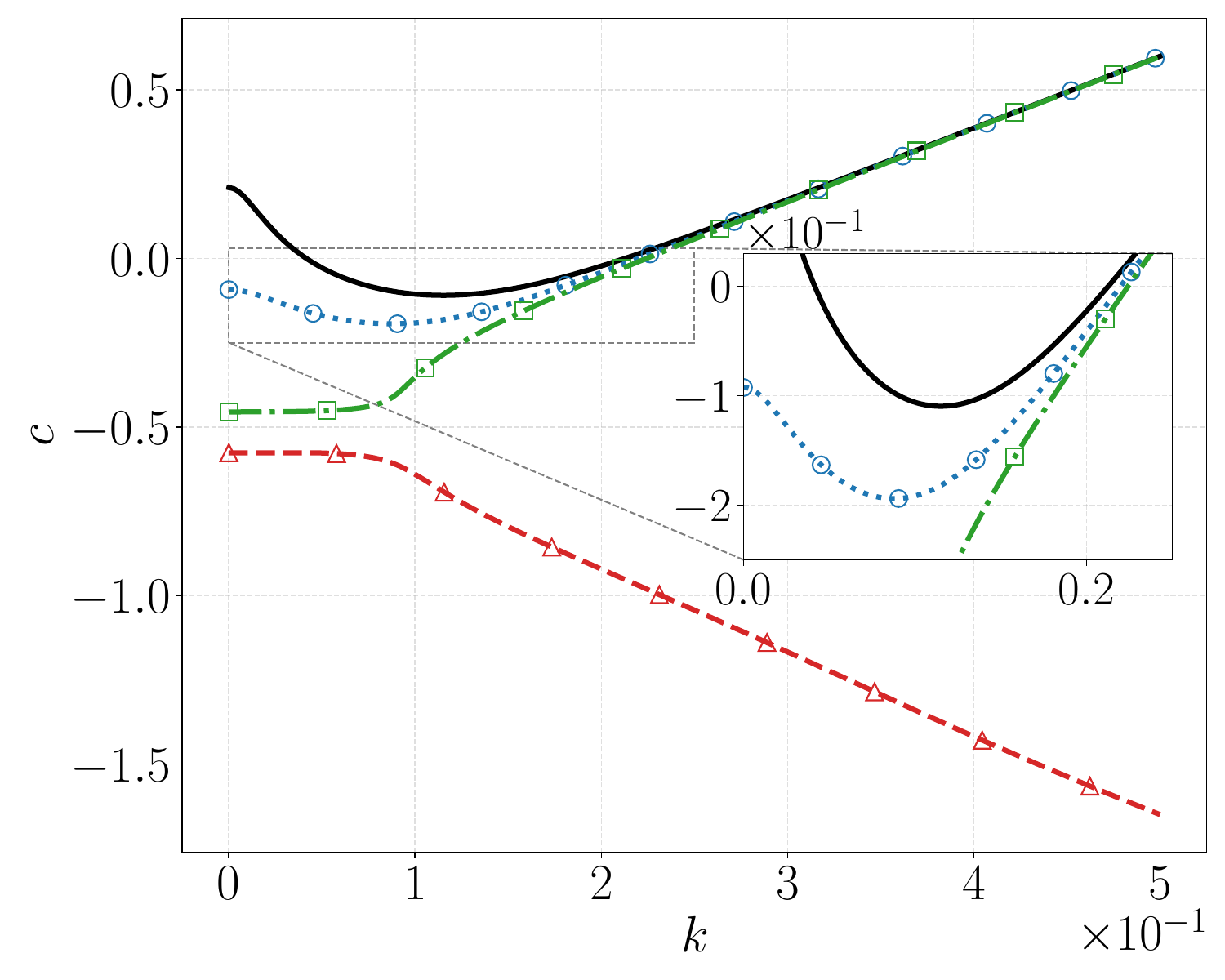}
    \caption{}
    \label{fig:omega_real_alpha_OS}
    \end{subfigure}
  \caption{Comparison of (a and c) the growth rate $\omega_i$ and (b and d) the real part $\omega_r$ of the complex pulsation $\omega$ as function of the wavenumber $k$ without (continuous black line) and with (coloured lines with markers) with (a and b) only pressure ($\alpha=0$) and (c and d) only shear ($\beta=0$)  feedback control, obtained solving the Orr-Sommerfeld eigenvalue problem for Reynolds number $\delta = 8$ ($\R=52$ and $\Ka=4000$) for $\bar{h}=1.1$.}
  \label{fig:disp_rel_contr_OS}
\end{figure}

A similar behaviour is observed for the solution of the OS eigenvalue problem. Figure~\ref{fig:disp_rel_contr_OS} shows the solution of the OS problem for $\delta=8$ ($\R=53$ and $\Ka=4000$) in terms of (a and b) the growth rate $\omega_i$ and (b and d) the phase speed $c$ as a function of the wavenumber $k$ in the uncontrolled (continuous black line) and in the controlled (coloured lines with markers) cases with (a and b) only pressure feedback ($\alpha=0$) varying $\beta$ and (c and d) only shear feedback ($\beta=0$) varying $\alpha$. For $\beta$ lying outside the stability boundary, as for $\beta=0.05$ or $\beta=-2$, the film is unstable. For $\beta=0.05$, the maximum growth rate is slightly larger than in the uncontrolled case, while the wavenumber remains unchanged. By contrast, for $\beta=-2,$ both the maximum growth rate and the cutoff wavenumber are reduced. For the value of $\alpha$ outside the stability boundary, as for $\alpha=1.3$, $\omega_i$ is again positive for $k<0.025$, although with a tiny magnitude. This suggests that in nonlinear and transient growth settings, the values of $\alpha$ and $\beta$ near the stability bounds may still guarantee the decay of small-amplitude perturbations. In numerical simulations or experiments, these modes might grow so slowly that they would require a long time to affect the nonlinear dynamics.

Turning our attention to the phase speed $c$, in the uncontrolled case, it is positive for both small and large values of $k$. Still, it becomes negative in the intermediate range $0.3 \lesssim k \lesssim 2.2$, reaching a minimum of $c \approx -0.2$ at $k \approx 0.1$. When $\beta$ is nonzero, $c$ reduces in magnitude for larger negative $\beta$, becoming positive for all wavenumbers for $\beta=-4.3$. When $\alpha$ is nonzero, the behaviour of $c$ changes significantly: for $\alpha = \in [0.5,1.1]$ it becomes negative at small $k$ while still recovering the same positive asymptote at large $k$, while for $\alpha = 1.3$, $c$ remains negative across all $k$, decreasing approximately linearly with $k$. 

The influence of $\alpha$ and $\beta$ on the linear dynamics can be understood using wave-hierarchy arguments. Instability occurs when the kinematic velocity $c_k$ \eqref{eq:kin_waves} lies outside the interval spanned by the dynamic wave speeds $c_{d\pm}$ \eqref{eq:dyn_waves}. The objective of the feedback control is therefore to adjust $\alpha$ and $\beta$ so that $c_k$ remains within this range for all wavenumbers.

In the case without control, the $c_k$ is negative (waves moving upwards) for thin film conditions $\bar{h}<1$, positive (waves moving downwards) for thick films $\bar{h}>1$ and zero for the \cite{derjaguin1993thickness}'s solution $\bar{h}=1$. The feedback shear stress affects both the magnitude and sign of $c_k$. The locus of critical liquid film thicknesses $\bar{h}_{\text{crs}}$ for which the kinematic wave speed $c_k$ change sign is obtained by solving the kinematic velocity equation \eqref{eq:kin_waves} setting $c_k = 0$, which yields:
\begin{equation}
\label{eq:zeros_hbar_kinematic_vel}
    \bar{h}_{\text{cr}} = \sqrt{\frac{2}{2 - \alpha}}.
\end{equation}

As we can notice, $\bar{h}_{\text{cr}}$ has a real solution only for $\alpha<2$. For $\alpha$ within the range $\alpha\in(-\infty,2]$, negative $\alpha$ moves the threshold thickness towards smaller values, while positive $\alpha$ moves the threshold thickness towards larger values. In particular, the critical value $\alpha_{\text{cr}}$, below which the kinematic velocity remains strictly negative (i.e., flows downward under gravity), is found by evaluating \eqref{eq:zeros_hbar_kinematic_vel} at $\bar{h}_{\text{cr}} = \sqrt{3}$, which corresponds to the maximum admissible flat film thickness in purely dip-coating conditions, which gives:
\begin{equation}
    \alpha_{\text{cr}} = \frac{4}{3}.
\end{equation}
This critical value lies outside of the range for stable conditions $1<\alpha<32/25$ for $\beta=0$. This means that, depending on the value of $\alpha$ inside the stability range, the phase speed changes sign depending on the liquid film thickness at leading order in $\varepsilon$.
\begin{figure}
  \begin{subfigure}[b]{0.5\textwidth}
    \includegraphics[width=\textwidth]{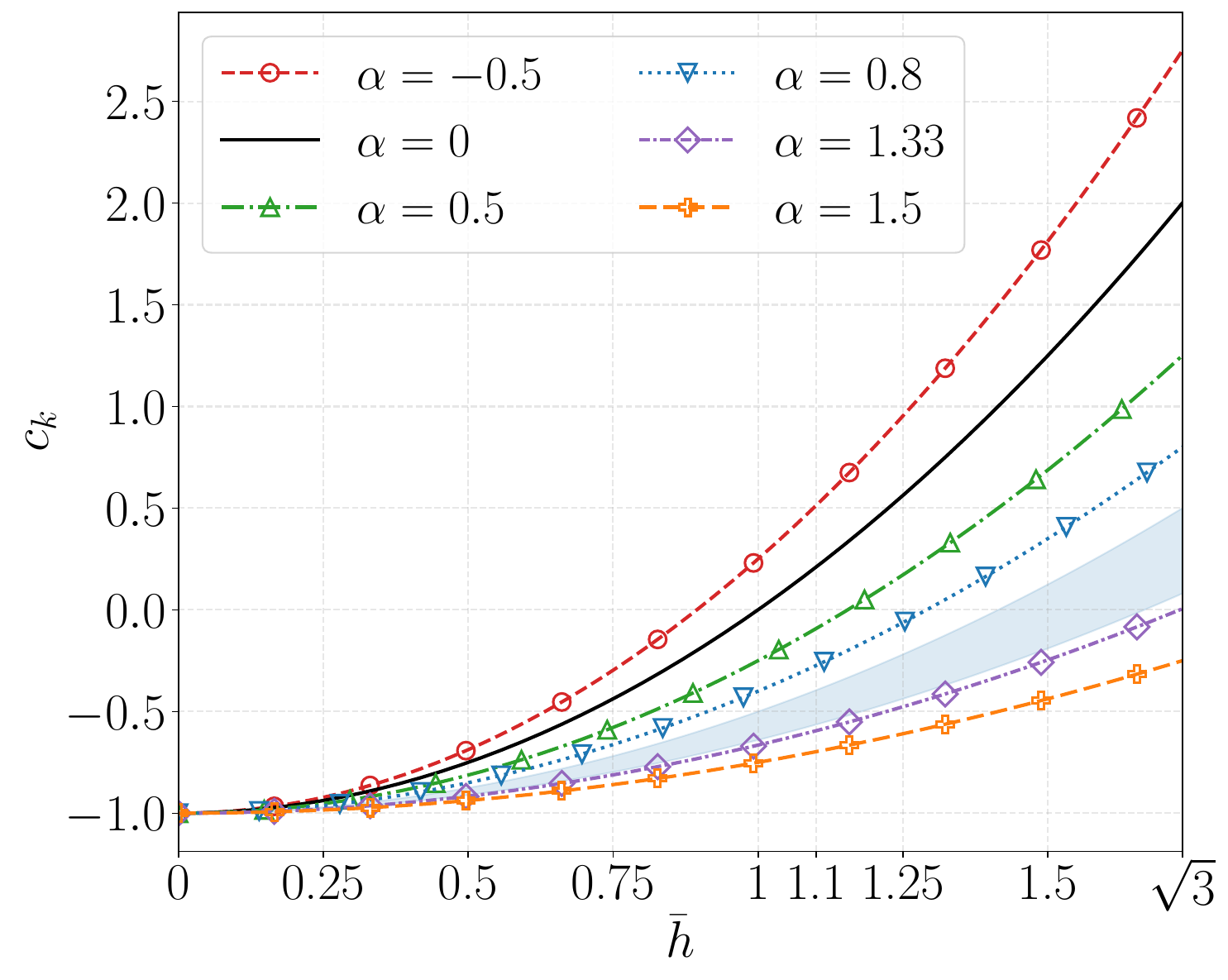}
    \caption{}
    \label{fig:stable_reg_OS_1_1}
  \end{subfigure}
  \hfill
  \begin{subfigure}[b]{0.5\textwidth}
    \includegraphics[width=\textwidth]{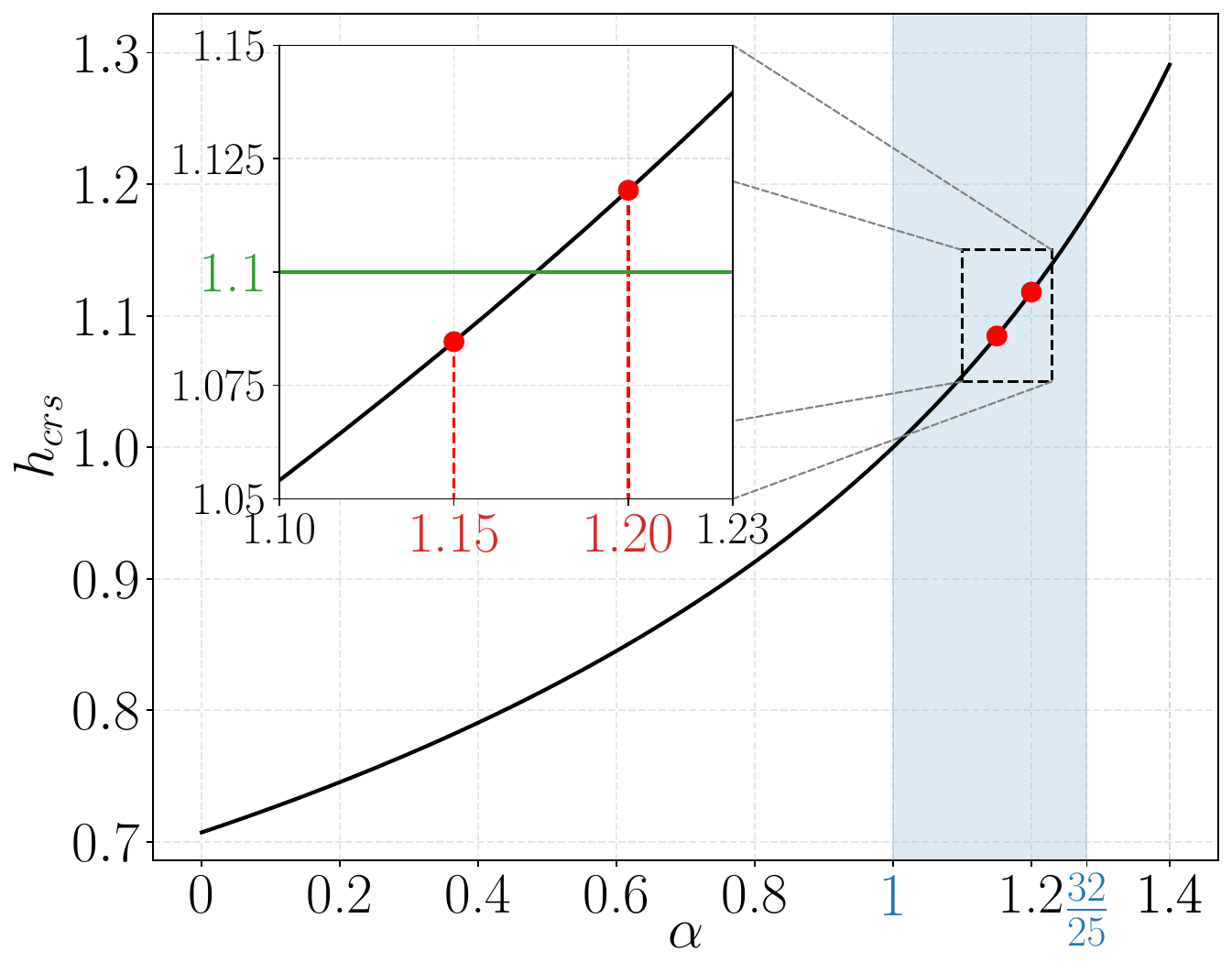}
    \caption{}
  \end{subfigure}
  \caption{(a) Kinematic velocity as a function of flat film thickness $\bar{h}$ for various values of $\alpha$ (coloured curves with markers). The shaded region highlights the range of $\alpha$ corresponding to the stability domain without pressure, and (b) Critical film thickness $\bar{h}_{crs}$, defined by the zero of the kinematic velocity, plotted against $\alpha$. The light blue area denotes the stability region. Red dots indicate the $\alpha$ values used in the nonlinear simulations, with a zoomed-in view of the corresponding region. The flat-film thickness used in these simulations, $\bar{h} = 1.1$ (horizontal, green, continuous line).}
  \label{fig:kin_vel}
\end{figure}
\begin{figure}
  \begin{subfigure}[b]{0.5\textwidth}
    \includegraphics[width=\textwidth]{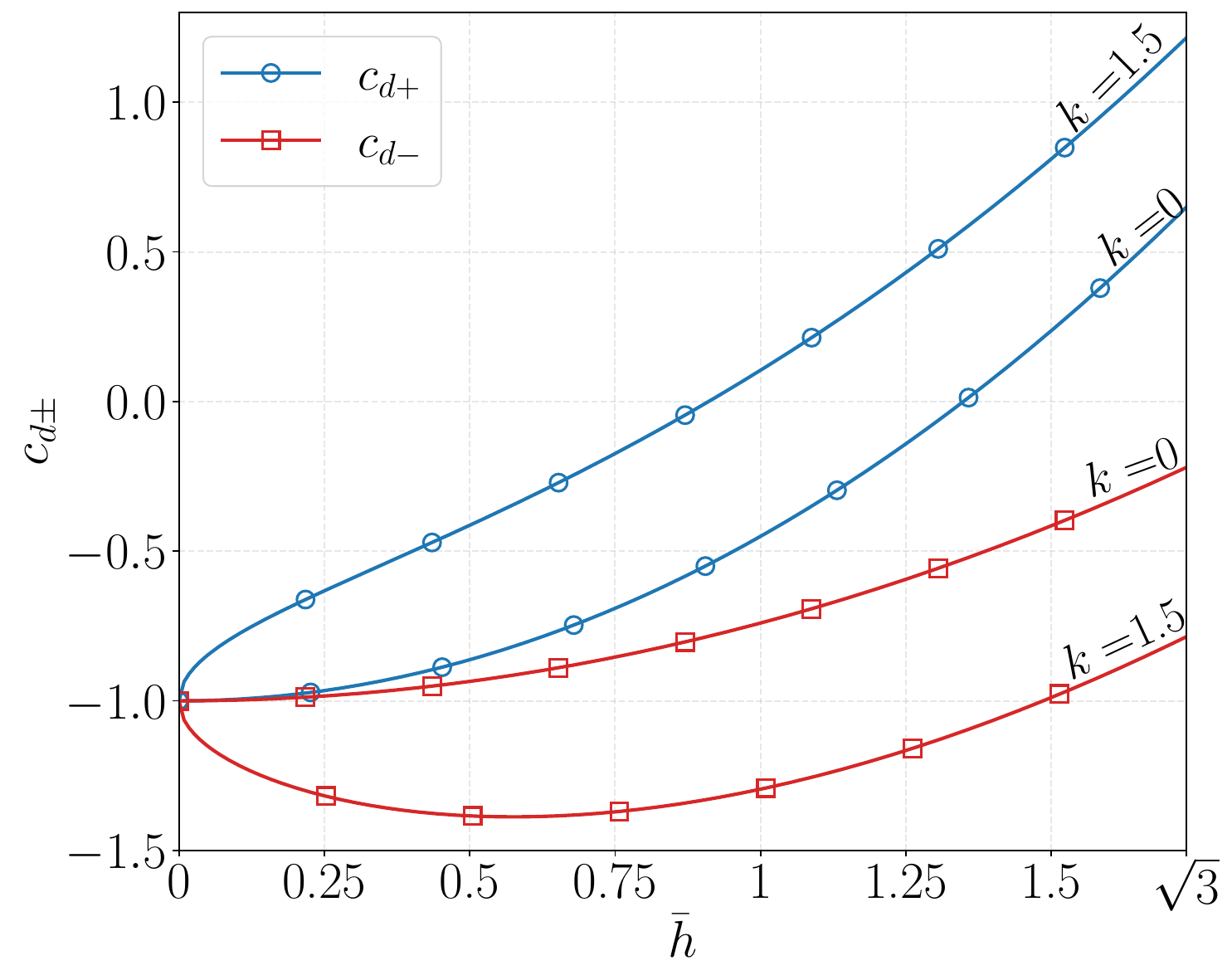}
    \caption{}
    \label{fig:stable_reg_OS_1_2}
  \end{subfigure}
  \hfill
  \begin{subfigure}[b]{0.5\textwidth}
    \includegraphics[width=\textwidth]{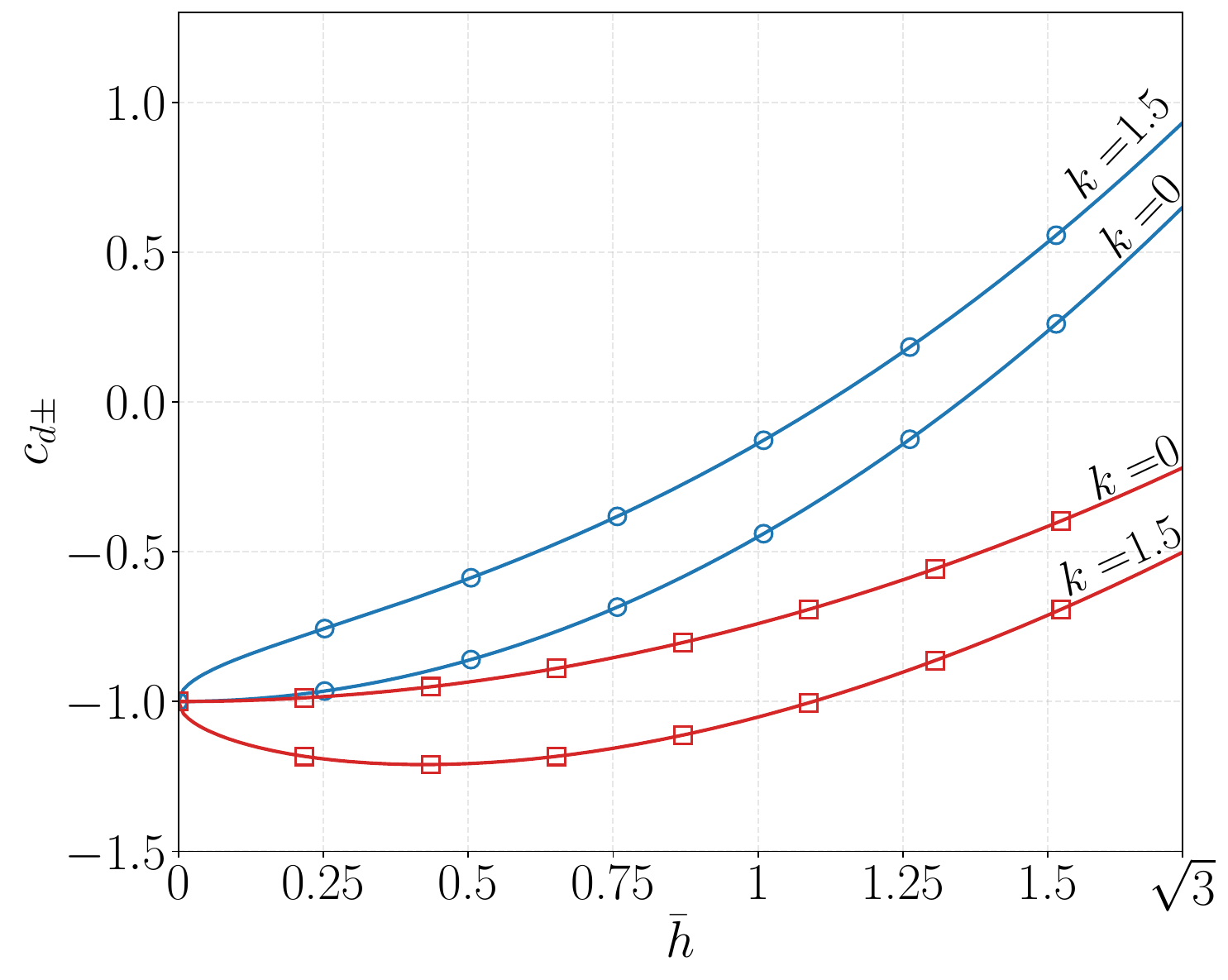}
    \caption{}
  \end{subfigure}
  \begin{subfigure}[b]{0.5\textwidth}
    \includegraphics[width=\textwidth]{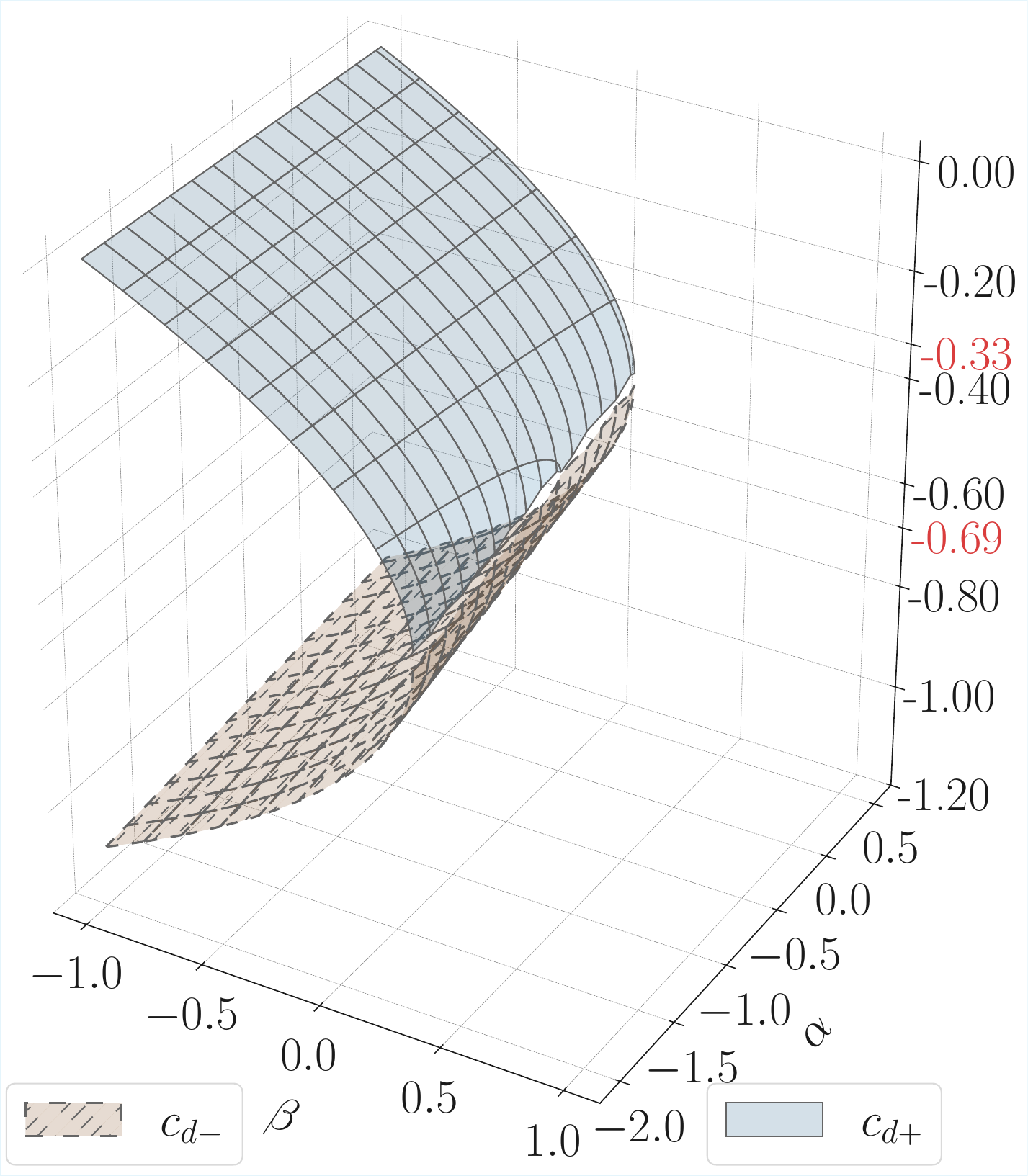}
    \caption{}
    \label{}
  \end{subfigure}
  \hfill
  \begin{subfigure}[b]{0.5\textwidth}
    \includegraphics[width=\textwidth]{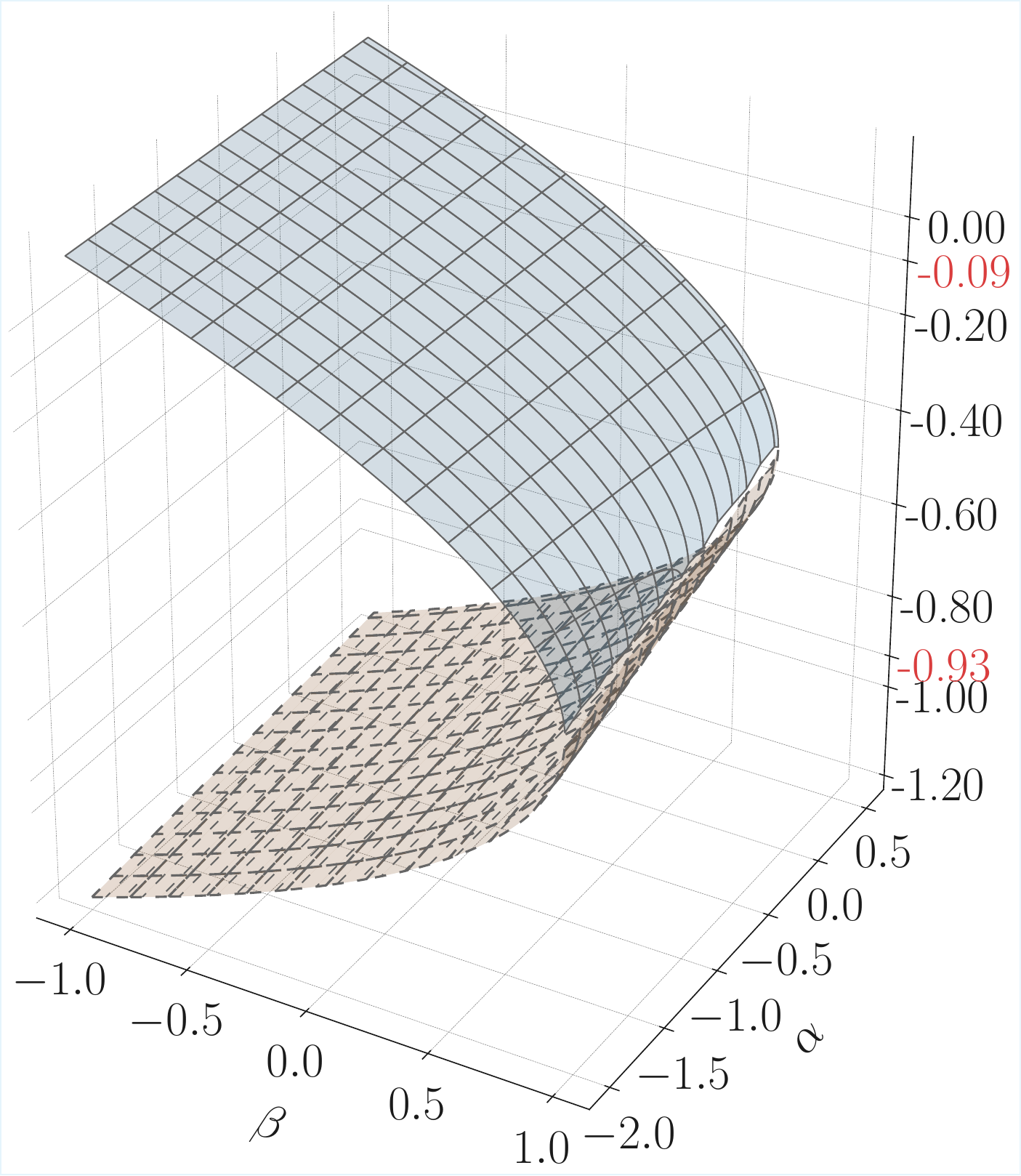}
    \caption{}
  \end{subfigure}
  \caption{Evolution of the two solution of the dynamic velocity $c_{d+}$ (blue line with empty circles) and $c_{d-}$ (red line with empty squares)  as a function of $\bar{h}$ for $k=0$ and $k=1.5$ with (a) $\delta=4$ and (b) $\delta=10$. Dynamic velocities $c_{d\pm}$ as function of the shear $\alpha$ and pressure $\beta$ feedback coefficients and the velocities for $\alpha=\beta=0$ (red ticks on the z axis) for $delta = 4$, $\bar{h}= 1.1$ and (a) $k=0$ and (b) $k=0.8$}
  \label{fig:dyn_vel}
\end{figure}

Figure \ref{fig:kin_vel} shows (a) the kinematic velocity $c_k$ \eqref{eq:kin_dyn_waves} as a function of $\bar{h}$ for different values of $\alpha$ and (b) the critical flat film thickness $\bar{h}_{\text{crs}}$ as a function of $\alpha$ (black line) with a zoom window around the liquid film thickness of interest for our analysis ($\bar{h}=1.1$) with $\alpha = 1.20$ and $\alpha = 1.15$ (red dots) corresponding to the value which will be used later for the nonlinear control simulations. In both graphs, the range $1<\alpha<32/25$ is shaded in light blue. 

As we have seen in \eqref{eq:zeros_hbar_kinematic_vel}, negative values of $\alpha$ increase the kinematic velocity compared to the uncontrolled case ($\alpha = 0$), while positive values of $\alpha$ reduce it. As shown in the zoomed-in region, the values of $\alpha$ used in our nonlinear simulations (presented later) lie on both sides of the green curve, indicating the targeted flat-film thickness. This implies that for $\alpha = 1.15$, kinematic waves propagate downward (in the direction of gravity), while for $\alpha = 1.20$, they propagate upward (against gravity, along the substrate). The highlighted effects of $\alpha$ on $c_k$ are similar to those induced by the coupling between a shearing turbulent gas and a falling film. When the gas flow rate exceeds a threshold, the surface waves reverse direction and flow upward. This suggests that, as also pointed out by \cite{tseluiko2011nonlinear}, the coupling between turbulent stresses and film thickness can be described by a linear relation at least for small-amplitude waves.

The coefficients $\alpha$ and $\beta$ also affect the sign and magnitude of the speed of the dynamical waves. Figure~\ref{fig:dyn_vel} plots the dynamic‐mode velocities $c_{d\pm}$ for $\delta$ equals to (a) 4 and (b) 10 in uncontrolled conditions versus the mean film thickness $\bar{h}$ for two wavenumbers, $k=0$ and $k=1.5$, with (a) $\delta=4$ and (b) $\delta=10$, and in controlled conditions for $\bar{h}=1.1$, $\delta=4$, at (c) $k=0$ and (d) $k=0.8$. In the uncontrolled case, in the long‐wave limit ($k\to0$), both branches stem from $-1$ as $\bar{h}\to0$ and remain negative over most of the $\bar{h}$–range; the upper branch, $c_{d+}$, crosses into positive values only when $\bar{h}>1.25$. At finite wavenumber ($k=1.5$), the trend of both curves changes: $c_{d+}$ attains larger magnitudes than in the long‐wave case and becomes positive in the interval $0.75<\bar{h}<1$, while $c_{d-}$ presents a minimum near $\bar{h}\approx0.6$ and maintains larger absolute values than its long‐wave counterpart. The same qualitative behaviour persists for higher inertia conditions ($\delta=10$), although the divergence between the two branches is less pronounced. The uncontrolled values $(\alpha=\beta=0)$ are marked in red. We observe that variations in $\beta$ have a much stronger effect than those in $\alpha$; as $|\beta|$ increases, the stable band of kinematic-wave speeds narrows. A large negative value of $\beta$ increases the magnitude of both velocities. As $\beta$ becomes positive, the two surfaces converge toward a common value, and beyond a critical $|\beta|$, real eigenvalues cease to exist. Taken together with the mild dependence on $\alpha$ discussed above, these results identify the limiting $(\alpha,\beta)$ combinations that satisfy the linear‐stability criterion.

As described by \cite{smith1990mechanism} for a falling film and \cite{pino2024linear} for a film over a moving substrate, the imbalance of interfacial shear stress in the linearised equations triggers the development of unstable perturbations, fostering the transformation of energy from the base state towards the perturbation kinetic energy. To analyse the influence of $\alpha$, we introduce the external shear feedback into the linearised version of the tangential stress balance at the interface, which gives: 
\begin{equation}
\label{eq:shear_stress_new_interface_0}
D \hat{u}(\hat{h}) = D\Bar{u}(\Bar{h}) + D\Tilde{u}(\Bar{h}) + D^2\Bar{u}(\Bar{h})\Tilde{h} + \alpha\Tilde{h} = 0.    
\end{equation}

Since the base state is shear-free at the interface $D\Bar{u}(\Bar{h})=0$ and knowing that $D^2\Bar{u}(\Bar{h})=-1$, \eqref{eq:shear_stress_new_interface_0} reduces to:
\begin{equation}
\label{eq:shear_stress_new_interface_mod}
D\Tilde{u}(\Bar{h})  + (\alpha - 1)\Tilde{h}=0.    
\end{equation}

A value of $\alpha$ larger than 1 compensates the shear stress of the base state $\bar{h}$, reducing the contribution of the perturbation's gradient $D\Tilde{u}(\Bar{h})$. This impairs the inception mechanism and the development of the \textit{inertial stress} at $O(k)$, stabilising the perturbation. However, a too large value of $\alpha$ amplifies the triggering mechanism leading to large negative shear stress conditions, with the consequence of the development of instabilities as highlighted earlier in the growth rates trends for $\alpha>1.15$ in figure \ref{fig:disp_rel_contr_OS}.

To summarise, we have seen via wave hierarchy arguments and by analysing the structure of the OS equations that the feedback gains $\alpha$ and $\beta$ influence the linear dynamics by broadening the interval of dynamic‐mode speeds—primarily through the pressure actions ($\beta$) and by shifting the kinematic wave speed via shear‐stress feedback through the shear actions ($\alpha$). This interplay creates a window of linear stability that also includes positive $\beta$ values.
\begin{figure}
  \begin{subfigure}[b]{0.5\textwidth}
    \includegraphics[width=\textwidth]{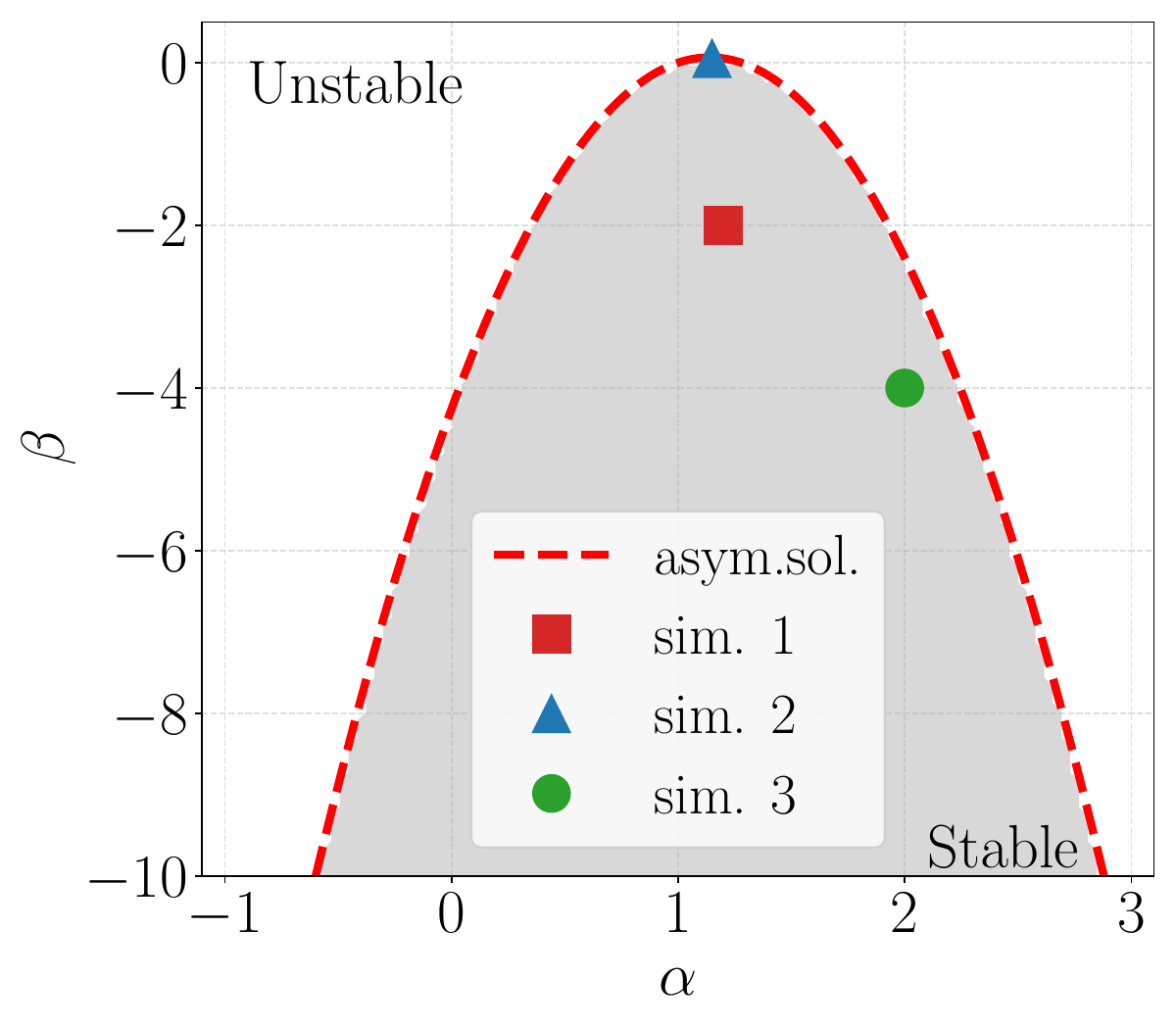}
    \caption{}
    \label{fig:stable_reg_OS_1_3}
  \end{subfigure}
  \hfill
  \begin{subfigure}[b]{0.5\textwidth}
    \includegraphics[width=\textwidth]{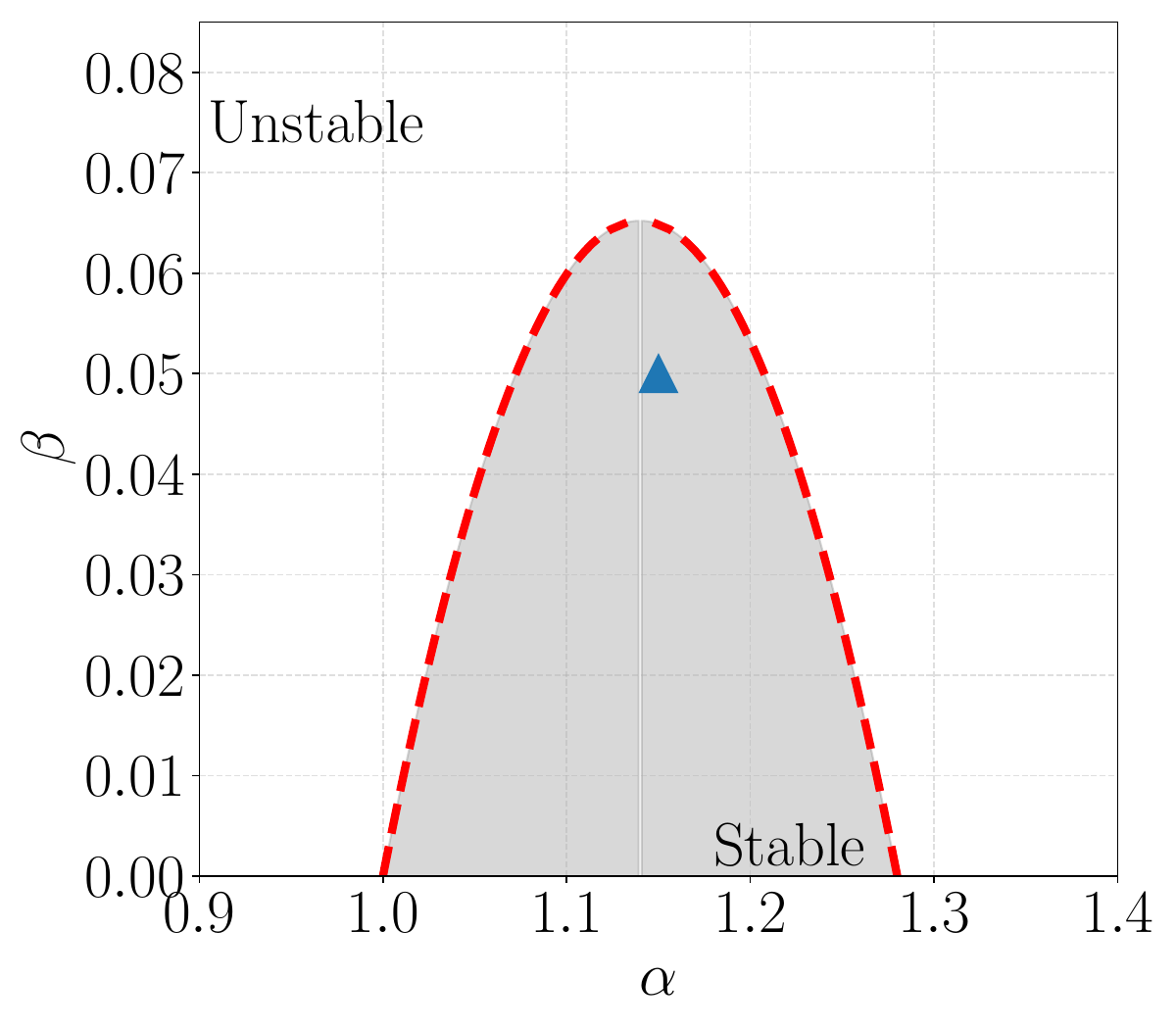}
    \caption{}
  \end{subfigure}
  \caption{(a) Stability region for the space of feedback coefficients $(\alpha,\beta)$, and (b) zoomed view of the positive-$\beta$ half-plane. The region is depicted using slow scaling, obtained from the numerical solution of the Orr-Sommerfeld eigenvalue problem (grey area) and a long-wave asymptotic expansion (red dashed line). The three parameter combinations used in the nonlinear control test cases are also indicated: simulation 1 with $\alpha = 1.2$, $\beta = -2$ (red square), simulation 2 with $\alpha = 1.15$, $\beta = 0.05$ (blue triangle), and simulation 3 with $\alpha = 2$, $\beta = -4$ (green circle).}
  \label{fig:OS_alpha_beta}
\end{figure}

Figure~\ref{fig:OS_alpha_beta} shows (a) the stability regions for $\alpha$ and $\beta$ in the WIBL scaling and (b) a zoom in the positive $\beta$ half-plane, obtained from the Orr–Sommerfeld eigenvalue problem (grey area) and from the long-wave asymptotic analysis at $k=0$ (red dashed line) \eqref{eq:cont_OS_feedback}, for $\bar{h}=1.1$ with the value of the feedback coefficient used in the nonlinear simulations $\alpha=1.2$ and $\beta=-2$ (red square) and $\alpha=1.15$ and $\beta=0.05$ (blue triangle). The grey region is consistent with the asymptotic prediction, thereby confirming the validity of the analysis also for all $k$. The stable region has a bell-shaped structure, centred around $\alpha=1.14$, and extends predominantly into the negative $\beta$ half-plane with a small region in the positive $\beta$ half-plane. 

In practice, the stabilising action of $\beta$ counterbalances the destabilising effect of $\alpha$ outside the interval $[1,32/25]$, thereby broadening the stability domain. Similarly, the stabilising effect of $\alpha$ enlarges the admissible range of $\beta$ compared to its isolated contribution, enabling positive (destabilising) values of $\beta$. 

A comparable stabilising mechanism based on the balance between shear and pressure gradient was reported by \cite{lavalle2019suppression}, who showed that the phase shift between the pressure gradient, the shear stress at the free surface, and the film thickness can lead to linear stability. Their study focused on a shearing gas flowing over a falling liquid film on an incline within a strongly confined channel. Linear stability analysis revealed that the amplitude of the free-surface stresses depends on both the confinement ratio and the gas flow rate. In particular, under strong confinement, the shear stress and pressure gradient act in opposition but remain in phase with the free-surface displacement. In contrast, in our case, the pressure gradient is consistently shifted by $\pi/2$ relative to the free-surface displacement. At the same time, the magnitude of the feedback coefficients determines the stress amplitudes. Furthermore, \cite{vellingiri2015absolute} demonstrated that for a turbulent shearing gas, shear stress can either stabilise or destabilise the film depending on its intensity. Similarly, we observed that excessively large shear stresses render the film linearly unstable, with growth rates exceeding those in the uncontrolled configuration.

\subsection{Nonlinear control case}
\label{subsec:nonlin_control_case}
Moving to the nonlinear test case, we selected three combinations of feedback gains, coloured markers in figure \ref{fig:OS_alpha_beta}, from the region where $\alpha$ and $\beta$ are both stabilising, referred to as the \textit{stable case} (sim.1 with a red square), one from the region of unstable pressure, referred to as \textit{unstable pressure case} (sim.2 with a blue triangle) and one from the region of unstable shear, referred to as \textit{unstable pressure case} (sim.3 with a green circle). 
\begin{figure}
  \begin{subfigure}[b]{0.5\textwidth}
    \includegraphics[width=\textwidth]{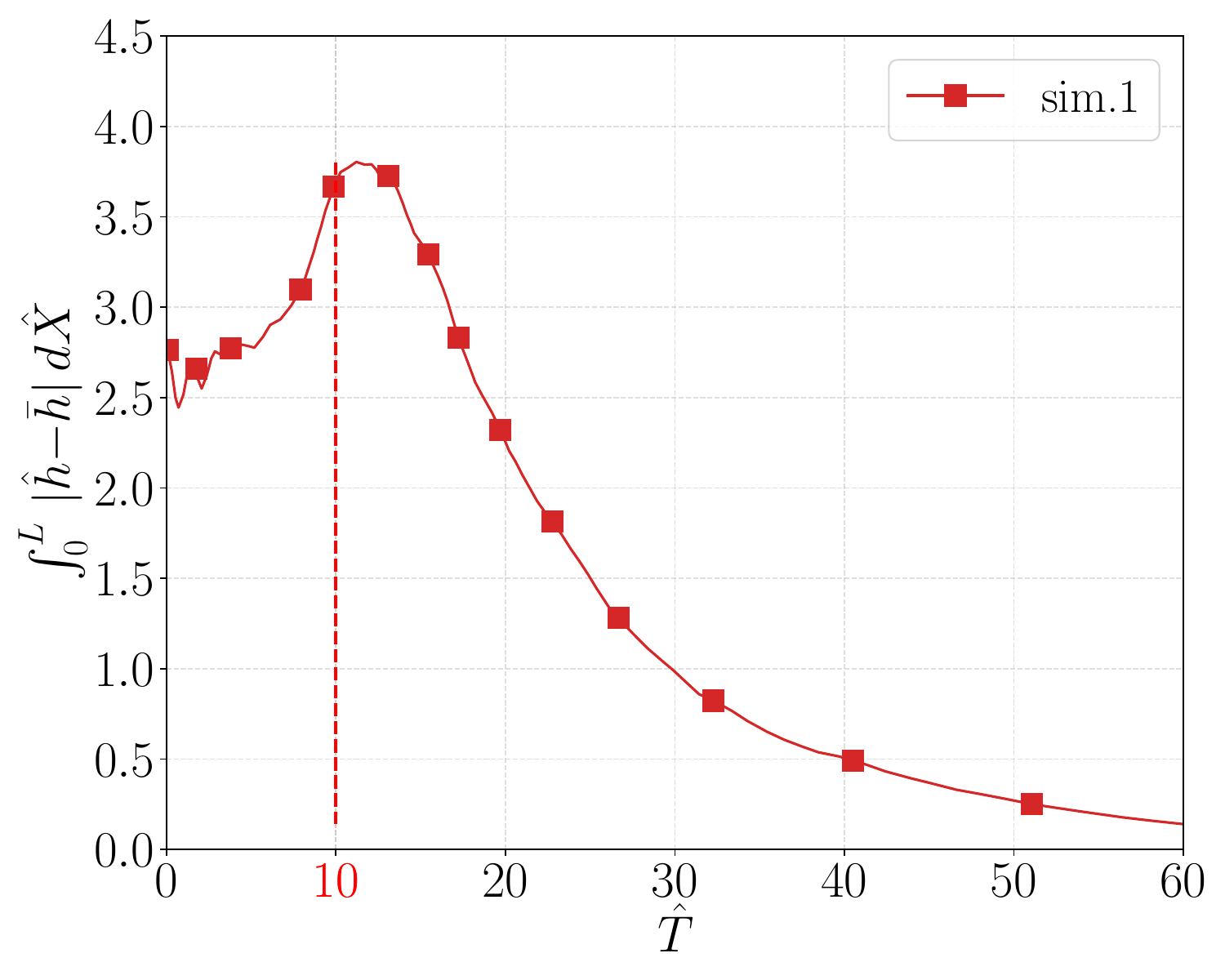} % error_log_ind_stresses_1
    \caption{}
  \end{subfigure}
  \hfill
  \begin{subfigure}[b]{0.5\textwidth}
    \includegraphics[width=\textwidth]{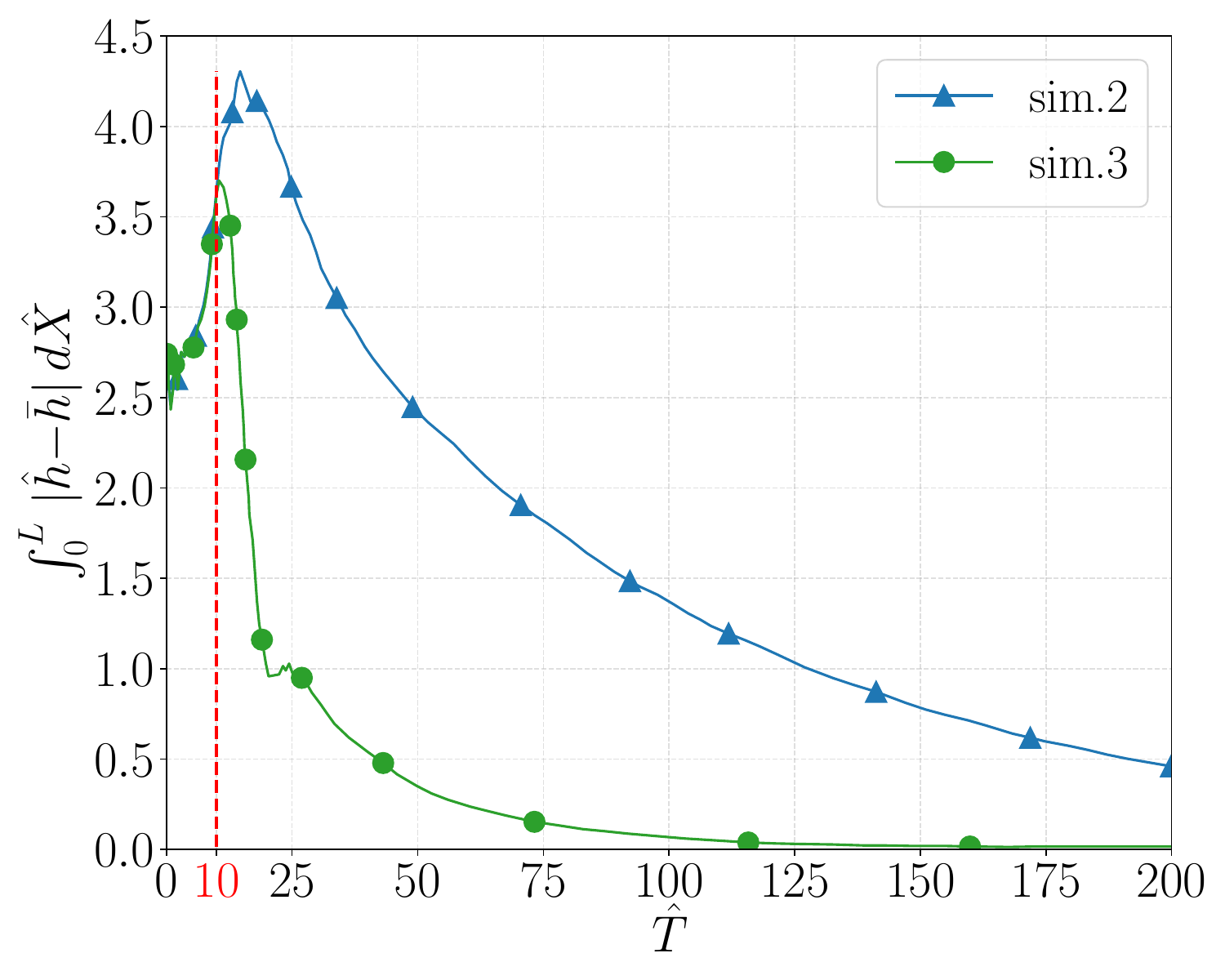} % error_log_ind_stresses_2
    \caption{}
  \end{subfigure}  
  \caption{Evolution of the error between the film thickness $\hat{h}$ and the targeted flat state $\bar{h}$ for independent free-surface stresses with: (a) simulation 1 with $\alpha = 1.2$, $\beta = -2$ (red line with square) and (b) simulation 2 with $\alpha = 1.15$, $\beta = 0.05$ (blue line with triangle), and simulation 3 with $\alpha = 2$, $\beta = -4$ (green line with circle)}
\label{fig:res_nonlinear_control_ind_press_shear_error}
\end{figure}

Because the control actions in the three cases operate on different time scales, the stable simulation was run until $\hat{T}=60$ while the unstable simulations were run until $\hat{T}=200$. Figure~\ref{fig:res_nonlinear_control_ind_press_shear_error} shows the time series of the integrated thickness error for (a) the stable case (red continuous line with squares) and (b) the unstable cases (continuous line with squares for unstable pressure and triangles for unstable shear). In all simulations, the error decreases after the controls are activated at $\hat{T}=10$. By the end of the stable run, the error has fallen below $0.2$; its decay follows a quadratic trend for $\hat{T}<30$ and then becomes approximately linear.

In the unstable cases, the error initially increases upon control application, peaking at approximately $4$ for the unstable pressure case and $3.7$ for the unstable shear case. Following this overshoot, the unstable pressure case exhibits a slow decay that reaches approximately $0.5$ at $\hat{T}=200$. By contrast, in the shear-controlled case, the error plummets to about $1$ at $\hat{T}\approx 22$ after the initial peak and then decays exponentially, reaching values near zero for $\hat{T}>125$.
\begin{figure}
  \begin{subfigure}[b]{0.5\textwidth}
    \includegraphics[width=\textwidth]{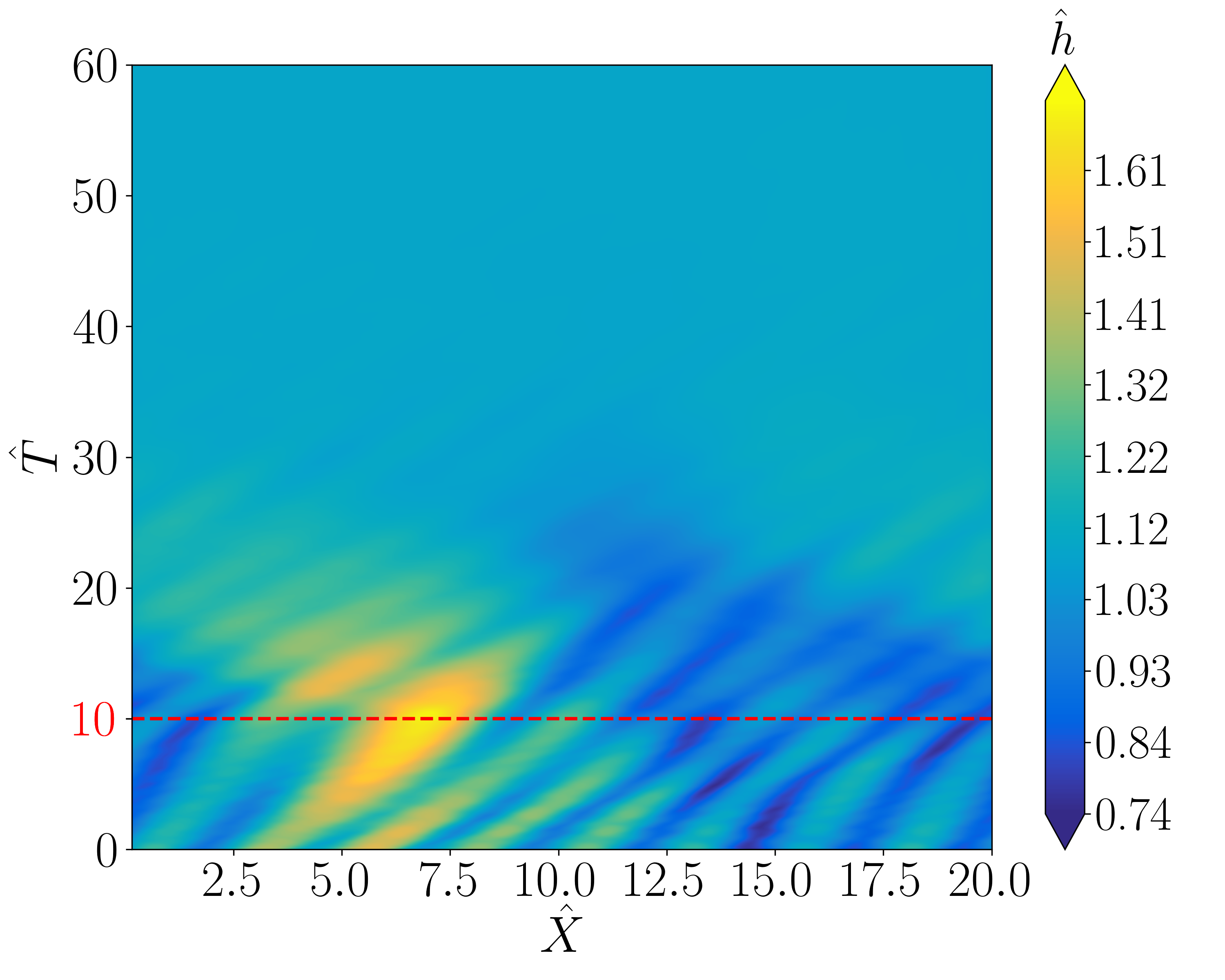}
    \caption{}
  \end{subfigure}
  \hfill
  \begin{subfigure}[b]{0.5\textwidth}
    \includegraphics[width=\textwidth]{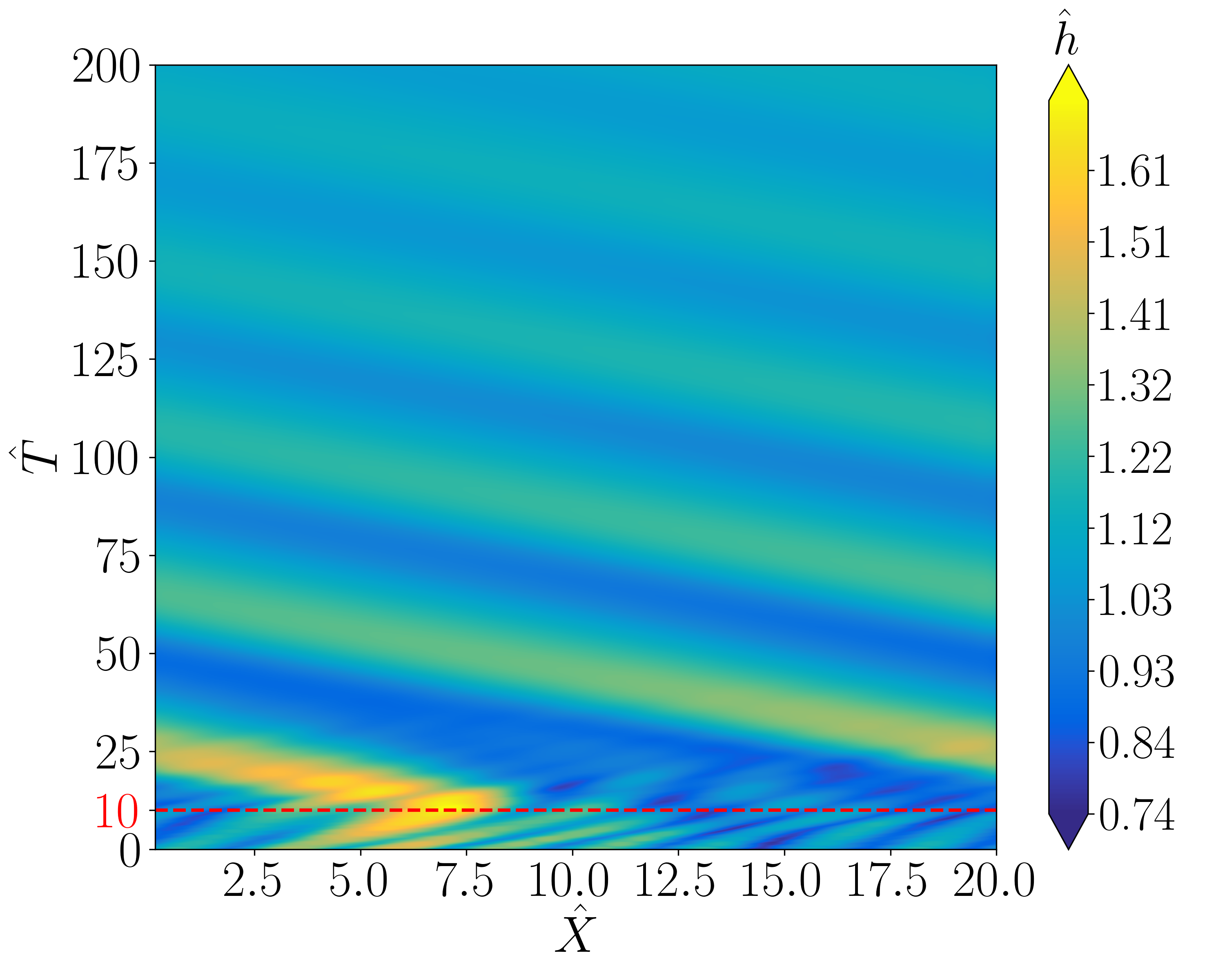}
    \caption{}
  \end{subfigure}
  \begin{subfigure}[b]{0.5\textwidth}
    \includegraphics[width=\textwidth]{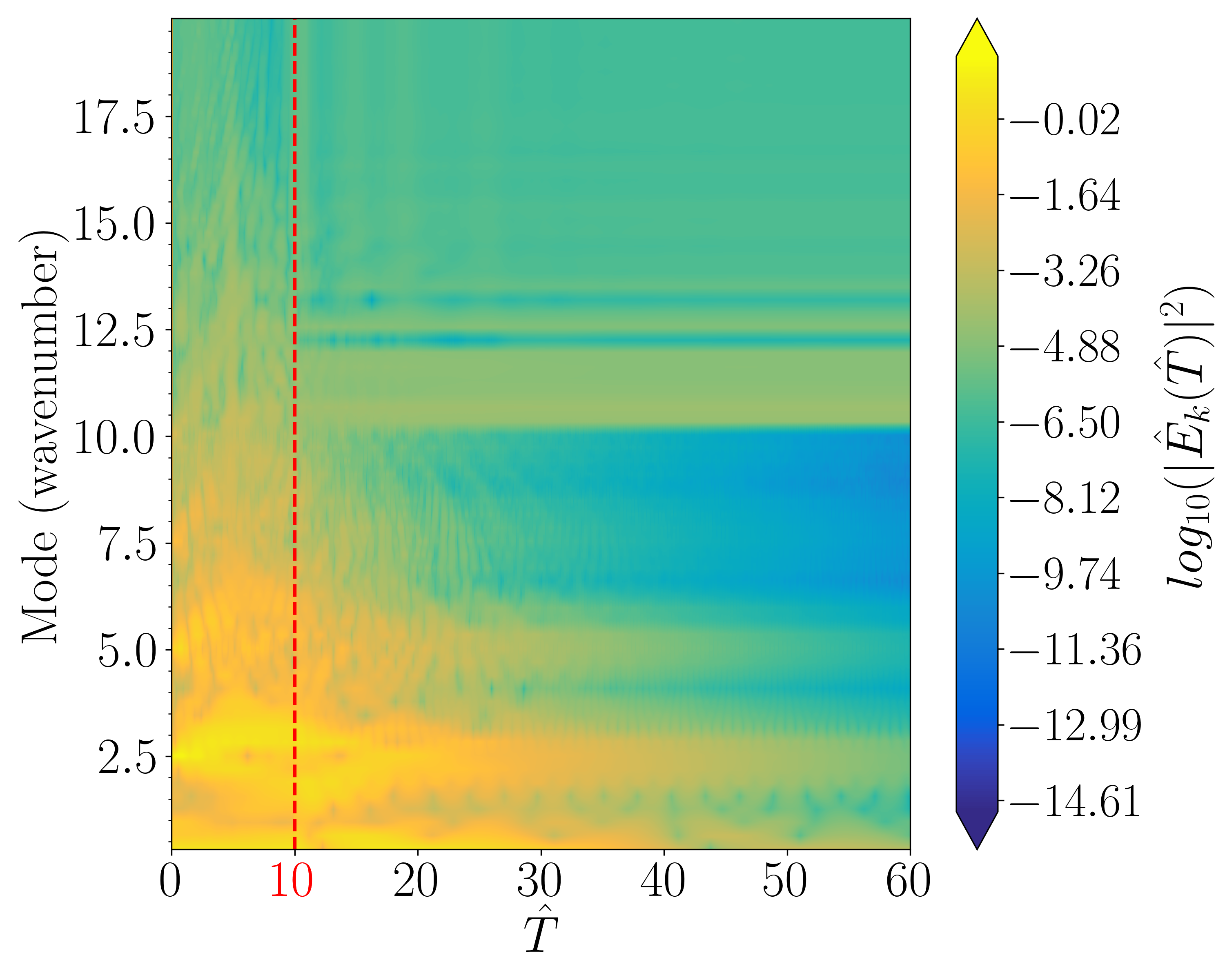}
    \caption{}
  \end{subfigure}
  \hfill
  \begin{subfigure}[b]{0.5\textwidth}
    \includegraphics[width=\textwidth]{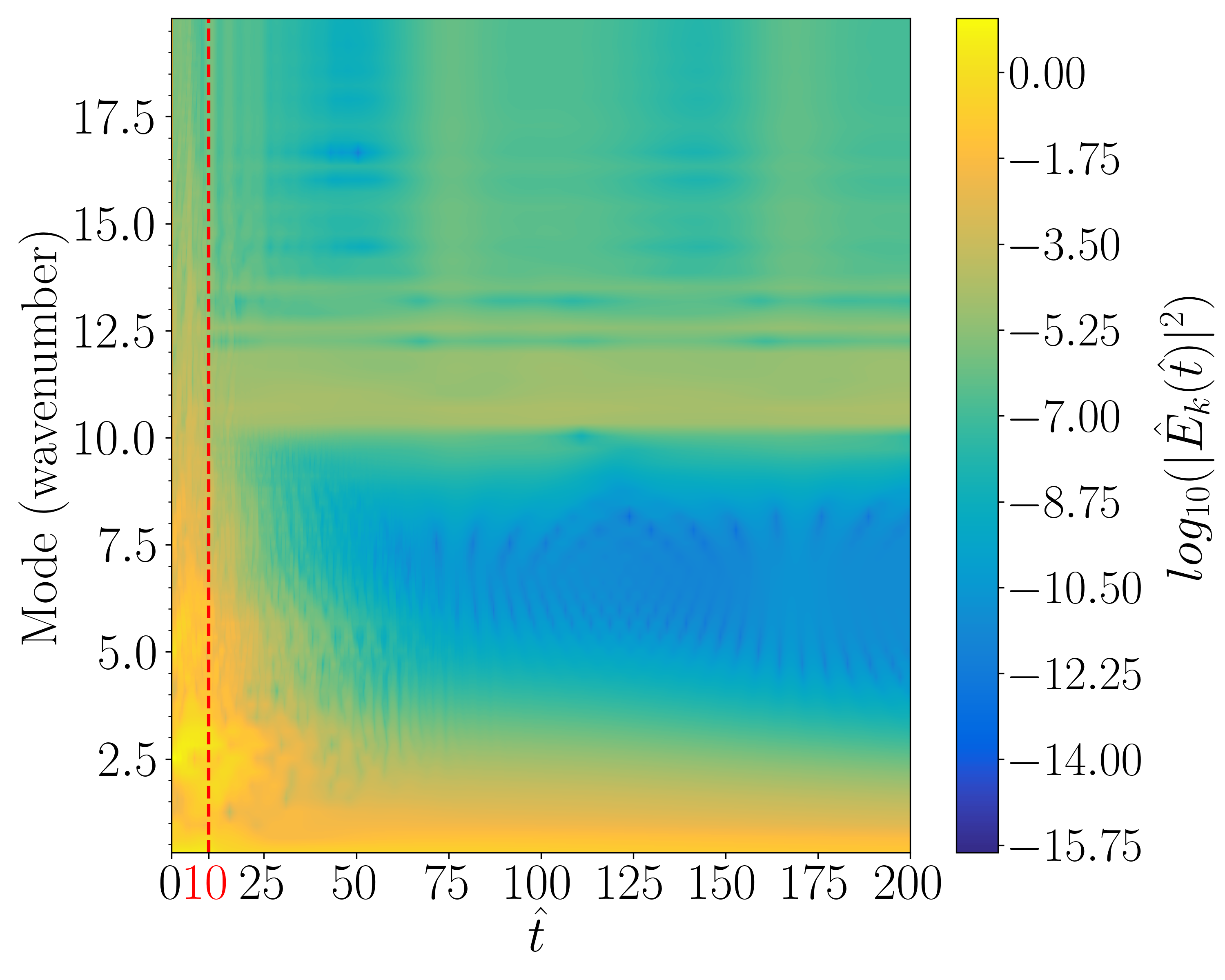}
    \caption{}
  \end{subfigure}
  \caption{Evolution of (a and b) the controlled liquid film and (c and d) the spectral energy during the control test case with $\alpha = 1.2$ and $\beta = -2$ (a and c) and with $\alpha = 1.15$ and $\beta = 0.05$ (b and d).}
\label{fig:res_nonlinear_control_ind_press_shear_colourmaps}
\end{figure}

Figure~\ref{fig:res_nonlinear_control_ind_press_shear_colourmaps} presents spatio‐temporal colour maps of (a,b) the film thickness, $\hat{h}(\hat{x},\hat{T})$, and (c,d) the spectral kinetic‐energy density as a function of wavenumber, for (a,c) the stable case and (b,d) the unstable pressure case case. In both cases, the shear stress shows only weak variations over time, thereby remaining in the range of validity of the assumptions \eqref{eq:ass_ROM}. 

Consistent with the evolution of the error norm, the two control setups follow distinct trajectories. In the stable case, the controller flattens the film by damping wave amplitudes while preserving the original characteristic directions. In the unstable case, for $10<\hat T<25$, short waves are eliminated, and the film goes towards a limit cycle with long-wave harmonics. Under these conditions, the imbalance between shear and pressure gradients reduces the amplitude of the waves over very long timescales as they propagate along new characteristics against gravity. A notable feature of these characteristics is that the wave travels at a much slower speed than its characteristic phase speed under uncontrolled conditions.

The dynamics of the stable and unstable simulations are reflected in the kinetic‐energy spectra: in the stable run, energy is transferred toward low wavenumbers and dissipated there, whereas in the unstable run, energy is first removed from high wavenumbers and then accumulates in the long‐wave modes. This energy redistribution is achieved through active pressure and shear feedback rather than solely through surface tension or viscous dissipation.
\begin{figure}
  \begin{subfigure}[b]{0.5\textwidth}
    \includegraphics[width=\textwidth]{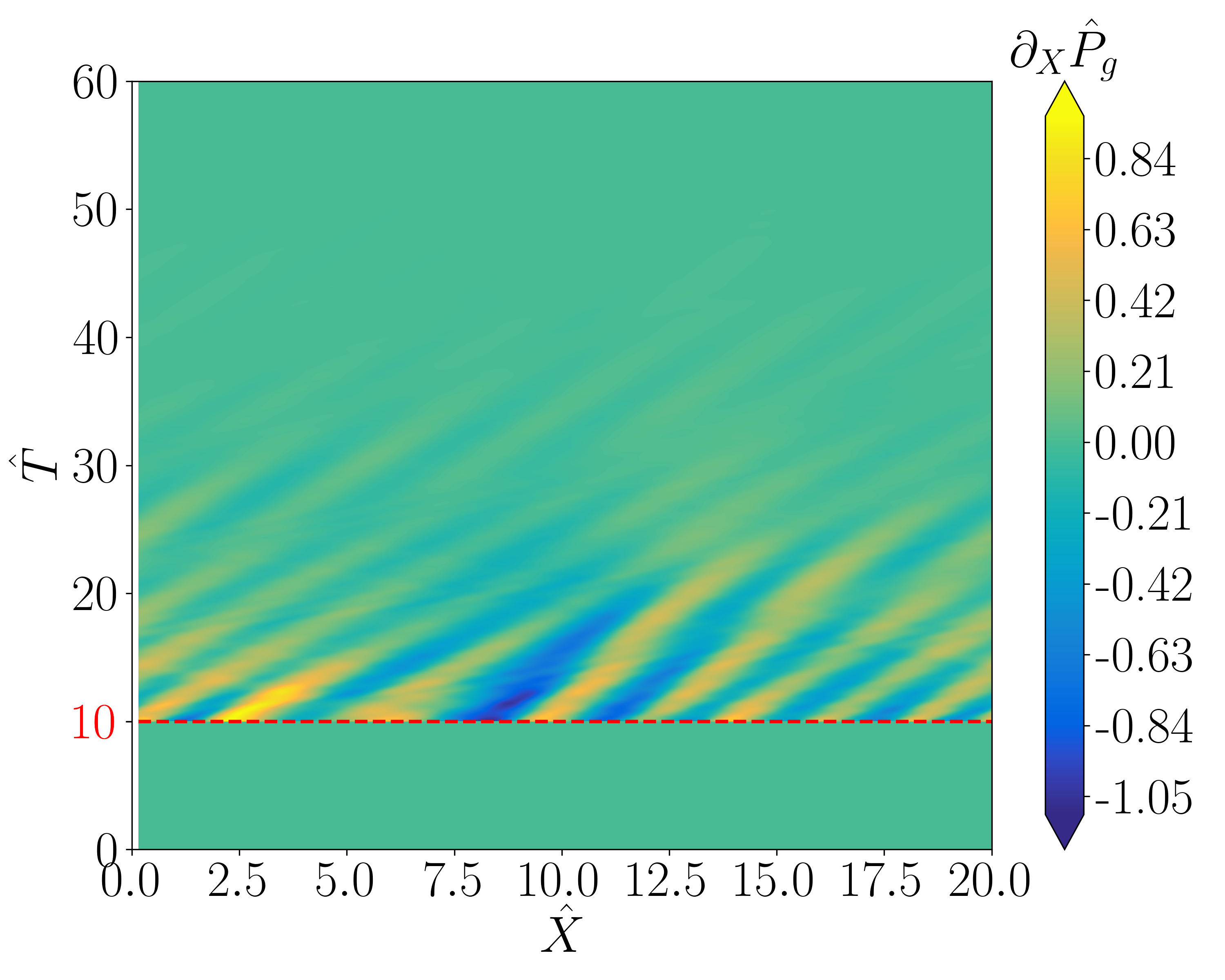}
    \caption{}
  \end{subfigure}
  \hfill
  \begin{subfigure}[b]{0.5\textwidth}
    \includegraphics[width=\textwidth]{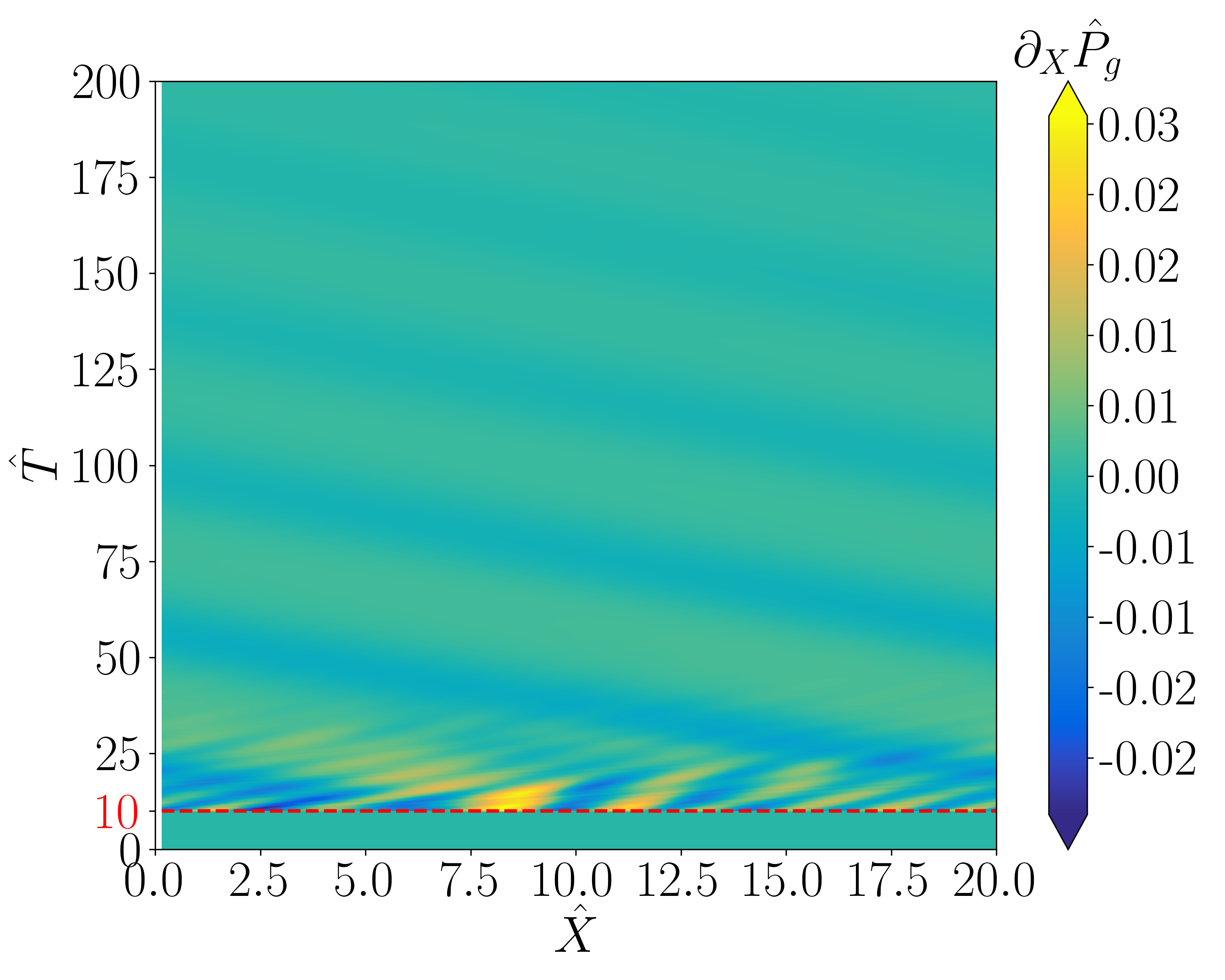}
    \caption{}
  \end{subfigure}
  \begin{subfigure}[b]{0.5\textwidth}
    \includegraphics[width=\textwidth]{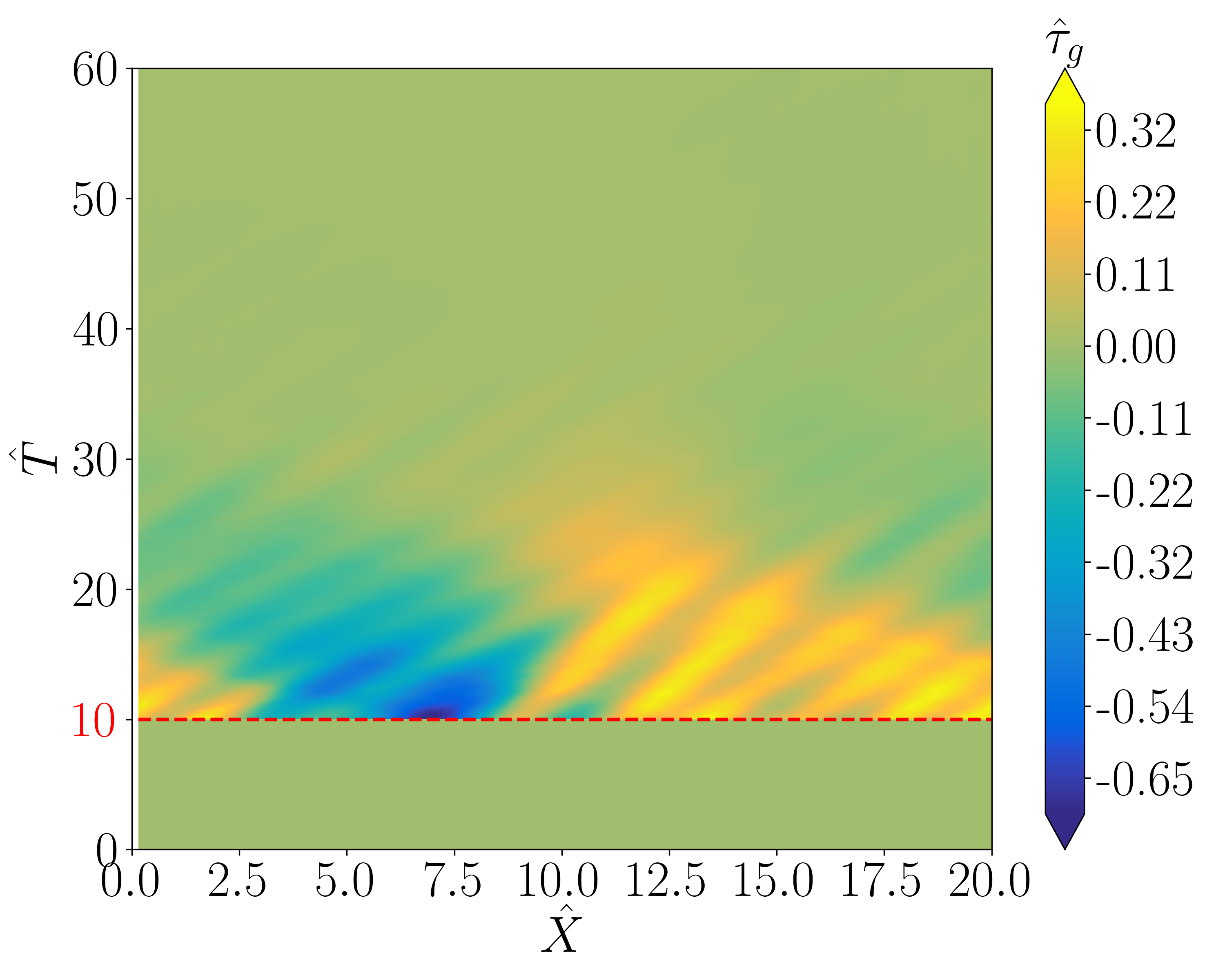}
    \caption{}
  \end{subfigure}
  \hfill
  \begin{subfigure}[b]{0.5\textwidth}
    \includegraphics[width=\textwidth]{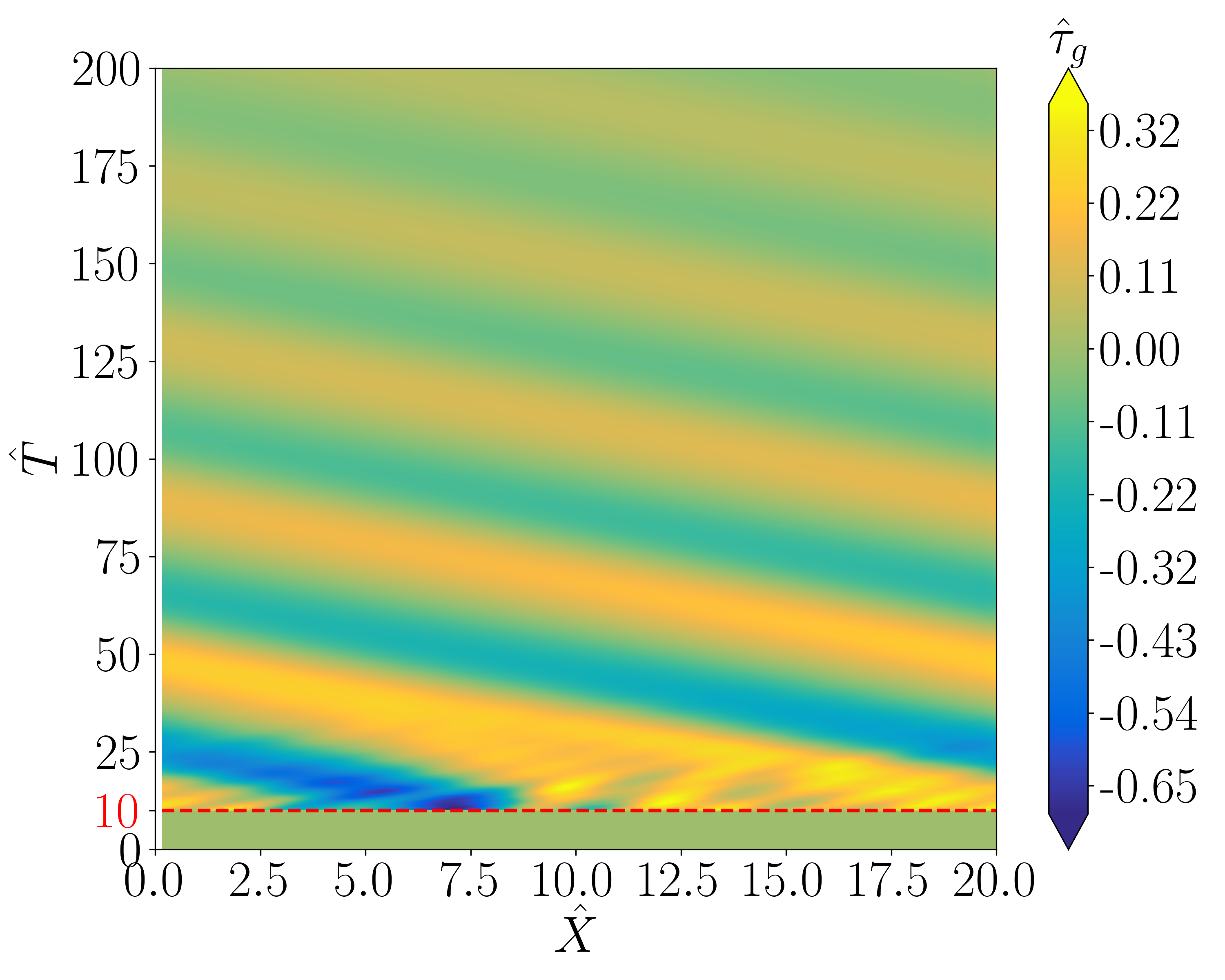}
    \caption{}
  \end{subfigure}
  \caption{Evolution of the nondimensional (a and b) pressure gradient and (b and c) shear stress at the free surface with independent stresses with feedback coefficients (a and c) $\alpha=1.2$ $\beta=-2$ and (b and d) $\alpha=1.15$ and $\beta=0.05$}
  \label{fig:evolution_DPg_taug_ind_stresses}
\end{figure}

Moving to the distribution of stresses at the free surface, figure \ref{fig:evolution_DPg_taug_ind_stresses} shows the evolution of (a and b) the pressure gradient and (c and d) the shear stress at the free surface for (a and c) the stable and (b and d) the unstable pressure cases. A clear difference emerges between the two cases. In the stable case, the pressure gradient is substantially larger than in the unstable case. In contrast, the magnitudes of shear stress are comparable. Hence, in the stable case, the control mechanism is dominated by the pressure gradient. In the unstable case, the control mechanism is instead dominated by shear-stress effects, with only a modest contribution from the pressure gradient. 

From a physical perspective, the pressure-gradient term depends on the gradient of the film-thickness disturbance. It is therefore proportional to the wavenumber $k$, whereas the shear stress term acts at $O(1)$ and scales directly with the disturbance amplitude. Consequently, the two mechanisms act on different scales. As a result, the pressure gradient is more effective at stabilising short, high $k$ disturbances, whereas shear stress is more effective at damping large-amplitude waves.
\begin{figure}
  \begin{subfigure}[b]{0.5\textwidth}
    \includegraphics[width=\textwidth]{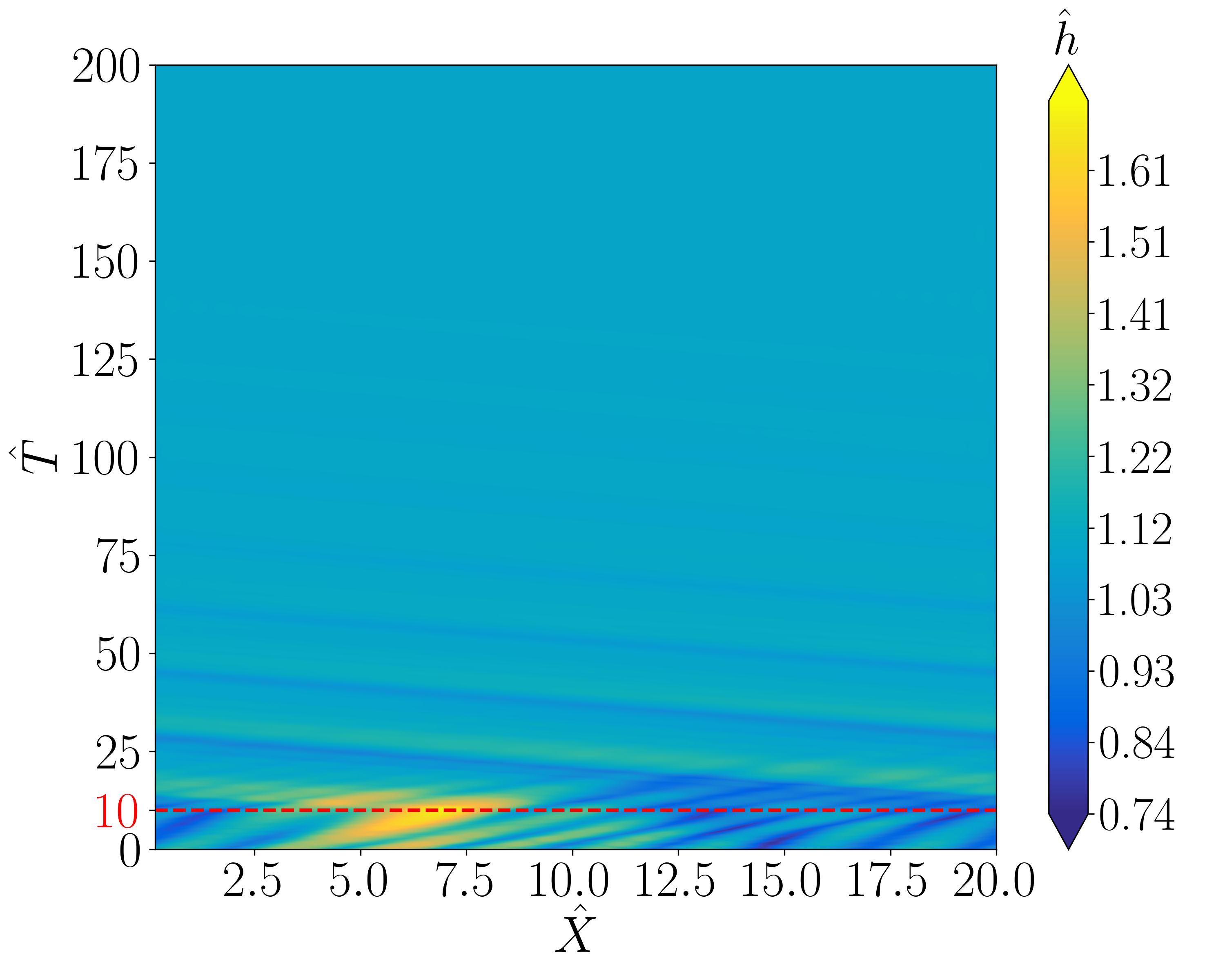}
    \caption{}
  \end{subfigure}
  \hfill
  \begin{subfigure}[b]{0.5\textwidth}
    \includegraphics[width=\textwidth]{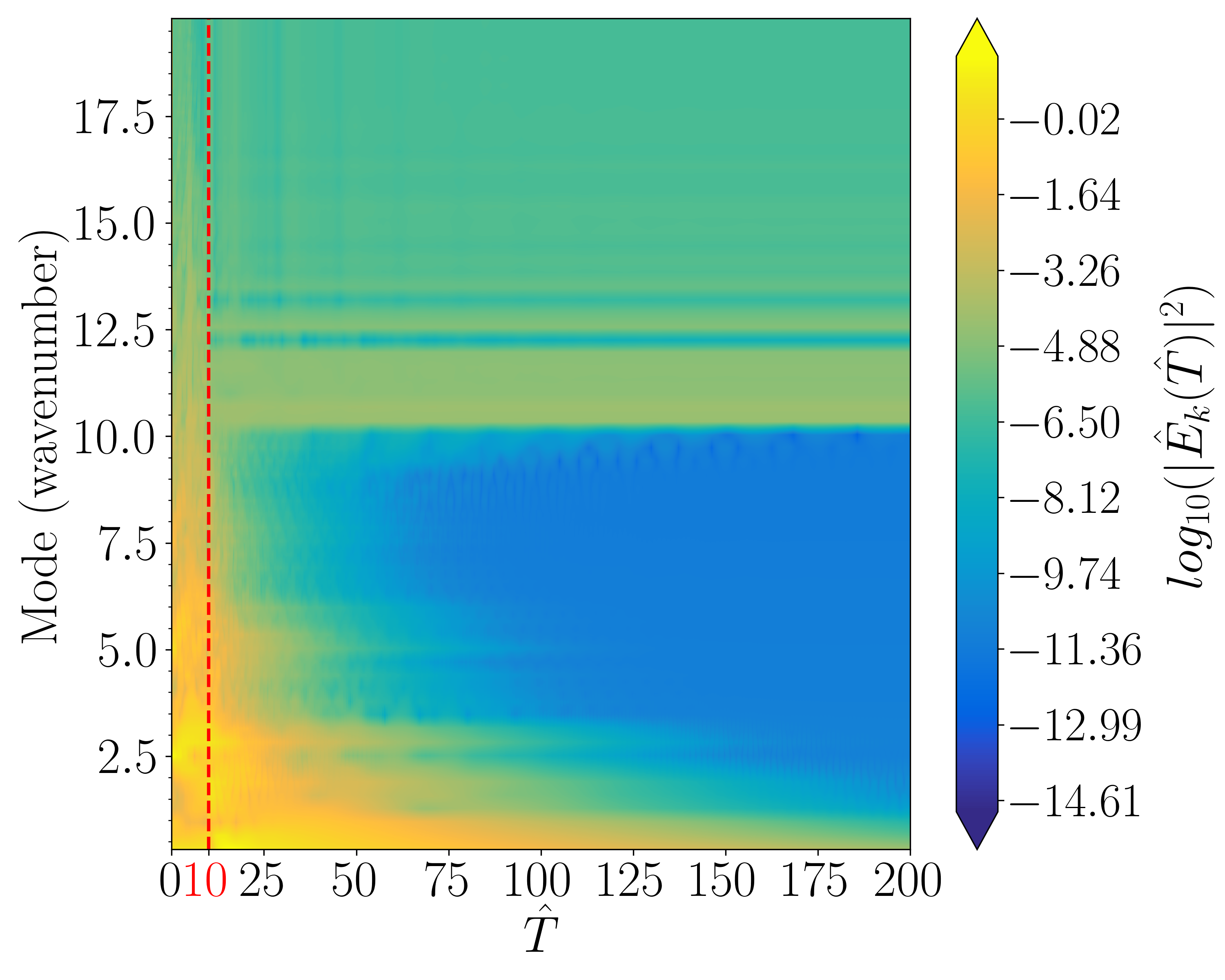}
    \caption{}
  \end{subfigure}
  \begin{subfigure}[b]{0.5\textwidth}
    \includegraphics[width=\textwidth]{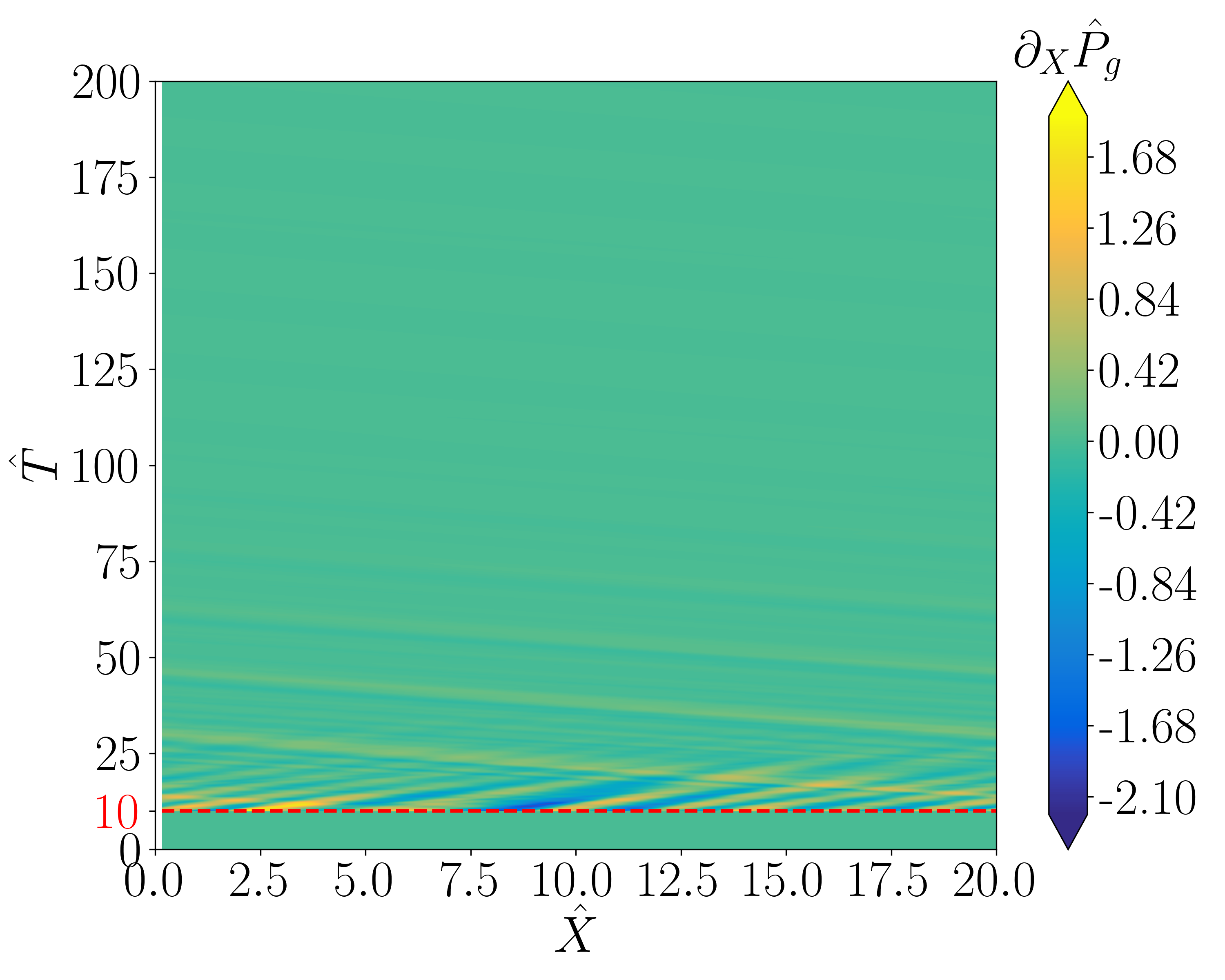}
    \caption{}
  \end{subfigure}
  \hfill
  \begin{subfigure}[b]{0.5\textwidth}
    \includegraphics[width=\textwidth]{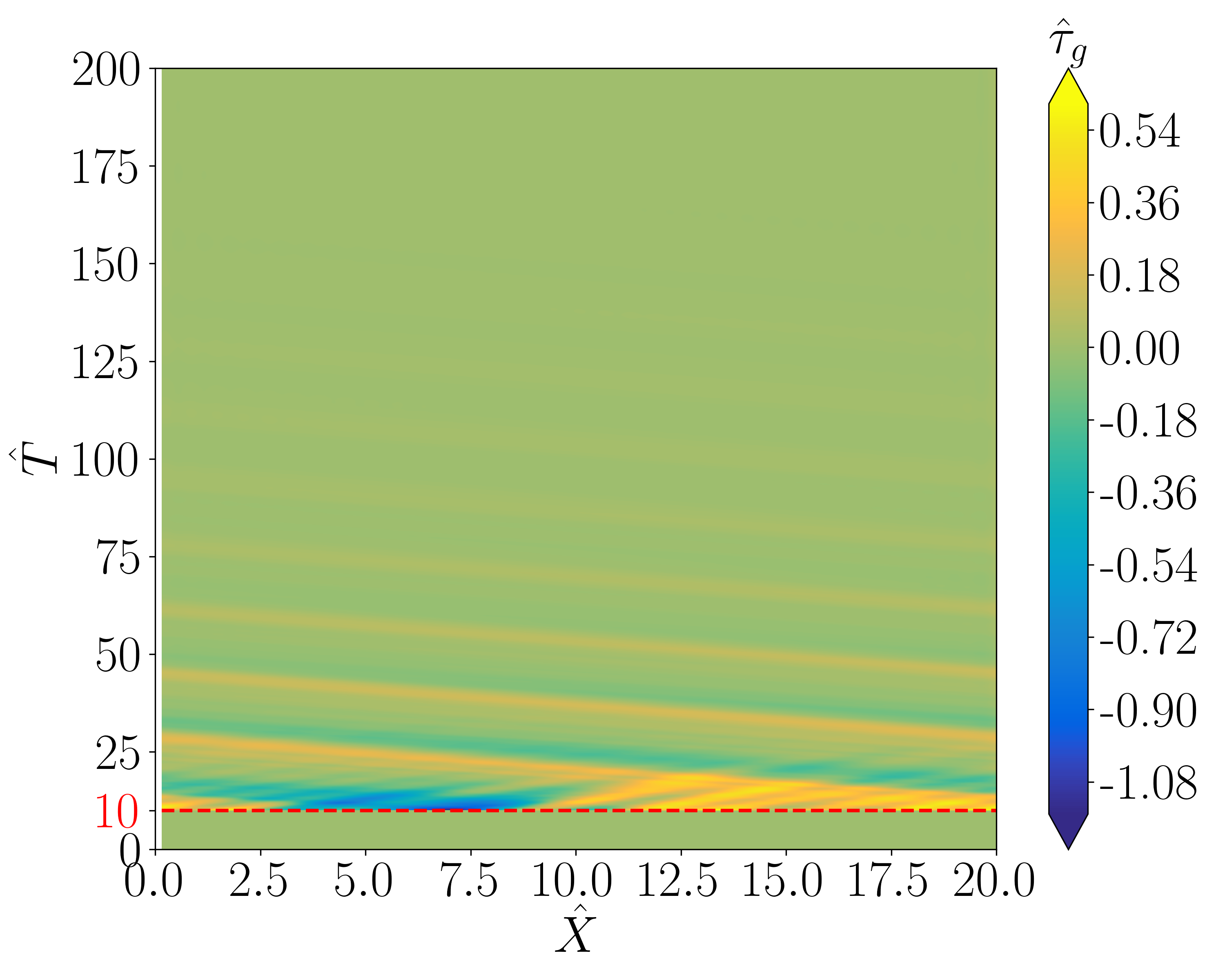}
    \caption{}
  \end{subfigure}
  \caption{Evolution of the nondimensional (a) thickness, (b) kinetic energy, (c) pressure gradient and (d) shear stress at the free-surface for independent shear stresses with feedback coefficients $\alpha=2$ $\beta=-4$.}
  \label{fig:res_stab_press_des_shear}
\end{figure}

A similar balance between shear and pressure-gradient forces also determines the long-term solution in the unstable shear case. Figure \ref{fig:res_stab_press_des_shear} shows the results for the unstable shear case, showing the evolution of (a) the liquid film free thickness $\hat{h}$, (b) the kinetic energy distribution across different wavenumbers, (c) the pressure gradient, and (d) the shear stress. As with destabilising pressure, the liquid film evolves toward a travelling-wave solution dominated by a long wave. However, in this scenario, small-amplitude waves remain, and they appear to propagate in the opposite direction to the large-amplitude wave. This observation is consistent with the earlier analysis of wave speed as a function of the wavenumber $k$ and its dependence on the feedback coefficients. Additionally, the maximum pressure is approximately twice that in the destabilising-pressure case, whereas the shear stress remains of the same order of magnitude. 
\begin{figure}
  \begin{subfigure}[b]{0.5\textwidth}
    \includegraphics[width=\textwidth]{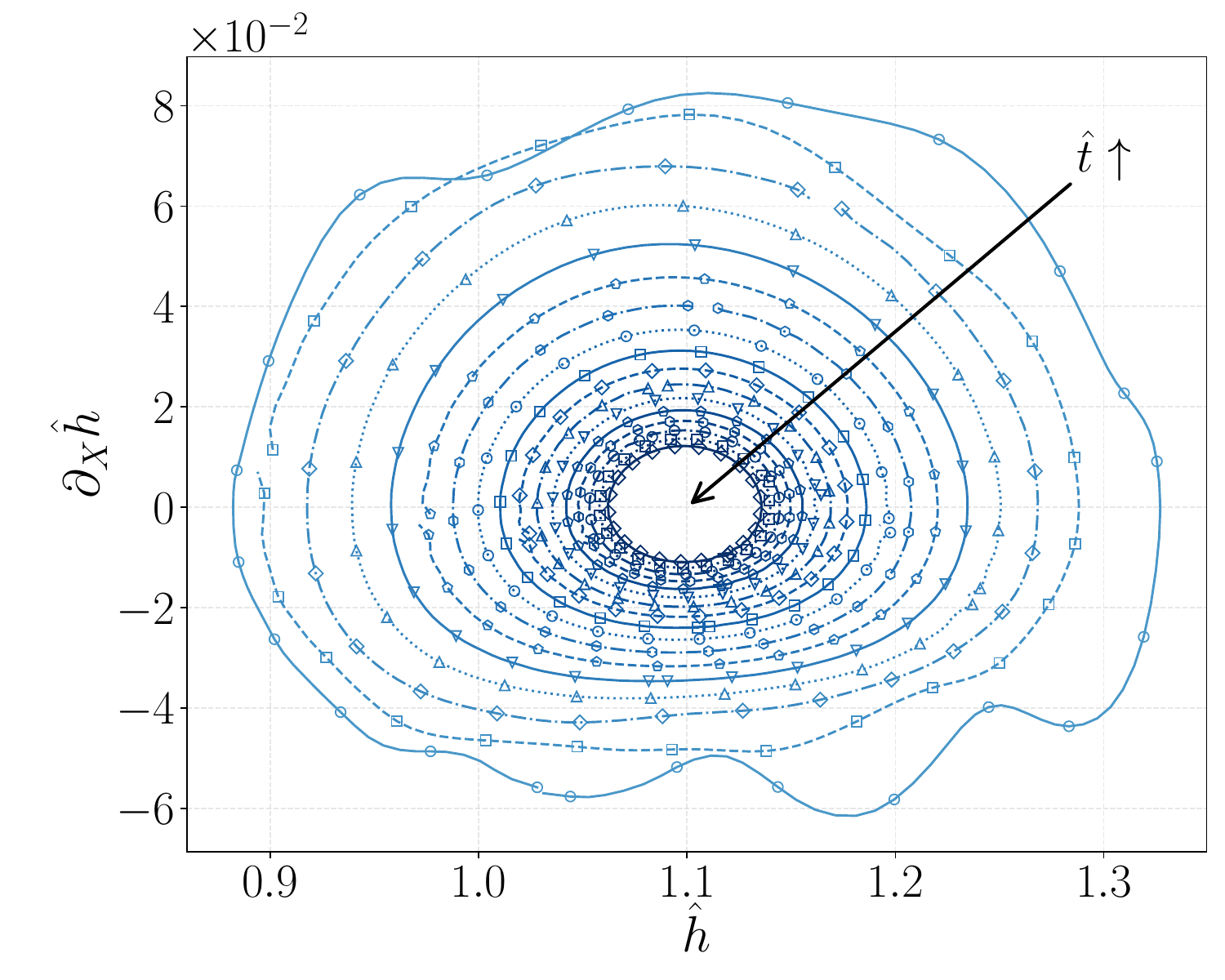}
    \caption{}
  \end{subfigure}
  \hfill
  \begin{subfigure}[b]{0.5\textwidth}
    \includegraphics[width=\textwidth]{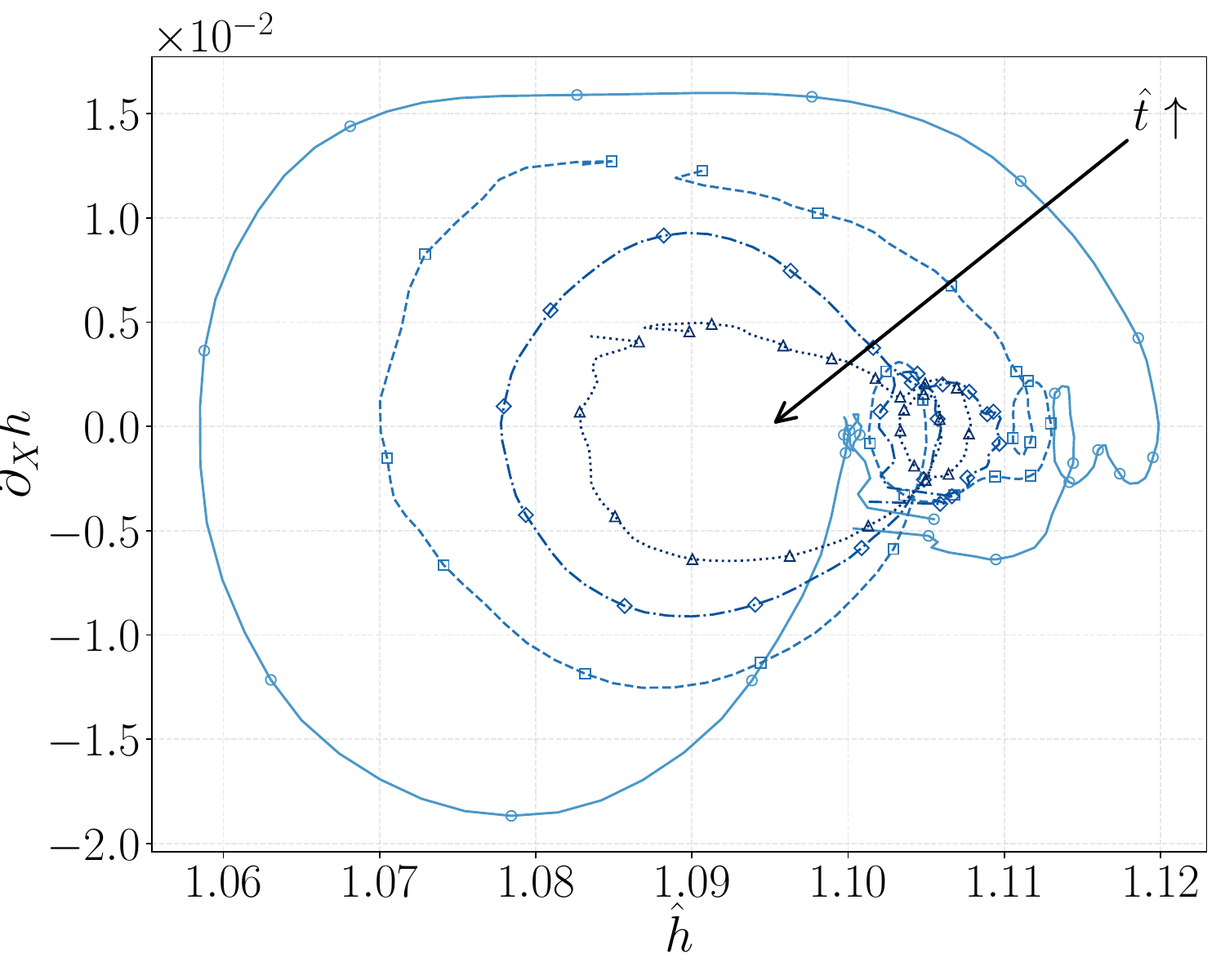}
    \caption{}
  \end{subfigure}
  \caption{Evolution of the controlled liquid film in the space ($\hat{h},\partial_X\hat{h}$) at different instant of time (colours going from light to dark with different makers and line styles) for the case with independent shear and pressure distribution with destabilising (a) pressure ($\alpha=1.15$ and $\beta=0.05$) and (b) shear ($\alpha=2$ and $\beta = -4$).}
  \label{fig:evolution_h_Dh_ind_stresses}
\end{figure}

As we have seen, in both unstable cases, the system reaches a saturated, nonlinear travelling-wave solution. As the wave amplitude increases under the destabilising action of pressure or shear feedback, the corresponding pressure or shear feedback strengthens and begins to oppose further growth. This competition leads to a quasi-equilibrium state characterised by progressively longer wavelengths and reduced amplitudes. The selected dominant harmonic stems from the combined effect of the feedback coefficient on the linear and nonlinear parts of the WIBL operator. In particular, the feedback terms modify the linear dispersion relation (producing a band-pass with a peak at a preferred wavenumber) and alter the nonlinear saturation mechanism. Together, these effects favour a single, mono-harmonic travelling wave when the balance is achieved.

To have a better perspective on this balance and on the travelling wave solutions, figure~\ref{fig:evolution_h_Dh_ind_stresses} shows the evolution of the controlled film in the $(\hat{h},\partial_X\hat{h})$ phase space at different times (colour going from light to dark with various markers and line styles) for the two destabilising scenarios: (a) pressure-driven and (b) shear-driven. In the unstable-pressure case, the system transitions to a single dominant frequency, depicted in phase space as a single closed loop that gradually shrinks in amplitude while retaining its shape. By contrast, in the shear-driven case, the solution does not converge to a single travelling wave: several waves, mostly with small wavenumbers, persist and appear as multiple distinct loops in phase space. This difference suggests that, under destabilising pressure, the dynamics are dominated by the interplay among gravity, induced pressure gradients, shear stress, and viscous dissipation. In contrast, for destabilising shear, the influence of surface tension is more pronounced, producing the small, distorted loops visible in the phase portrait and allowing multi-modal behaviour.

From a broader perspective, in the context of a falling film sheared by a turbulent gas \cite{tseluiko2011nonlinear}, the slowly decaying amplitude observed here in the unstable case resembles the \emph{flooding} phenomenon, characterised by large-amplitude, nearly stationary surface waves and a reduction in phase speed. The emergence of almost-stationary, large-amplitude waves in our simulations with unstable feedback coefficients strengthens the connection between our results and the coupled dynamics of a turbulent gas interacting with a deformable free surface.
\section{Conclusion}
\label{sec:conclusion}
We investigated the feedback control of 2D finite-amplitude waves in liquid films flowing over a moving substrate, modulating the free-surface pressure and shear-stress distributions. The feedback control coefficients were derived analytically from the linearised governing equations and tested in a regulation problem involving a finite-amplitude wave simulated with a WIBL model.

We found a stability range for the feedback coefficients, with a subregion in which both shear and pressure exert stabilising effects, and another in which either shear or pressure exerts a destabilising effect. Tests on finite-amplitude control with representative coefficient pairs in both regions demonstrated that perturbations were effectively controlled and that the film returned to its flat state within a finite time. When pressure or shear acted in a destabilising manner, the film entered a slowly decaying travelling-wave limit cycle, reflecting the imbalance between stabilising shear stresses on the thickness and destabilising pressure gradients. The saturated wave solution features a single dominant long-wave harmonic for destabilising pressure and multiple harmonics for destabilising shear. Wave-hierarchy arguments explain this mechanism: shear alters the phase speeds of both kinematic and dynamic waves, whereas pressure primarily affects the dynamic modes, with implications for the control of absolute versus convective instabilities.

Several directions need further investigation. Coupling thin liquid films with a turbulent shear gas is a key extension that enables turbulence statistics to serve as time-dependent control parameters and incorporates multiphysical effects, such as thermal gradients and oxidation, relevant to industrial coating applications. A fundamental open problem is the nonlinear interaction between free-surface stresses and turbulent gas forcing, which governs both control authority and potential heat-transfer optimisation. Feedback coefficients linking shear-stress distributions to film displacement can provide a compact characterisation of the coupled air–liquid dynamics. Finally, while carefully designed white-box feedback laws stabilise thin films in idealised settings, practical constraints on efficiency, robustness, and scalability highlight the need for data-assisted controllers for resilient, real-time operation.

\begin{acknowledgments}
F.Pino was supported by an F.R.S.-FNRS FRIA grant, and Arcelor-Mittal funded his PhD research project. B.Scheid is Research Director at F.R.S.-FNRS.
\end{acknowledgments}

\bibliographystyle{jfm}
\bibliography{Fabio_et_al_2024}

\end{document}